\documentclass[sigconf, nonacm]{acmart}

\newcommand\vldbdoi{XX.XX/XXX.XX}
\newcommand\vldbpages{XXX-XXX}
\newcommand\vldbvolume{19}
\newcommand\vldbissue{1}
\newcommand\vldbyear{2026}
\newcommand\vldbauthors{\authors}
\newcommand\vldbtitle{\shorttitle} 
\newcommand\vldbavailabilityurl{https://github.com/ZJU-DAILY/Allan-Poe}
\newcommand\vldbpagestyle{plain} 

\usepackage{graphicx}
\usepackage{subfigure}
\usepackage{multirow}
\usepackage{amsthm}
\usepackage[explicit]{titlesec}
\usepackage[noend, linesnumbered, ruled, lined]{algorithm2e}
\usepackage{hyperref}
\usepackage{balance}
\usepackage{color}
\usepackage{pifont}
\usepackage{amsthm} 
\usepackage{amsmath}
\newtheorem{theorem}{Theorem}
\newtheorem{definition}{Definition}
\usepackage{flushend}
\usepackage{bm}
\usepackage{array}
\usepackage{setspace}

\begin{document}
\title{All-in-one Graph-based Indexing for Hybrid Search on GPUs}

\author{Zhonggen Li}
\affiliation{%
  \institution{Zhejiang University}
}
\email{zgli@zju.edu.cn}

\author{Yougen Li}
\affiliation{%
  \institution{Zhejiang University}
}
\email{yougenli@zju.edu.cn}

\author{Yifan Zhu}
\affiliation{
  \institution{Zhejiang University}
}
\email{xtf_z@zju.edu.cn}

\author{Congcong Ge}
\affiliation{
    \institution{Zhejiang University}
}
\email{gcc@zju.edu.cn}

\author{Zhaoqiang Chen}
 \affiliation{
  \institution{Huawei Cloud}
}
\email{chenzhaoqiang1@huawei.com}

\author{Yunjun Gao}
\affiliation{
  \institution{Zhejiang University}
}
\email{gaoyj@zju.edu.cn}

\begin{abstract}
  Hybrid search has emerged as a promising paradigm that combines lexical and semantic retrieval, enhancing accuracy for applications such as recommendations, information retrieval, and Retrieval-Augmented Generation. 
  However, existing methods are constrained by a trilemma: they sacrifice flexibility for efficiency, suffer from accuracy degradation, or incur prohibitive storage overhead for flexible combinations of retrieval paths. 
  
  This paper introduces \textsf{Allan-Poe}, a novel \textbf{\underline{All}}-in-one gr\textbf{\underline{a}}ph i\textbf{\underline{n}}dex accelerated by G\textbf{\underline{P}}Us f\textbf{\underline{o}}r \textbf{\underline{e}}fficient hybrid search. 
  We first analyze the limitations of existing retrieval paradigms and extract key design principles for an effective hybrid index. 
  Guided by the principles, we architect a unified graph-based index that flexibly integrates three retrieval paths—dense vector, sparse vector, and full-text—within a single, cohesive structure. 
  To enable efficient construction, we design a GPU-accelerated pipeline featuring a warp-level hybrid distance kernel, RNG-IP joint pruning, and keyword-aware neighbor recycling. 
  For query processing, we introduce a dynamic fusion framework that supports any combination of retrieval paths and weights without index reconstruction, flexibly leveraging logical structures from the knowledge graph to resolve complex multi-hop queries. 
  Extensive experiments on 6 real-world datasets demonstrate that \textsf{Allan-Poe} achieves superior end-to-end query accuracy and outperforms state-of-the-art methods by 1.5$\times$-186.4$\times$ in throughput, while significantly reducing storage overhead. 
\end{abstract}

\maketitle
\vspace{-2mm}
\pagestyle{\vldbpagestyle}
\begingroup\small\noindent\raggedright\textbf{PVLDB Reference Format:}\\
\vldbauthors. \vldbtitle. PVLDB, \vldbvolume(\vldbissue): \vldbpages, \vldbyear.
\href{https://doi.org/\vldbdoi}{doi:\vldbdoi}
\endgroup
\begingroup
\renewcommand\thefootnote{}\footnote{\noindent
This work is licensed under the Creative Commons BY-NC-ND 4.0 International License. Visit \url{https://creativecommons.org/licenses/by-nc-nd/4.0/} to view a copy of this license. For any use beyond those covered by this license, obtain permission by emailing \href{mailto:info@vldb.org}{info@vldb.org}. Copyright is held by the owner/author(s). Publication rights licensed to the VLDB Endowment. \\
\raggedright Proceedings of the VLDB Endowment, Vol. \vldbvolume, No. \vldbissue\ %
ISSN 2150-8097. \\
\href{https://doi.org/\vldbdoi}{doi:\vldbdoi} \\
}\addtocounter{footnote}{-1}\endgroup

\vspace{-1.5mm}

\ifdefempty{\vldbavailabilityurl}{}{
\vspace{.3cm}
\begingroup\small\noindent\raggedright\textbf{PVLDB Artifact Availability:}\\
The source code, data, and/or other artifacts have been made available at \url{\vldbavailabilityurl}.
\endgroup
}
\vspace{-2mm}

\section{Introduction}
\label{sec:intro}

Recent advancements in vector databases have substantially improved the accuracy and efficiency of dense vector retrieval~\cite{pan2024survey,gao2024rabitq,cai2024navigating}. State-of-the-art approximate nearest neighbor search algorithms now consistently achieve over 99\% recall for top-$k$ neighbor searches. 
However, the end-to-end accuracy—that is, the accuracy of retrieved documents rather than vector similarity—remains limited, lagging behind the vector recall by 10\%-30\%~(\S\ref{subsec:motivation}). 
This discrepancy arises because the minimal distance between query and answer vectors in the embedding space does not guarantee their semantic relevance in natural language~\cite{zhang2024efficient,sawarkar2024blended,liu-etal-2025-hoprag}. 
While dense vector retrieval excels at capturing broad semantic similarity, its inherent loss of fine-grained lexical details restricts the accuracy, hindering the broader adoption of vector databases in critical areas such as search engines~\cite{li2014enterprise,li2018design,gou2024semantic}, recommendation systems~\cite{miao2022het,parchas2020fast,yang2024revisitingkdd}, and Retrieval-Augmented Generation (RAG)~\cite{ang2024tsgassist,yu2025aquapipe,chen2025automatic,jiang2023chameleon}. 

\begin{figure}
    \centering
    \includegraphics[width=1\linewidth]{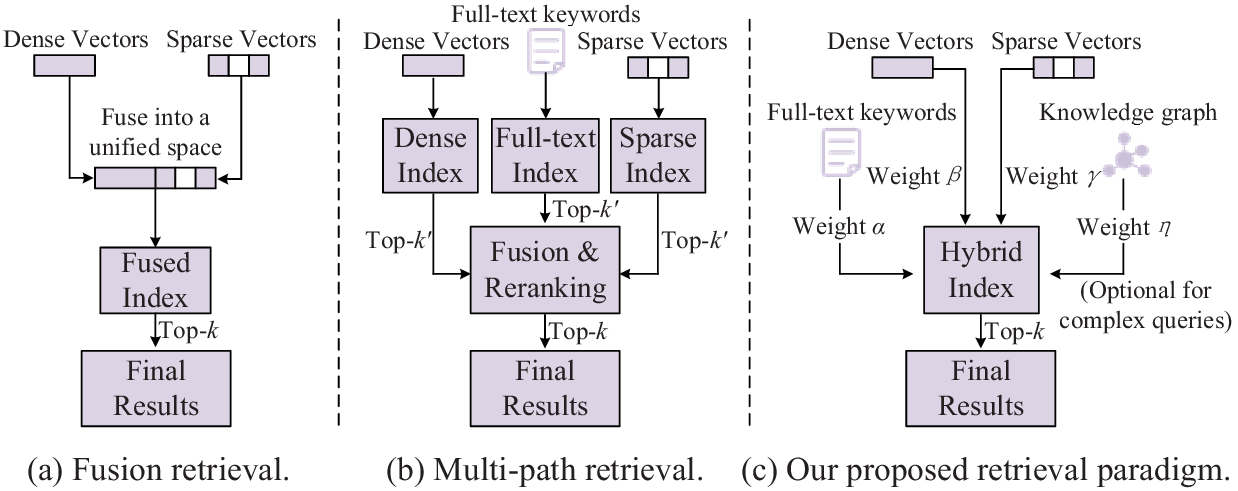}
    \vspace{-7mm}
    \caption{Comparison of existing hybrid search paradigms. }
    \vspace{-8mm}
    \label{fig:existing_methods}
\end{figure}

In addition to the popular use of dense vectors, alternative retrieval methods utilize lexicality-based~\cite{wang2017experimental,mallia2017faster} and learned-based~\cite{kong2023sparseembed,formal2024towards,bruch2024efficient} sparse vectors to improve the semantic relevance. 
While these sparse vectors offer superior interpretability and cross-domain robustness, their semantic representation capabilities are generally weaker than dense vectors. 
Consequently, dependence on any single retrieval path is often insufficient for achieving high end-to-end accuracy~\cite{weller2025theoretical, mala2025hybrid}. 
To overcome this limitation, hybrid search has recently emerged as a promising solution~\cite{ouyang2025adarag,mala2025hybrid,wang2025balancing}. 
As illustrated in Figure \ref{fig:existing_methods} (a) and (b), two primary methodologies have been proposed: \textbf{(1)} \textit{Fusion retrieval} integrates dense and sparse vectors by mapping them into a unified space via dimensionality reduction~\cite{bruch2024bridging} or by constructing the graph-based index with weighted distance calculations~\cite{zhang2024efficient}. 
However, these approaches often suffer from low efficiency or exhibit poor flexibility for arbitrary fusion weights. Furthermore, they are limited to dense and sparse vectors, failing to accommodate the complex requirements of real-world applications. 
\textbf{(2)} \textit{Multi-path retrieval} constructs separate indexes for various retrieval paths~\cite{wang2025balancing}, including dense vector~\cite{malkov2018efficient,fu2017nsg,yang2024revisiting}, sparse vector~\cite{kong2023sparseembed,formal2024towards,bruch2024efficient}, and full-text search~\cite{wang2017experimental,mallia2017faster}.
For a given query, this paradigm retrieves the top-$k'$ neighbors from each index independently. These intermediate results are subsequently fused using re-ranking methods, such as Reciprocal Rank Fusion (RRF)~\cite{cormack2009reciprocal,bruch2023analysis}, Weighted Sum~\cite{infinity_hybrid,milvus_rerank}, or ColBERT~\cite{formal2024splate,khattab2020colbert}, to produce the final top-$k$ ($k\le k'$) list. 
Due to its effectiveness and flexibility, this paradigm is widely adopted in modern databases~\cite{milvus_hybrid, infinity_hybrid, weaviate_hybrid, elastic_hybrid}.

Despite the effectiveness of the multi-path retrieval paradigm, it introduces two issues due to the decoupled architecture. 
\textbf{(1)} {\em Index storage overhead.} Each retrieval path necessitates the construction, storage, and maintenance of a dedicated index. This not only increases system complexity but also incurs significant storage overhead\footnote{For instance, Infinity~\cite{infinity_hybrid} requires 5GB to store the indexes of three retrieval paths for a dataset with 1M documents, where the HNSW index size for dense vector retrieval only accounts for 23\%.}~\cite{zhang2024efficient,liang2024unify}. 
\textbf{(2)} {\em Retrieval accuracy.} The optimal results for a hybrid query may not be present within the top-$k'$ results from any single path~\cite{wang2021milvus,wang2024must,zhu2024hjg}, leading to decreased accuracy. 



Collectively, these limitations constrain existing paradigms within a trilemma of flexibility, efficiency, and accuracy.
In this paper, we argue that resolving this trilemma requires rethinking hybrid retrieval from a structural perspective. 
We observe that heterogeneous retrieval paths can be naturally modeled as complementary relationships over a unified space with arbitrary weights, which motivates us to unify diverse retrieval paths within an all-in-one hybrid index.
As illustrated in Figure \ref{fig:existing_methods}(c), this unified architecture significantly reduces storage overhead. Furthermore, it enhances query performance by integrating three mainstream retrieval paths into a single index while optionally leveraging logical structures from knowledge graphs to augment retrieval for complex queries. 
Constructing such an integrated framework requires careful design. 
Given the superior efficiency and accuracy of the graph-based index in vector search~\cite{voruganti2025mirage,jiang2025digra,xu2024irangegraph,wang2021comprehensive}, it represents a promising foundation for building the unified framework. 
However, incorporating these heterogeneous retrieval paths into a graph index is non-trivial.
We identify two primary challenges that must be addressed:

{\em \textbf{Challenge I}: How to design a flexible graph-based hybrid index for hybrid search?} 
The fundamental heterogeneity of retrieval methods presents the primary obstacle: dense vectors utilize graph-based indexes, sparse vectors and full-text search rely on inverted indexes, and knowledge graphs employ entity-relationship structures. Fusing these disparate architectures into a single graph-based index without compromising performance is non-trivial. 
Furthermore, optimal fusion weights for the similarity from each retrieval path are inherently dynamic, varying by query context and user preference. Pre-computing and storing indexes for all possible weight combinations is infeasible. 
Finally, exhaustive use of all available paths for every query is suboptimal, which unnecessarily diminishes the efficiency when fewer paths would suffice. Consequently, designing a flexible graph index that supports arbitrary path combinations without index reconstruction remains a critical challenge.

{\em \textbf{Challenge II}: How to achieve efficient construction and effective search on the hybrid index?} 
While Inner Product (IP) serves as the similarity metric~\cite{bruch2024bridging, wang2025balancing}, existing graph indexes optimized for it suffer from inefficient construction and redundant edge connectivity.
Moreover, the indexing and querying processes of the hybrid index incur significant overhead and are inefficient on CPUs due to the massive hybrid distance calculations and the pruning of edges from various retrieval paths. 
Recent studies have demonstrated that accelerating index construction and query processing on GPUs is a promising solution~\cite{ootomo2024cagra, li2025scalable}. 
However, they do not apply to the hybrid index, where dense vectors and sparse vectors comparisons require vector dot product and set intersection operations, respectively. 
These disparate computational characteristics create challenges for efficient memory access and effective GPU parallelization. 
Furthermore, the computational demands of edges from various retrieval paths in the hybrid index introduce multiple efficiency bottlenecks. 
Semantic matching alone proves inadequate for complex queries. Although knowledge graphs provide complementary logical similarity~\cite{zhou2025depth, cao2024lego}, a fundamental granularity mismatch exists: our index operates on document-level representations while knowledge graphs model fine-grained entity relationships. This disparity complicates the effective integration of logical information into the vector search process.

To address these challenges, we propose {\sf Allan-Poe}, a unified and flexible graph-based hybrid index accelerated by GPUs. 
{\sf Allan-Poe} is built upon a unified lexical-semantic metric space, a framework that fuses diverse vector representations into a cohesive similarity metric, supporting retrieval under arbitrary fusion weights.
To support this framework, we design an isolated heterogeneous edge storage mechanism that seamlessly integrates dense, sparse, and full-text retrieval, while optionally accommodating logic structures from knowledge graphs. 
To achieve efficient hybrid index construction, we develop a GPU-accelerated indexing pipeline featuring: (1) a warp-level hybrid distance computation kernel optimizing both dense and sparse operations parallelized on GPUs; (2) RNG-IP joint pruning that maintains search quality while reducing index complexity by combining Relative Neighborhood Graph (RNG) and Inner Product (IP) neighbor pruning; (3) keyword-aware neighbor recycling that preserves keyword search functionality by efficiently recycling the pruned neighbors to ensure keyword-based navigation on the index; and (4) an optional logical edge augmentation module that integrates the entity-level knowledge graph into the document-level hybrid index for complex queries. 
To deliver a high-performance search service, we design a dynamic query framework on GPUs that decouples computation from storage, supporting any combination of retrieval paths at runtime without requiring index reconstruction.
The framework also incorporates: (1) dynamic heterogeneous edges loading for efficient traversal on the index; and (2) entity-document joint traversal for knowledge graph integration. 

In summary, this paper makes the following contributions. 
\begin{itemize}
\vspace{-1mm}
    \item \textit{Holistic hybrid indexing architecture}. We propose a unified hybrid index that seamlessly integrates dense, sparse, and full-text retrieval paths with arbitrary weights, and optionally augment complex queries with knowledge graphs (\S~\ref{sec:hybridindex}).
    \item \textit{Hardware-optimized construction pipeline}. We present a novel GPU-accelerated indexing pipeline to overcome the inefficiency of hybrid computations and heterogeneous edge pruning with fine-grained parallelization (\S~\ref{subsec:construction}). 
    \item \textit{Dynamic and flexible query framework}. We introduce a dynamic query framework that decouples computation from storage, enabling runtime path combinations with zero index reconstruction cost. (\S~\ref{subsec:search}).
    \item \textit{Extensive experiments}. We conduct comprehensive experiments on 6 real-world datasets, demonstrating that {\sf Allan-Poe} outperforms existing methods by 1.5$\times$-186.4$\times$ (\S~\ref{sec:experiments}). 
\end{itemize}
\vspace{-1mm}

The paper is organized as follows. Section \ref{sec:preliminaries} reviews the related work and illustrates the motivation. Section \ref{sec:hybridindex} introduces the structure of the hybrid graph-based index. Section \ref{sec:cons_and_search} describes the index construction and query framework of {\sf Allan-Poe}. Section \ref{sec:experiments} presents the experimental results. We conclude this paper in Section \ref{sec:conclusion}. 
\section{Background and Motivation}
\label{sec:preliminaries}

In this section, we first review related work on single-path and hybrid retrieval methods. We then establish the motivation for designing an all-in-one hybrid index by analyzing the limitations of existing approaches. 

\subsection{Related Work}
\label{subsec:relatedwork}
\subsubsection{\textbf{Single-path Retrieval.}} 
In the field of information retrieval, there are 4 distinct mainstream retrieval strategies. 

\noindent (1) \textbf{Full-text search} is a lexical search method based on exact keyword matching. It evaluates the term importance through frequency-based models such as TF-IDF~\cite{wu2008interpreting,paik2013novel,yahav2018comments} and BM25~\cite{robertson2025bm25, lv2011documents,blanco2012extending}. The inverted index is always used to achieve full-text search, employing retrieval algorithms such as WAND~\cite{khattab2020finding} and Block-Max WAND~\cite{mallia2017faster}. However, the exact term matching limits the recall of semantically relevant documents that lack the specific query keywords.

\noindent (2) \textbf{Sparse vector search} is another modern lexical approach that retrieves documents based on learned semantic representations of keywords. It utilizes models such as SPLADE~\cite{formal2021splade} to encode documents into high-dimensional sparse vectors, where each dimension corresponds to the importance of a term from an expanded vocabulary. The inverted index and various pruning strategies are used to retrieve similar documents via vector similarity~\cite{mallia2024faster,bruch2024efficient}. While the sparse vector addresses the issues caused by exact term matching, it still lacks comprehensive semantic understanding.

\noindent (3) \textbf{Dense vector search} constitutes a semantic retrieval paradigm that employs deep language models like BERT~\cite{devlin2019bert} to generate dense vector representations capturing overall document semantics. Similarity is evaluated using hash-based~\cite{zhang2020continuously,lu2020vhp}, tree-based~\cite{zeng2023litehst,echihabi2022hercules}, or graph-based indexes~\cite{malkov2018efficient,fu2017nsg}. Despite its popularity, this approach is limited by embedding space constraints and the absence of explicit term matching, which can compromise retrieval accuracy. 

\noindent (4) \textbf{Knowledge graph search} implements logical and semantic retrieval by converting queries into subgraphs and identifying similar structures through subgraph matching algorithms~\cite{hu2017answering,yuan2021subgraph,sun2022depth}. Subsequent enhancements incorporate entity and relation embeddings for improved efficiency~\cite{zheng2020dgl,wang2017knowledge}. Recent approaches like GraphRAG leverage dense vector search and community detection for document retrieval of global questions~\cite{edge2024local,cao2024lego,zhou2025depth,li2025neutronrag}. While GraphRAG excels at summarization tasks, it typically underperforms vanilla RAG for simple question answering~\cite{han2025rag}. Unlike GraphRAG, our work selectively integrates logical information from knowledge graphs to enhance vector search, establishing a distinct paradigm applicable to more general scenarios beyond RAG. 

\vspace{-2mm}
\subsubsection{\textbf{Hybrid Retrieval.}} 
Given the individual limitations of single-path retrieval methods, hybrid retrieval has emerged as a prominent search paradigm. Existing hybrid retrieval approaches can be categorized into two primary types.

\noindent (1) \textbf{Fusion retrieval} integrates multiple retrieval methods within a unified index structure. Current fusion methods are primarily restricted to two-path combinations~\cite{wang2025balancing}. For instance, DS-ANN~\cite{zhang2024efficient} employs a pre-defined fusion weight to combine dense and sparse vectors, constructing an HNSW index~\cite{malkov2018efficient} for efficient querying. While efficient, this method requires complete index reconstruction if the fusion weights change. IVF-Fusion~\cite{bruch2024bridging} addresses this by reducing the dimensionality of sparse vectors before combining them with dense vectors, then using an IVF index for retrieval. Although this eliminates weight-dependent reconstruction, it limits the flexibility to select different retrieval paths for varying scenarios. 

\noindent (2) \textbf{Multi-path retrieval} represents a more flexible paradigm that executes searches separately across different indexes and subsequently fuses the results~\cite{milvus_hybrid, infinity_hybrid, elastic_hybrid}. 
This paradigm is widely used in modern databases. 
However, the flexibility adversely affects query efficiency and accuracy while complicating index management. 


\subsection{Motivation}
\label{subsec:motivation}
\subsubsection{\textbf{Limitations of Single-path Search.}} 
Recently, dense vector search has become the most popular paradigm in vector databases among single-path retrieval methods, distinguished by its capacity for comprehensive semantic representation~\cite{pan2024survey,wang2021comprehensive}. 
Numerous specialized indexes have been developed to enhance their query efficiency and accuracy~\cite{gao2025practical, gou2025symphonyqg, peng2023efficient}. However, vector similarity alone does not guarantee end-to-end relevance between queries and documents. To investigate this limitation empirically, we conduct experiments using two established real-world QA datasets: \textbf{NaturalQuestions (\textit{NQ})}~\cite{wang-etal-2024-rear} for simple question answering and \textbf{2WikiMultiHopQA (\textit{WM})}~\cite{ho-etal-2020-constructing} for multi-hop question answering. For each dataset, we evaluate the first 1,000 queries, each associated with 1-3 ground-truth documents. We employ the BGE-M3 model~\cite{chen2024bge} for embedding both documents and queries. The ground-truth baselines of vector similarity are established by computing the top-10 nearest neighbors via brute-force vector similarity search. We assess the retrieval accuracy of two state-of-the-art graph-based indexes: HNSW (CPU-based)~\cite{malkov2018efficient} and CAGRA (GPU-based)~\cite{ootomo2024cagra}. 

\begin{figure}
    \centering
    \includegraphics[width=0.38\textwidth]{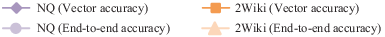}
    \subfigcapskip = -2pt
    \vspace{-2mm}
    \subfigure[HNSW]{
    \includegraphics[width=0.23\textwidth]{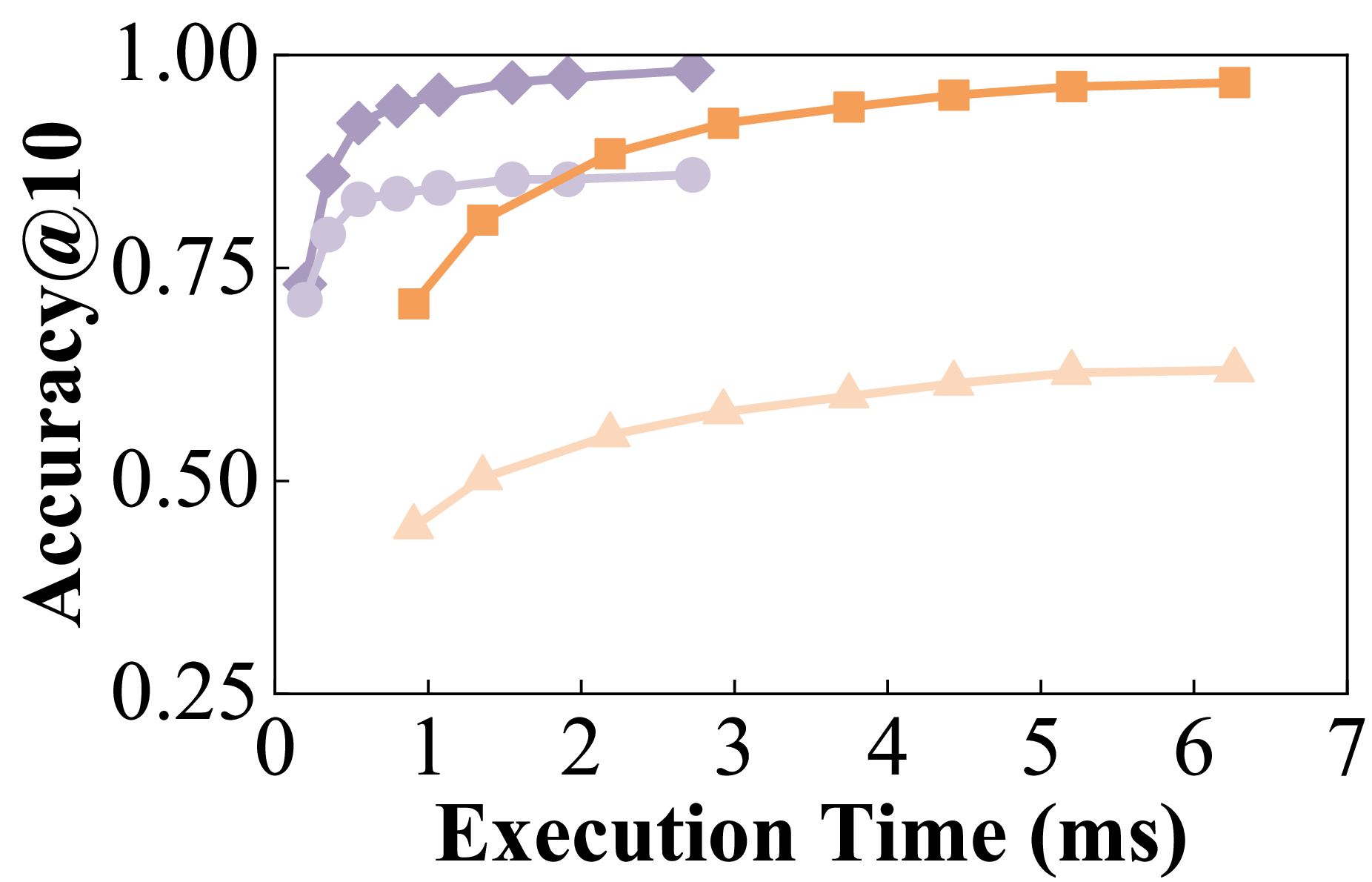}}
    \hspace{-2mm}
    \subfigure[CAGRA]{
    \includegraphics[width=0.24\textwidth]{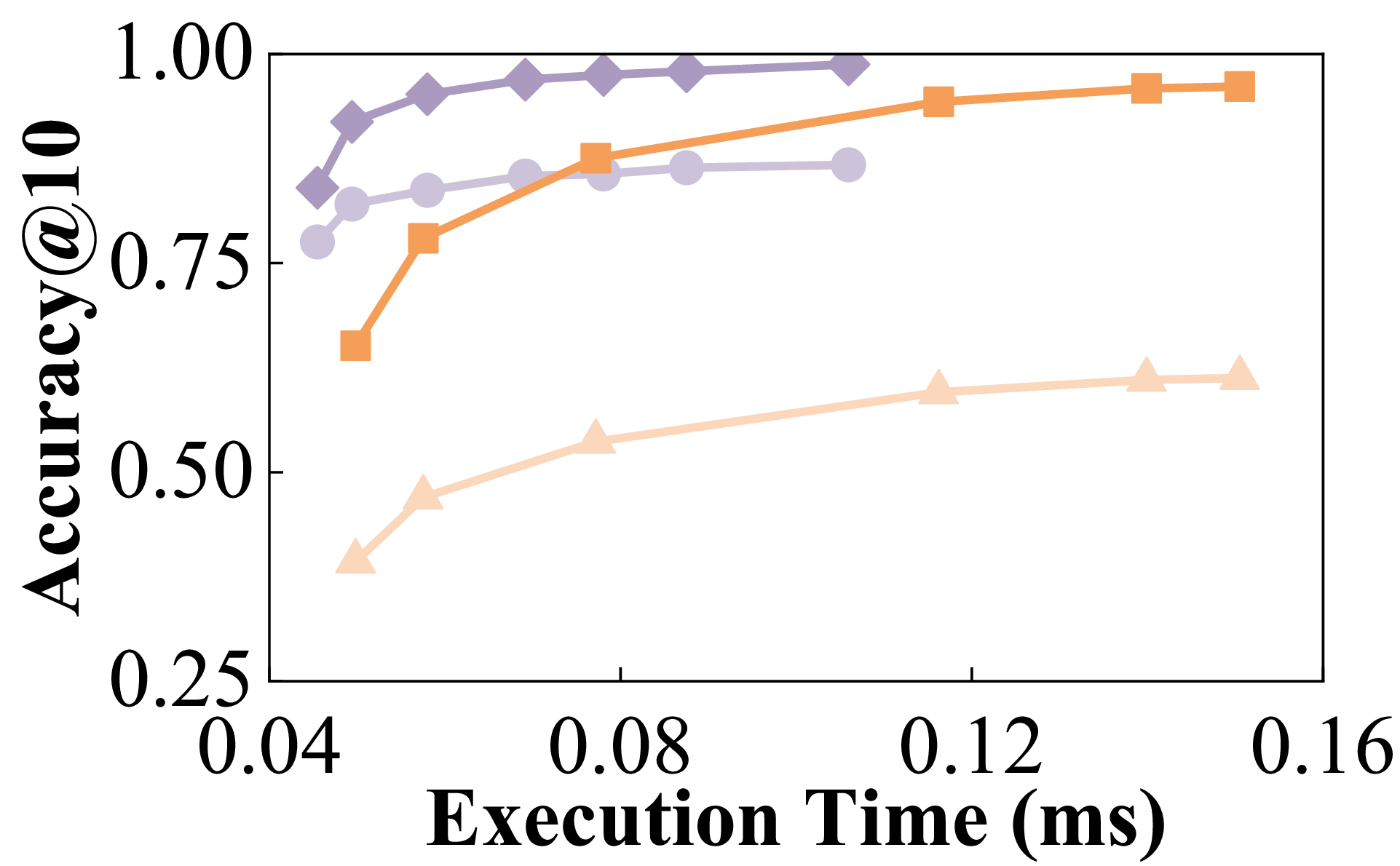}}
    \vspace{-3mm}
    \caption{Gaps between the vector similarity and end-to-end document similarity of two graph-based indexes.}
    \vspace{-5mm}
    \label{fig:motivation_dense}
\end{figure}

As shown in Figure \ref{fig:motivation_dense}, while vector-similarity accuracy can easily reach 99\% within 4ms on the CPU and 0.1ms on the GPU, the corresponding end-to-end accuracy is substantially lower. In practical document retrieval systems, it is this end-to-end accuracy—not vector-similarity accuracy—that dictates performance in downstream tasks. 
Furthermore, Figure \ref{fig:motivation_dense} reveals that the end-to-end accuracy for the \textit{WM} dataset is lower than that for \textit{NQ}, whereas their maximal vector-based accuracies are comparable. This discrepancy underscores the limitations of dense vector search in handling complex queries. Consequently, reliance on single-path retrieval methods is insufficient for achieving satisfactory end-to-end performance, restricting their utility in downstream applications.

\vspace{-2mm}
\subsubsection{\textbf{Effectiveness of Hybrid Search}.} 
Hybrid search has emerged as a powerful strategy to mitigate the limitations of single-path retrieval and improve end-to-end accuracy~\cite{infinity_hybrid, wang2025balancing}. To evaluate its effectiveness, we measure the retrieval quality using Infinity~\cite{infinity_hybrid}, a modern database featuring efficient hybrid search. To better assess the quality of retrieved documents, we employ Normalized Discounted Cumulative Gain at rank $k$ (nDCG@$k$)~\cite{wang2013theoretical} with $k=5$, which evaluates both the recall and positional ranking of relevant documents in the retrieved results. 

\begin{figure}
    \centering
    \subfigcapskip = -5pt
    \subfigure[Comparison of retrieval accuracy.]{
    \includegraphics[width=0.23\textwidth]{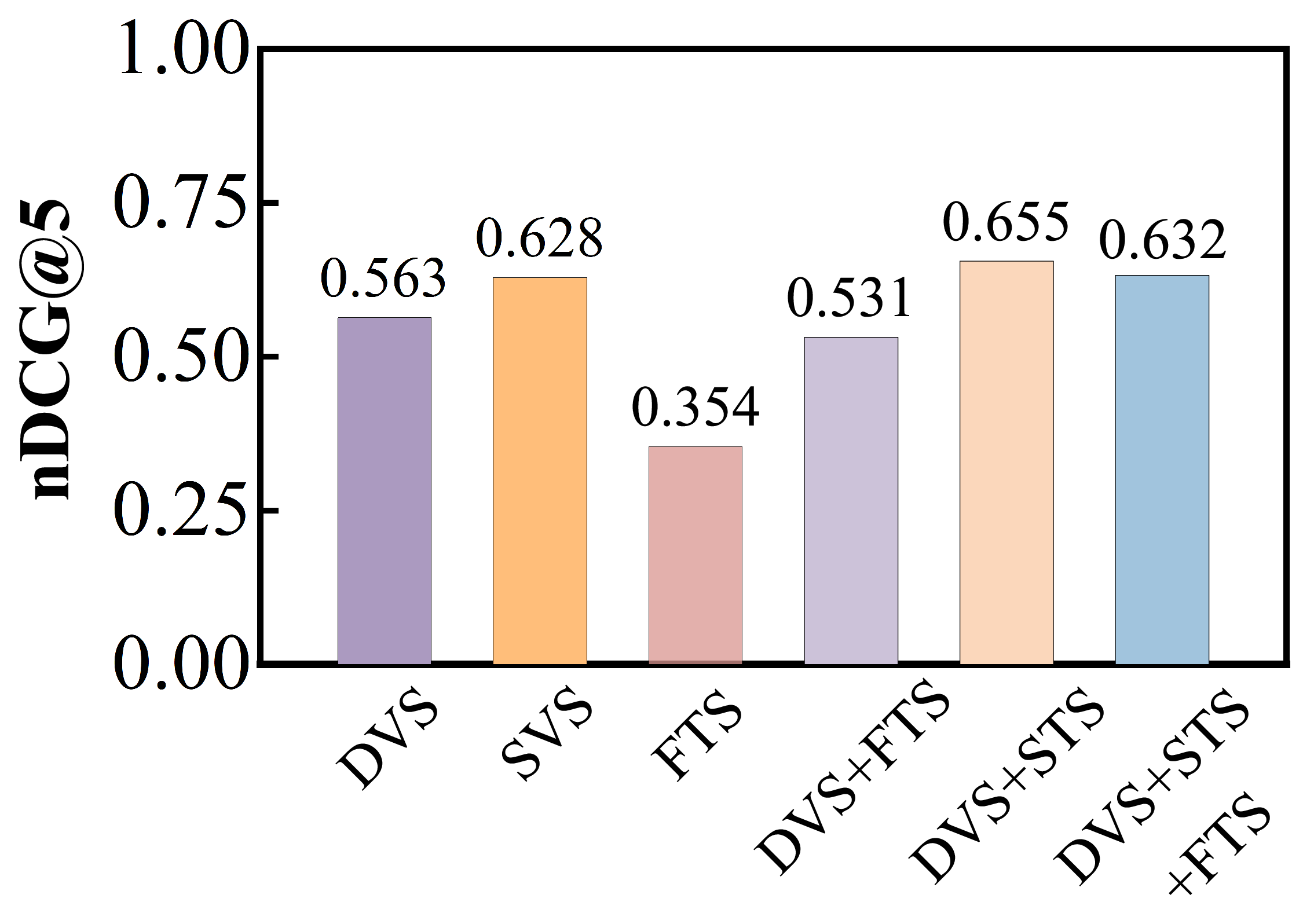}}
    \hspace{-2mm}
    \subfigure[Comparison of retrieval latency.]{
    \includegraphics[width=0.23\textwidth]{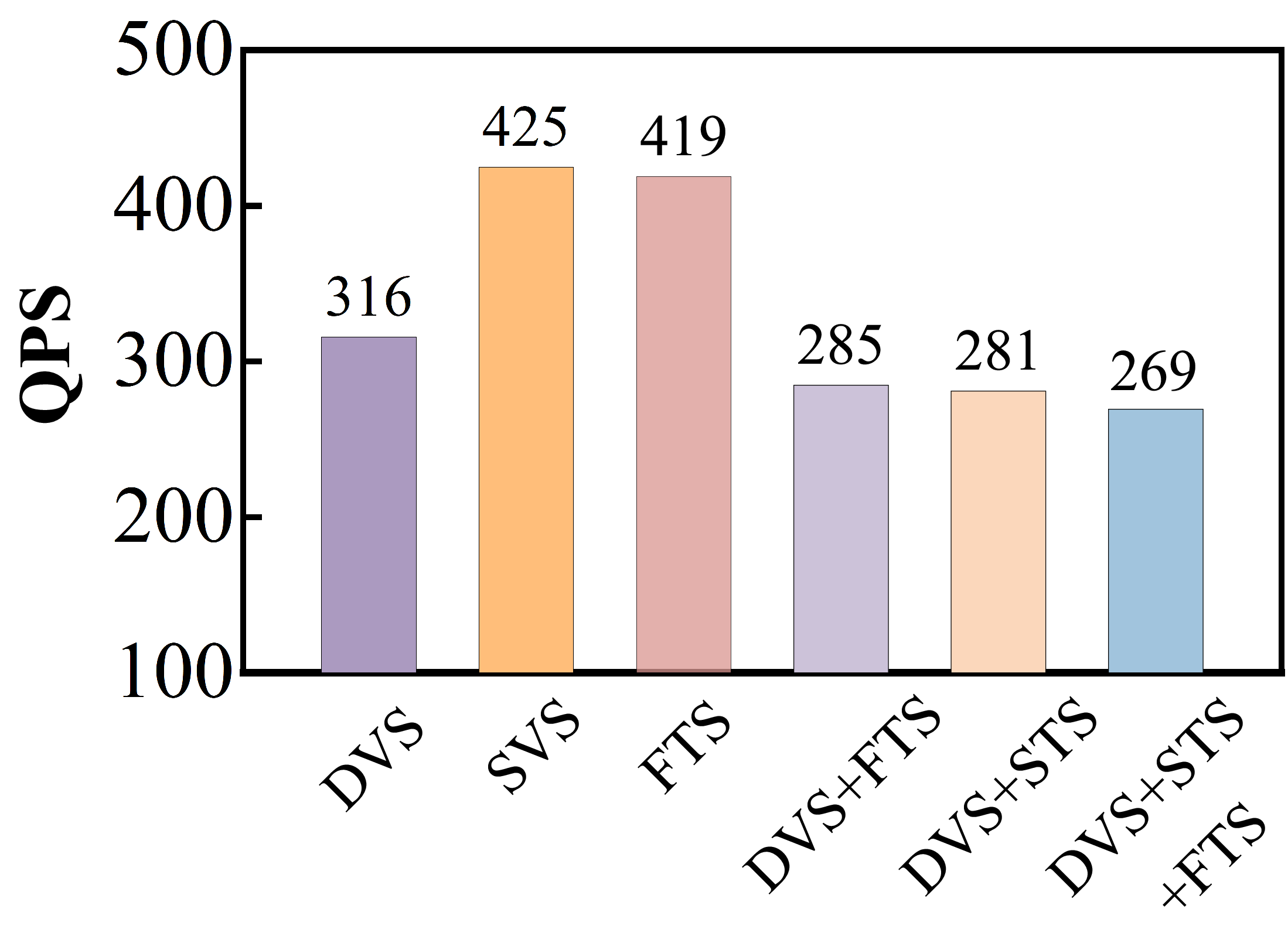}}
    \vspace{-5mm}
    \caption{Comparison of various retrieval paths on \textit{NQ} using Infinity~\cite{infinity_hybrid}. DVS, SVS, and FTS denote dense vector, sparse vector, and full-text search, respectively. }
    \vspace{-4mm}
    \label{fig:motivation_multipath_com}
\end{figure}

\begin{figure}
    \centering
    \includegraphics[width=0.95\linewidth]{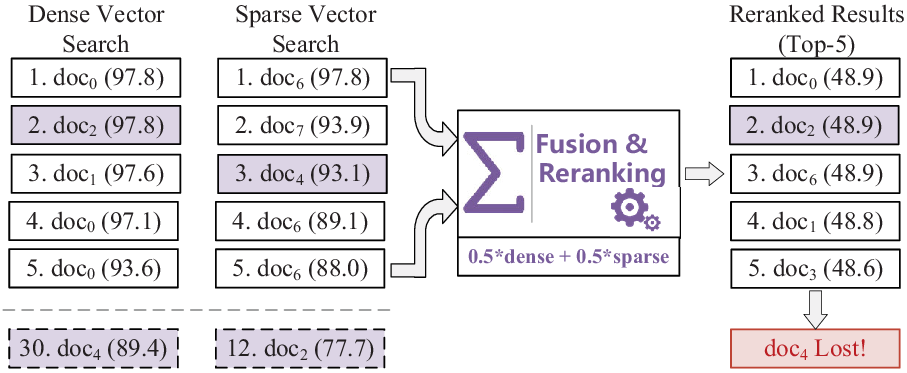}
    \vspace{-4mm}
    \caption{Example of retrieval in separate paths on \textit{NQ}. The ground truth documents are doc$_2$ and doc$_4$. }
    \vspace{-6mm}
    \label{fig:motivation_multi_path}
\end{figure}

As shown in Figure \ref{fig:motivation_multipath_com}, multi-path retrieval methods such as DVS+STS and DVS+STS+FTS generally exhibit higher nDCG than single-path retrieval. This demonstrates that multi-path retrieval can leverage the complementary strengths of individual paths to enhance result quality.

However, no single configuration is optimal for all scenarios. Retrieval with more paths does not consistently outperform fewer or single paths. 
For example, in Figure \ref{fig:motivation_multipath_com}(a), the two-path combination of dense vector and full-text search yields lower nDCG than the single-path methods using either dense or sparse vectors alone on dataset \textit{NQ}. And the three-path combination shows lower nDCG than the two-path retrieval of dense and sparse vector search. On other datasets, the accuracy ranking may be completely different. 
Furthermore, different path combinations present a distinct trade-off between accuracy and efficiency. Although the single-path methods typically achieve lower accuracy than the multi-path approaches, they are more efficient and incur less overhead (Figure \ref{fig:motivation_multipath_com}(b)), sometimes making it preferable for a few real-time applications. 
For instance, the sparse vector retrieval method achieves comparable accuracy with the three-path combination while maintaining lower latency. 
These findings underscore the necessity for flexible path combination in hybrid search.
The trade-off strategies have been studied in~\cite{wang2025balancing}, which is orthogonal to our work. 

 \vspace{-2mm}
\subsubsection{\textbf{Limitations of Separate Multi-path Search}.} 
Multi-path retrieval is widely adopted for flexible hybrid search in modern vector databases~\cite{infinity_hybrid, milvus_hybrid, elastic_hybrid}. Although it achieves superior accuracy compared to single-path approaches, this comes at the cost of increased time overhead.
Moreover, this paradigm performs separate retrievals across individual indexes before fusing the results. The optimal results for a hybrid query may not be present within the top-$k$ results from any single path or may be excluded after fusing the results~\cite{wang2024must,wang2021milvus}. 
To illustrate this, we examine the retrieval of top-5 documents from the \textit{NQ} dataset using dense and sparse vectors independently. 
Figure \ref{fig:motivation_multi_path} presents a representative example. 
Using dense vectors, ground-truth documents doc$_2$ and doc$_4$ are ranked 2$^{nd}$ and 30$^{th}$, respectively, while sparse vectors rank them 3$^{rd}$ and 12$^{th}$. If we fuse the top-5 results from both paths using equal weights (i.e., $0.5 \times$dense similarity$+ 0.5 \times$sparse similarity), doc$_4$ will be excluded because it only involves the similarity score from the sparse path. 
To include doc$_4$, more candidates should be included for each retrieval path (e.g., top-30).
However, expanding candidates entails longer retrieval times, and it does not guarantee higher accuracy. This is because low-ranking noisy documents can be promoted during the fusion, disrupting the final accuracy.

\subsection{Design Principles of Hybrid Index} 
\begin{figure}
    \centering
    \hspace{-8mm}
    \includegraphics[width=0.7\linewidth]{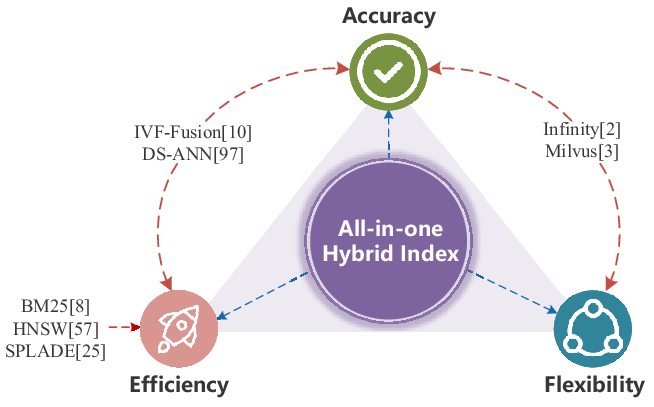}
    \vspace{-5mm}
    \caption{Trilemma of existing retrieval methods. }
    \vspace{-6mm}
    \label{fig:trillemma}
\end{figure}

The preceding analysis reveals that existing retrieval methods face a fundamental trilemma, being unable to simultaneously achieve high accuracy, efficiency, and flexibility.  
As illustrated in Figure \ref{fig:trillemma}, single-path search methods (e.g., HNSW~\cite{malkov2018efficient}, BM25~\cite{blanco2012extending}, and SPLADE~\cite{formal2021splade}) achieve high efficiency but suffer from the semantic gap between vector similarity and end-to-end relevance, resulting in limited accuracy. 
Multi-path retrieval methods (e.g., Infinity~\cite{infinity_hybrid} and Milvus~\cite{milvus_hybrid}) enable flexible hybrid search through separate-then-fuse strategies but incur storage and efficiency costs, as illustrated previously. 
Fusion retrieval methods (e.g., IVF-Fusion~\cite{bruch2024bridging} and DS-ANN~\cite{zhang2024efficient}) improve accuracy over single-path approaches while maintaining intermediate efficiency. However, existing approaches are restricted to two-path combinations of dense and sparse vector search, limiting the potential accuracy gains. Moreover, as the combination of dense and sparse vectors is not always effective, they also face concerns about the adaptability to diverse scenarios. 

Consequently, an effective hybrid index should satisfy the following three key requirements: 
\begin{itemize}
    \item \textbf{Flexibility}: Supporting arbitrary combinations of retrieval paths and fusion weights without index reconstruction to accommodate diverse application needs.
    \item \textbf{Accuracy}: Leveraging complementary information from multiple paths to maximize end-to-end relevance. 
    \item \textbf{Efficiency}: Maintaining low-latency retrieval and low storage overhead despite multiple paths and heterogeneous distance computations.
\end{itemize}

Guided by the three principles, we propose \textsf{Allan-Poe}, which provides flexible integration of three retrieval paths augmented by knowledge graphs via a well-designed isolated heterogeneous edges mechanism. Unlike separate-then-fuse approaches, \textsf{Allan-Poe} performs path fusion during query processing within a unified hybrid index, thereby avoiding the associated accuracy limitations. Additionally, we leverage massive GPU parallelism to achieve both efficient index construction and real-time query performance, meeting the requirement of efficiency. 
\section{Hybrid Index Structure}
\label{sec:hybridindex}
This section presents the structure of our proposed hybrid index, demonstrating how it resolves the trilemma by simultaneously achieving accuracy, efficiency, and flexibility.

\subsection{Overview}
\label{subsec:overview}
The hybrid index of {\sf Allan-Poe} integrates the dense vectors, sparse vectors, full-text, and knowledge graph retrieval within a Unified Semantic Metric Space (\textit{USMS}). 
\begin{definition}[\textbf{Unified Semantic Metric Space - USMS}]
    A USMS is a tuple $H=\{D,F,M_w\}$ defined as follows:
    \begin{itemize}
        \item $D$: The set of all documents in the corpus. 
        \item $F$: A set of feature extractors $\{f_\text{dense},f_\text{sparse},f_\text{full},f_\text{kg}\}$ that maps each document $d\in D$ to its respective feature representation. For example, $f_\text{dense}(d)\in \mathbb{R}^m$ and $f_\text{kg}(d)\subseteq E$, where $m$ is the dense vector dimension and $E$ denotes the entity set. 
        \item $M_w$: A composite similarity metric $M_w:D\times D\rightarrow \mathbb{R}$, defined for any weight vector $w=[w_d,w_s,w_f]\in\mathbb{R}^3$ as $M_w(q,d)=w_d\cdot \text{sim}_d(q,d)+w_s\cdot \text{sim}_s(q,d)+w_f\cdot \text{sim}_f(q,d)$, where $\text{sim}_d$, $\text{sim}_s$, and $\text{sim}_f$ denote the inner product similarities. Furthermore, $w_k\cdot \text{sim}_k(q,d)$ is used to augment $M_w$ for complex queries, where $\text{sim}_k$ represents the path length between query and document entities in the knowledge graph. 
    \end{itemize}
\end{definition}

Notably, $\text{sim}_k$ employs path length as its measurement, contrasting with the inner product metrics used by other similarities in the composite similarity metric $M_w$. 
Furthermore, the knowledge graph retrieval operates on fine-grained entities, while other paths function at the document level, which reflects a fundamental granularity difference. 
Additionally, $\text{sim}_k$ captures logical relationships, whereas the other represents semantic similarity. Direct integration of these dissimilar metrics could compromise the structural integrity of semantic edges. Consequently, we maintain separate semantic and logical edges within our graph-based index. 

\begin{figure}
    \centering
    \includegraphics[width=1\linewidth]{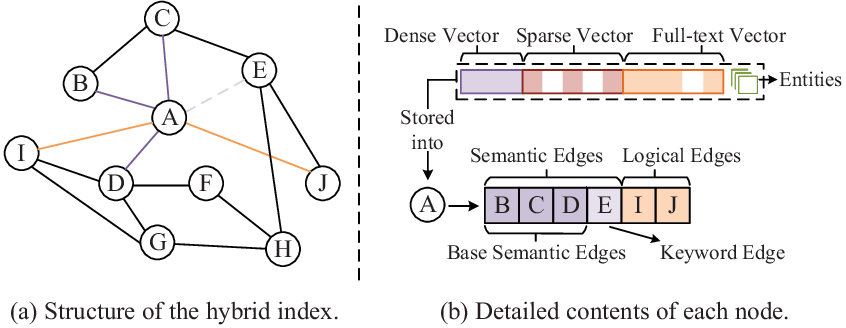}
    \vspace{-8mm}
    \caption{Overview of the hybrid index in {\sf Allan-Poe}. }
    \vspace{-6mm}
    \label{fig:index_overview}
\end{figure}

Figure \ref{fig:index_overview} illustrates the hybrid index architecture. 
As shown in Figure \ref{fig:index_overview}(b), each node in the graph index (representing a corpus document) stores four data types corresponding to \textit{USMS} features: dense vector, sparse vector, full-text vector, and entities. 
We classify the heterogeneous edges connecting these nodes into two categories based on the aforementioned incompatibility: semantic edges and logical edges.  
Semantic edges are further categorized into: (1) base semantic edges connecting nodes with similar fused vector semantics (detailed in Section \ref{subsec:hybrid_vector}), and (2) keyword edges connecting nodes sharing common keywords (detailed in Section \ref{subsec:keyword_edge}). 
These edges guide the traversal toward nearest neighbors in vector space. Logical edges, established from knowledge graph relations (detailed in Section \ref{subsec:kg_edge}), complement semantic edges by connecting nodes that are distant in vector space but logically related. 
The isolated heterogeneous edge storage guarantees the flexibility of {\sf Allan-Poe} when dealing with any combination of retrieval paths. 

\subsection{Hybrid Vector Representation}
\label{subsec:hybrid_vector}
To integrate dense vector, sparse vector, and full-text retrieval, we employ a vector fusion technique~\cite{wang2021milvus} that maps these representations into the \textit{USMS}. 
Specifically, dense and sparse vectors are naturally represented in vector form, while keywords used in full-text search can be encoded as sparse vectors where each dimension represents the importance of a term in the vocabulary. For clarity, the sparse vectors generated by full-text search models such as BM25~\cite{robertson2025bm25} are denoted as lexical sparse vectors, while those produced by learning models such as SPLADE~\cite{formal2021splade} are called learned sparse vectors. 
The vector fusion process concatenates these three vector types into a unified high-dimensional representation. Formally, for a document $d$, the concatenated vector is defined as $f_\text{concat}(d)=[f_\text{dense}(d), f_\text{sparse}(d), f_\text{full}(d)]$. 

\vspace{-0.5mm}
\begin{theorem}
    \label{theorem1}
    The hybrid retrieval problem with any dynamic weight vector $w=[w_d,w_s,w_f]\in \mathbb{R}^3$ is mathematically equivalent to a Maximum Inner Product Search (MIPS) problem in USMS. Consequently, graph-based indices are applicable to the weighted hybrid search without structural modification. 
\end{theorem}

\vspace{-3mm}
\begin{proof}
   Let the fused vector of a corpus be $f_\text{concat}(d)=[w_d*f_\text{dense}(d),w_s*f_\text{sparse}(d),w_f*f_\text{full}(d)]$. The objective of the hybrid search with weight vector $w=[w_d,w_s,w_f]$ is to maximize the composite similarity score $M_w(q,d)$. 
   We construct a transformed query vector $q'$ in the Unified Semantic Metric Space by scaling the individual components with their corresponding weights: $q'=[w_d*f_\text{dense}(q),w_s*f_\text{sparse}(q),w_f*f_\text{full}(q)]$. 
   Now, we perform the inner product between the transformed query $q'$ and the stored corpus vectors $f_{concat}(d)$ in the Unified Semantic Metric Space: $\langle q',f_\text{concat}(d)\rangle=\langle [w_d*f_\text{dense}(q),w_s*f_\text{sparse}(q),w_f*f_\text{full}(q)], [f_\text{dense}(d), f_\text{sparse}(d), f_\text{full}(d)]\rangle=w_d*f_\text{dense}(q)\cdot f_\text{dense}(d)+w_d*f_\text{sparse}(q)\cdot f_\text{sparse}(d)+w_f*f_\text{full}(q)\cdot f_\text{full}(d)=M_w(q,d)$. 
   This derivation proves that finding $d$ that maximizes the weighted hybrid similarity $M_w(q, d)$ is exactly equivalent to finding the corpus that maximizes the inner product $\langle q', f_\text{concat}(d) \rangle$ in USMS. 
\end{proof}

\vspace{-1mm}
To achieve efficient maximum inner product search on GPUs, we employ the framework CAGRA~\cite{ootomo2024cagra} as a backbone of the hybrid index, which has been proven effective and efficient for approximate neighbor retrieval on GPUs~\cite{ootomo2024cagra}. 

\subsection{Keyword Edges Supplement}
\label{subsec:keyword_edge}
While the vector fusion approach described in Section~\ref{subsec:hybrid_vector} enables flexible hybrid retrieval within a unified index, it inherently compromises the keyword-based search function in the original full-text search. 
In many applications, users explicitly require certain keywords to appear in retrieved documents to enhance accuracy~\cite{liu2006effective,li2008effective}. 
Although the function of keyword search can be easily achieved using the traditional inverted index for full-text search, it is non-trivial for graph-based indexes. 
Recent research has explored graph-based indexes with attribute filtering capabilities~\cite{ait2025rwalks, patel2024acorn, gollapudi2023filtered}, which can be employed to achieve keyword search in our hybrid index. 
However, these approaches typically integrate attribute constraints directly into the primary graph structure, limiting flexibility.
To address this limitation, we propose a dual-assessment mechanism that selectively preserves pruned edges as dedicated keyword edges.


During construction of the hybrid index, we first prune the graph using the strategy in CAGRA~\cite{ootomo2024cagra} to leverage GPU computational power. Recall that the pruning strategy used in CAGRA prunes edges according to the number of detourable routes, where edges with more detourable routes will be pruned. 
Specifically, for a node $A$ and its neighbors $X$, if there exists another neighbor $Y$ such that $max(dis(A,X),dis(X,Y))<dis(A,Y)$, then the path $A\rightarrow X\rightarrow Y$ constitutes a detourable route for the edge $A\rightarrow Y$. CAGRA retains the $d$ neighbors with the fewest such detourable routes.  
Our dual-assessment mechanism operates during the above pruning phase of CAGRA by evaluating keyword overlap between nodes.
For each neighbor $X$ of node $A$ that would normally be pruned by CAGRA, if there exists another neighbor $Y$ of $A$ such that $K(A)\cap K(X)\subseteq K(Y)$, then $X$ is pruned. This is because the navigation from $A$ to $X$ can be replaced by $Y$ to $X$ for any keywords. 
However, if a neighbor scheduled for pruning does not satisfy this condition, it is preserved as a keyword edge.
As depicted in Figure \ref{fig:index_overview} previously, keyword edges are maintained separately from base semantic edges to ensure clear separation. This distinct edge organization guarantees the pluggable nature of keyword functionality. The incorporation of keyword edges facilitates efficient traversal to semantically relevant neighbors sharing common keywords, significantly enhancing keyword-aware search performance. 

\vspace{-2mm}
\subsection{Logical Edges Augmentation}
\label{subsec:kg_edge}
While previous sections integrated dense vector, sparse vector, and full-text search through semantic edges in our hybrid index, semantic-based graph search still faces two fundamental challenges: 
(1) \textit{Semantic search retrieves semantically similar but logically unrelated documents}. For example, for a query "\textit{Where was John's mother born?}", two documents containing "\textit{John's father was born in the US}" and "\textit{Linda's mother was born in the US}" can be retrieved, as they exhibit high semantic similarity due to the similar keywords "John", "mother", or "born" despite describing logically distinct relationships.
(2) \textit{Semantic search struggles with complex queries involving multiple entities or multi-hop reasoning}. The query "\textit{Who is younger, Linda or John?}" contains multiple entities, often causing graph traversal to settle in local optima and retrieve documents about only one entity. Similarly, for the multi-hop query "\textit{Where was Linda's mother born?}", if information about Linda and her mother is distributed across different documents, semantic search only returns documents about Linda, missing crucial contextual information. 
To address these limitations, \textsf{Allan-Poe} augments semantic search with logical edges utilizing knowledge graphs.

Knowledge graphs can be constructed from the corpus using deep language models such as BERT~\cite{devlin2019bert} or LLMs~\cite{fu2026llm}. For each node in the hybrid index, we store associated entities alongside the vector and maintain an entity-to-node mapping. We then extract inter-entity relations from the knowledge graph and represent them as logical edges.
Formally, let $V(X)$ denote the entity set of node $X$, and $G(V,R)$ represent the knowledge graph with entities $V$ and relations $R$. The logical edges for node $X$ comprise triplets $\{(s, r, t) \mid s \in V(X), r \in R, t \in V \setminus V(X)\}$, where $s$ and $t$ are entities connected by relation $r$. Thus, any two entities from different nodes that are related in the knowledge graph establish a logical edge between their corresponding document nodes. 
During search, we dynamically leverage logical edges through a fine-grained entity-document unified strategy that enhances query capability while preserving efficiency (detailed in Section~\ref{subsec:search}). 
Notably, logical edge augmentation is optional, representing a trade-off between potential accuracy improvements and the substantial computational cost of knowledge graph construction~\cite{yang2025graphusion}. Therefore, selectively applying logical edges to small-scale, high-value data shards serves as a strategic approach to enhance query performance. 

\vspace{-1mm}
\section{Hybrid Index Construction and Query}
\label{sec:cons_and_search}

Section \ref{sec:hybridindex} presents the basic structure of the hybrid index in {\sf Allan-Poe}. The integration of multiple retrieval paths in the index structure introduces significant complexity to both construction and query processing, creating substantial efficiency challenges. 
To address these issues, this section describes our approach to efficient hybrid index construction and high-performance retrieval leveraging GPU acceleration. 

\subsection{Efficient Index Construction on GPU}
\label{subsec:construction}
\begin{figure*}
    \centering
    \includegraphics[width=1\textwidth]{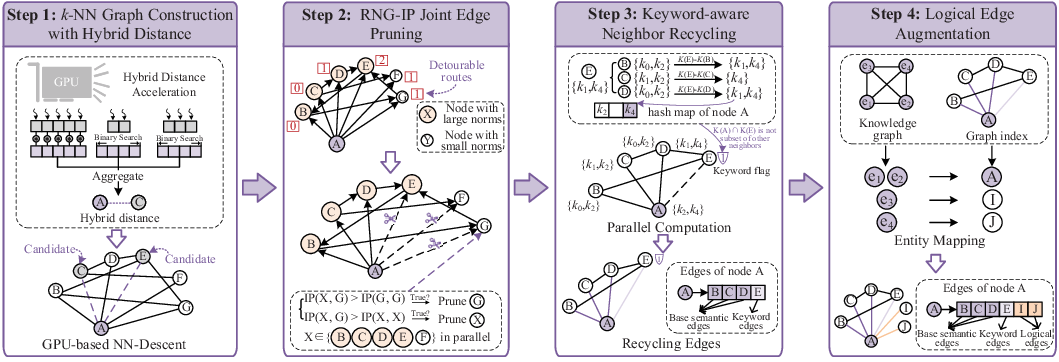}
    \vspace{-8mm}
    \caption{Index construction process in {\sf Allan-Poe}. }
    \vspace{-6mm}
    \label{fig:index_construction}
\end{figure*}

The hybrid index construction presents additional complexity due to the challenges brought by hybrid distance computation and heterogeneous edge establishment. This substantial overhead motivates our use of GPUs to accelerate the construction. As illustrated in Figure~\ref{fig:index_construction}, the indexing pipeline comprises four key steps.

\noindent \textbf{\underline{Step 1}}: \textit{$k$-NN Graph Construction with Hybrid Distance}. We employ the NN-Descent algorithm~\cite{dong2011efficient} to construct an approximate $k$-nearest neighbor ($k$-NN) graph from the fused vectors. NN-Descent operates on the principle that "a neighbor's neighbors are likely neighbors"—it iteratively refines the graph by evaluating 2-hop neighbors and updating connections based on distance comparisons. This approach has demonstrated superior efficiency to incremental methods on GPUs~\cite{li2025scalable,wang2021fast}. 
Within this framework, hybrid distance calculation between fused vectors represents the most computationally intensive operation, involving two distinct computation patterns. To accelerate this process on GPUs, we assign an entire warp to compute each distance collaboratively. For dense vector computation, each thread fetches four operands using CUDA vectorized instructions~\cite{gale2020sparse}, maximizing memory bandwidth utilization. Threads then multiply their operands in parallel, storing intermediate results in registers. 
After processing all dense dimensions, the algorithm computes distances for both learned and lexical sparse vectors using identical processing. Sparse vectors are stored in CSR format~\cite{greathouse2014efficient}, with non-zero values and indices maintained in separate arrays. 
Distance calculation is transformed into a parallel set intersection: each thread fetches an index from the candidate's sparse vector and searches for it in the explored node's sparse vector using parallel binary search. This design enables the frequently used explored nodes' vectors to be cached in the shared memory, minimizing access overhead. When matches occur, the corresponding values are multiplied and accumulated with the dense vector results. Finally, per-thread values are aggregated using warp-level reduction operations in CUDA.  

\noindent \textbf{\underline{Step 2}}: \textit{RNG-IP Joint Edge Pruning}. For the inner product metric, vectors with large norms (lengths) frequently appear in maximum inner product search results, making connections to such vectors an effective strategy for improving search efficiency~\cite{tan2019efficient}. Recent approaches either apply inner product (IP) pruning directly or augment RNG-pruned edges with high-norm nodes in graph indexes~\cite{chen2025stitching, chen2025maximum}. However, these methods increase index size and often create uneven node degrees, which hinders aligned parallel processing on GPUs. 
To overcome these limitations, we combine RNG and IP pruning strategies with GPU-specific optimizations, achieving efficient inner product search while preserving both the original index size and aligned degree structure. 
Our joint pruning approach operates in two phases. The first phase employs RNG pruning from CAGRA to refine neighbors and reduce search complexity. As shown in Figure \ref{fig:index_construction}, we calculate the detourable routes for neighbors in the $k$-NN graph in parallel and sort neighbors by their detourable route counts. 
The second phase applies the IP pruning strategy\footnote{For node $A$ with neighbor set $R$, candidate $G$ is excluded if $\exists X\in R$ such that $IP(G,X)>IP(G,G)$. Similarly, $X$ is filtered if $IP(G,X)>IP(X,X)$~\cite{tan2019efficient}.} to remove neighbors with small vector norms. 
To efficiently achieve the neighbor pruning, We parallelize the inner product distance calculations between the candidate neighbor and current neighbors (e.g., between candidate $G$ and neighbors $X\in\{B,C,D,E,F\}$ in Figure \ref{fig:index_construction}), where each warp is responsible for the distance calculation between a current neighbor and the candidate neighbor. 
Finally, $d/4$ neighbors and $d/4$ reverse neighbors are concatenated to fill each node's edge list, where $d$ represents the target index degree.
For the remaining $d/2$ neighbors, we append the fused semantic neighbors with single-path neighbors. 
This composite structure effectively approximates the Pareto frontiers, thereby ensuring retrieval robustness under arbitrary weights.

\noindent \textbf{\underline{Step 3}}: \textit{Keyword-aware Neighbor Recycling}. This procedure recycles pruned edges from \textit{Step 2} according to the strategy established in Section \ref{subsec:keyword_edge}. 
A brute-force implementation would incur substantial computational overhead, as checking the condition $K(A)\cap K(X)\subseteq K(Y)$ for each node $A$ and its pruned neighbors $X$ against all current neighbors $Y$ requires numerous set intersection operations. 
To optimize this process, we fully utilize the results calculated in previous steps. Specifically, we assign a Boolean type keyword flag to each neighbor of each node. During the second phase of \textit{Step 2}, while computing lexical sparse vector similarities via set intersection between neighbors of node $A$, we check whether non-intersecting keywords exist in $A$'s keyword set $K(A)$. If so, we set the corresponding neighbor's keyword flag to 1. For efficient $K(A)$ lookups, we maintain it in a hash map within GPU shared memory. 
For example, in Figure \ref{fig:index_construction}, node $E$ is scheduled for pruning during \textit{Step 2} due to its numerous detourable routes. During the second phase's distance calculations between $E$ and current neighbors ($B$, $C$, $D$) in \textit{Step 2}, we examine non-intersecting keywords from $E$ (e.g., $K(E) \setminus K(B) = \{k_1,k_4\}$) and verify their presence in $K(A)$ using the hash map. Here, $k_4$ exists in $K(A)$ but not in any current neighbors, violating the subset condition. Consequently, $E$'s keyword flag is set to 1. Finally, we traverse all pruned neighbors and recycle those with activated keyword flags as keyword edges. 

\noindent \textbf{\underline{Step 4}}: \textit{Logical Edge Augmentation}. Finally, we establish logical edges by mapping knowledge graph entities to their corresponding document nodes in the index and creating connections between nodes whose entities share relationships in the knowledge graph. 

Algorithm~\ref{alg:construct} summarizes the complete index construction pipeline. First, the initial $k$-NN graph is constructed (lines 1-4). Subsequently, for each node in the graph, the neighbors are sorted by their detourable route counts (lines 6-7), and the IP pruning strategy is applied to filter neighbors (lines 10-13). During IP pruning distance calculations, we simultaneously set keyword flags to enable neighbor recycling (lines 14-15). 
Finally, we augment the graph index with logical edges extracted from the knowledge graph (line 18). 

\setlength{\textfloatsep}{0.5pt}
\begin{algorithm}[tbp]
    \caption{Construction of the Hybrid Index}
    \label{alg:construct}
    \LinesNumbered
    \KwIn{fused vector data $V$, knowledge graph $KG$, numbers of iterations $it$ for NN-Descent, degree $d$}
    \KwOut{graph-based hybrid index $G$}
    $G\leftarrow$ Randomly initialize $k$ neighbors for each node;\\
    \For{($curIt\leftarrow 0; curIt<it; curIt$++)}{
        \ForEach{$u\in G$ in parallel}{
            Explore $u$'s 2-hop neighbors and update $N(u)$;\\
        }
    }
    \ForEach{$u\in G$ in parallel}{
        Calculate detourable routes for $v\in N(u)$;\\
        Sort $N(u)$ according to the number of detourable routes;\\
        $SE\leftarrow$ the first node in $N(u)$, $KE\leftarrow\emptyset$;\\
        \ForEach{$v\in N(u)$}{
            Calculate inner product between $v$ and nodes in $SE$;\\
            Set $v$.\textit{keywordFlag} based on the intersection results;\\ 
            \If{$|SE|<d \land \forall w\in SE$ s.t. $IP(w,v)<IP(v,v)$}{
                $SE\leftarrow SE\cup \{v\}$;\\
            }
            \ElseIf{$v$.\textit{keywordFlag} $=1$}{
                $KE\leftarrow KE\cup \{v\}$;\\
            }
        }
        $SE\leftarrow$ $d/4$ nodes in $SE$ and $d/4$ reverse neighbors;\\
        Append $SE$ with $d/2$ single path neighbors;\\
        $G[u].semanticEdge\leftarrow SE, G[u].keywordEdge\leftarrow KE$;\\
        $G[u].logicalEdge\leftarrow$ extend based on $KG$;\\
    }
    \Return{$G$}
\end{algorithm}
\setlength{\textfloatsep}{12pt plus 2pt minus 2pt}

\vspace{1.2mm}
\noindent \textbf{Updates of the Hybrid Index.} For the insertion of new nodes, the $k$ nearest neighbors of each newly inserted node are determined by merging two candidate sets: (1) the $k$-NN retrieved from the existing index using base semantic edges, and (2) the $k$-NN identified by performing NN-Descent among the newly inserted nodes. The $k$-nearest neighbors of each new node are transmitted to the subsequent procedures (pruning and edge augmentation), which remain identical to the initial construction. 
For node deletion, {\sf Allan-Poe} adopts the mark-deletion strategy where removed nodes remain in the index during search but are filtered from final results. 

\setlength{\textfloatsep}{10pt}
\begin{algorithm}[tbp]
    \caption{Query with the Hybrid Index}
    \label{alg:query}
    \LinesNumbered
    \KwIn{fused vector data $V$, graph-based hybrid index $G$, query $q$, specified keyword set $KW$, specified entity set $E$, multi-path weights $[w_d,w_s,w_f,w_k]$}
    \KwOut{$q$'s approximate $k$ nearest neighbors}
    Generate the query vector $q_v=f_\text{concat}(q,w_d,w_s,w_f)$;\\
    \tcc{Initialize the entry points}
    \If{query with knowledge graph}{
        $cand\leftarrow$ nodes containing user-specified entities;\\
        \For{$v\in cand$}{
            $v.hop=0$; \tcp{initial entities in the query}
            $v.ent=e$; \tcp{$e\in E$}
        }
    }
    \Else{
        $cand\leftarrow$ nodes with large vector length;\\ 
    }
    Calculate $dis(q_v,v)$ where $v\in cand$ in parallel;\\
    \tcc{Begin to search}
    \While{$|cand|>0$}{
        $u\leftarrow$ the nearest unvisited neighbor to $q_v$ in $cand$;\\
        $z\leftarrow$ the furthest neighbor to $q_v$ in $topk$;\\
        $N(u)\leftarrow G[u].semanticEdge$;\\
        \If{$K(u)\cap KW\ne \emptyset$}{
            $N(u)\leftarrow N(u)\cup G[u].keywordEdge$;\\
        }
        \If{$u.ent\ne none$}{
            $N(u)\leftarrow N(u)\cup G[u].logicalEdge$;\\
        }
        \For{unvisited $o\in N(u)$ in parallel}{
            \If{$u.ent\ne none$}{
                $o.ent\leftarrow$ entity in $o$ having relations with $u.ent$;\\
                \If{$o.ent\ne none$}{
                    $o.hop\leftarrow u.hop+1$;\\
                    $dis(o,q_v)\leftarrow dis(o,q)-w_k/o.hop$;\\
                }
            }
            \If{$dis(o,q_v)<dis(z,q_v)$}{
                $cand\leftarrow cand\cup\{o\}$, push $o$ to $topk$;\\
                \If{$|topk|>k$}{
                    pop the furthest node $m$ from $topk$;\\
                    push $m$ to $kwCand$ if $K(m)\cap KW\ne \emptyset$;\\
                }
            }
        }
    }
    \While{$|res|<k$}{
        push $s\in topk\cup kwCand$ s.t. $K(s)\cap KW \ne \emptyset$ to $res$;\\
    }
    \Return{$res$}
\end{algorithm}
\setlength{\textfloatsep}{10pt plus 2pt minus 2pt}

\subsection{Flexible Query Processing on GPU}
\label{subsec:search}
Given the hybrid index constructed in Section \ref{subsec:construction}, the query algorithm needs to be carefully designed to enable efficient retrieval across flexible path combinations while maintaining performance. 
This subsection presents our GPU-accelerated query processing algorithm, which efficiently handles keyword and knowledge graph augmentations without compromising query latency.

\vspace{-2mm}
\subsubsection{\textbf{Query on the Base Semantic Edges}.} As established in Section \ref{subsec:hybrid_vector}, base semantic edges constructed from fused vectors support arbitrary weight combinations across retrieval paths. Given a weight vector $w=[w_d,w_s,w_f]$ for dense, learned sparse, and lexical sparse vectors respectively, the query vector is formulated as $f_\text{concat}(q, w_d,w_s,w_f)=[w_d*f_\text{dense}(q),w_s*f_\text{sparse}(q),w_f*f_\text{full}(q)]$. 
Single-path retrieval is achieved by setting the corresponding weight to 1 (or any non-zero value) and others to 0. For instance, the fused query vector $f_{concat}(q,1,0,0)$ retrieves documents using only dense vector similarity. This approach extends naturally to two-path and three-path configurations. 
To optimize search efficiency, we select entry points from nodes with the largest vector norms. 
The computationally intensive hybrid distance calculations during query processing are accelerated using the same GPU-optimized strategy described in Section~\ref{subsec:construction}. The only difference is that during the set intersection operation, each thread fetches an index from the document's sparse vector and searches for it in the query's sparse vector. This design reduces time complexity by searching the typically smaller query vector.

\vspace{-2mm}
\subsubsection{\textbf{Query with Keyword Augmentation}.} {\sf Allan-Poe} enables users to specify required keywords in queries, ensuring retrieved documents contain these terms. However, loading keyword edges during every node traversal would incur substantial overhead. To address this, we implement dynamic keyword edge loading: when expanding a node's neighbors into the candidate pool, we check for keyword commonality (already computed during distance calculation) and load keyword edges only for nodes sharing keywords with the query.  
Crucially, we do not restrict traversal exclusively to keyword-matched nodes, as this would impair accuracy by excluding potential pathway nodes. Instead, we employ a twin candidate pool approach~\cite{ait2025rwalks}, maintaining a secondary pool for keyword-satisfying nodes excluded from the primary pool due to larger distances. Upon query completion, we merge both pools and filter for nodes containing the required keywords. 

\vspace{-2mm}
\subsubsection{\textbf{Query with Knowledge Graph Augmentation}.} {\sf Allan-Poe} further enables users to specify key entities in queries to enhance retrieval through logical similarity. 
As discussed in Section \ref{subsec:kg_edge}, logical edges address two key challenges: (1) complex queries with multiple entities or multi-hop, and (2) semantically similar but logically unrelated results. 
To mitigate local optima in multi-entity queries, we employ entity-based entry point selection, choosing nodes containing user-specified entities as initial entry points via the entity-node mapping. 
For query processing augmented by logical edges, the core principle is that nodes containing entities related to user-specified entities exhibit higher logical similarity, which should reduce their effective hybrid distance. 
Based on this, during the query process, for each explored node, we first expand the candidate pool using base semantic edges from the current node. If that node is within $x$ hops of user-specified entities in the knowledge graph, we additionally expand the candidate pool using its logical edges. 
It's worth noting that not all the logical edges of this node are loaded, but only those edges whose source entities are within $x$ hops of the target entities. 
Each neighbor expanded via logical edges is annotated with its hop distance from the query entities, thereby avoiding the need to recalculate the hop distance from scratch. 
Furthermore, we verify whether candidates expanded via base semantic edges are knowledge graph neighbors of entities in the explored node. 
We incorporate logical similarity by rewarding nodes based on their hop distance from query entities: fewer hops yield greater reward (i.e., reduced effective distance). This approach integrates fine-grained entity relations into the document-level graph search, effectively addressing both logically unrelated results and multi-hop query challenges.

Algorithm~\ref{alg:query} presents the pseudo-code for \textsf{Allan-Poe}'s query processing. The algorithm begins by fusing retrieval vectors (line 1) and initializing the candidate pool with path-appropriate entry points (lines 2-9). During search, the neighbor list is initialized with base semantic edges (line 13), while keyword and logical edges are dynamically loaded based on the current node (lines 14-17). For each unvisited neighbor, we compute its distance from the query vector and incorporate logical similarity (lines 19-23), then expand the candidate pool accordingly (lines 24-28). Finally, results are filtered to ensure they contain the queried keywords (lines 29-30).
\vspace{-2mm}
\section{Experiments}
\label{sec:experiments}
In this section, we conduct comprehensive experiments to evaluate the performance of {\sf Allan-Poe} and compare it with existing state-of-the-art retrieval methods. 

\vspace{-2mm}
\subsection{Experiment Settings}
\label{subsec:exp_setting}
\vspace{-1.2mm}
\subsubsection{\textbf{Datasets}.} For comprehensive evaluations, we use 6 real-world datasets of varying scales, which have been widely used in related works~\cite{zhang2024efficient,wang2025balancing,jimenez2024hipporag,bruch2024bridging}. Among them, \textbf{NaturalQuestions} (\textit{NQ})~\cite{wang-etal-2024-rear} and \textbf{MS MARCO} (\textit{MS})~\cite{nguyen2016ms} include simple queries, while \textbf{2WikiMultiHopQA} (\textit{WM})~\cite{lv2011documents} and \textbf{HotpotQA} (\textit{HP})~\cite{yang2018hotpotqa} contain complex, multi-hop queries. Table \ref{tab:dataset} summarizes the detailed information of each dataset. 
We employ the BGE-M3 model~\cite{chen2024bge} to generate the dense vectors with a dimension of 1024, the SPLADE model~\cite{formal2021splade} to generate the sparse vectors, and the BM25 algorithm~\cite{robertson2025bm25} to generate the full-text vectors. Considering the high dimensionality of various types of vectors, we extract at most one million documents from each dataset, which aligns with the typical data segment size in vector databases~\cite{wang2024starling,wang2021milvus}. 

\begin{table}[tbp]
\belowrulesep=0pt
\aboverulesep=0pt
    \centering
    \small
    \caption{Statistics of Datasets. "D. Dim", "S. Dim", and "F. Dim" denote the dimensions of the dense, sparse, and full-text vectors, respectively. }
    \vspace{-3mm}
    \setlength{\tabcolsep}{.02\linewidth}{
        \begin{tabular}{lllllll}
        \toprule
        \textbf{Dataset} & \textbf{\#Corpus} & \textbf{\#Queries} & \textbf{D. Dim} & \textbf{S. Dim} & \textbf{F. Dim} \\
        \toprule
          \textit{NQ}~\cite{wang-etal-2024-rear} & 1,000,000 & 1,000 & 1,024 & 30,522 & 852,356\\
          \textit{MS}~\cite{nguyen2016ms} & 1,000,000 & 1,000 & 1,024 & 30,522 & 831,592\\
          \textit{WM}~\cite{lv2011documents} & 414,743 & 1,000 & 1,024& 30,522 & 529,931\\
          \textit{HP}~\cite{yang2018hotpotqa} & 509,176 & 1,000 & 1,024 & 30,522 & 699,002\\
          \textit{NQ-9633}~\cite{wang-etal-2024-rear} & 9,633 & 100 & 1,024 & 30,522 & 42,834\\
          \textit{WM-6119}~\cite{lv2011documents} & 6,119 & 100 & 1,024 & 30,522 & 33,357\\
        \bottomrule
        \end{tabular}
    }
    \vspace{-2mm}
    \label{tab:dataset}
\end{table}

\label{subsec:overall}
\begin{figure*}
    \centering
    \includegraphics[width=0.71\textwidth]{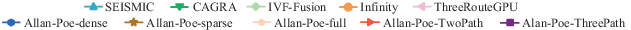}\\
    \vspace{-2mm}
    \subfigcapskip = -4pt
    \subfigure[\textit{NQ}]{
    \includegraphics[width=0.245\textwidth]{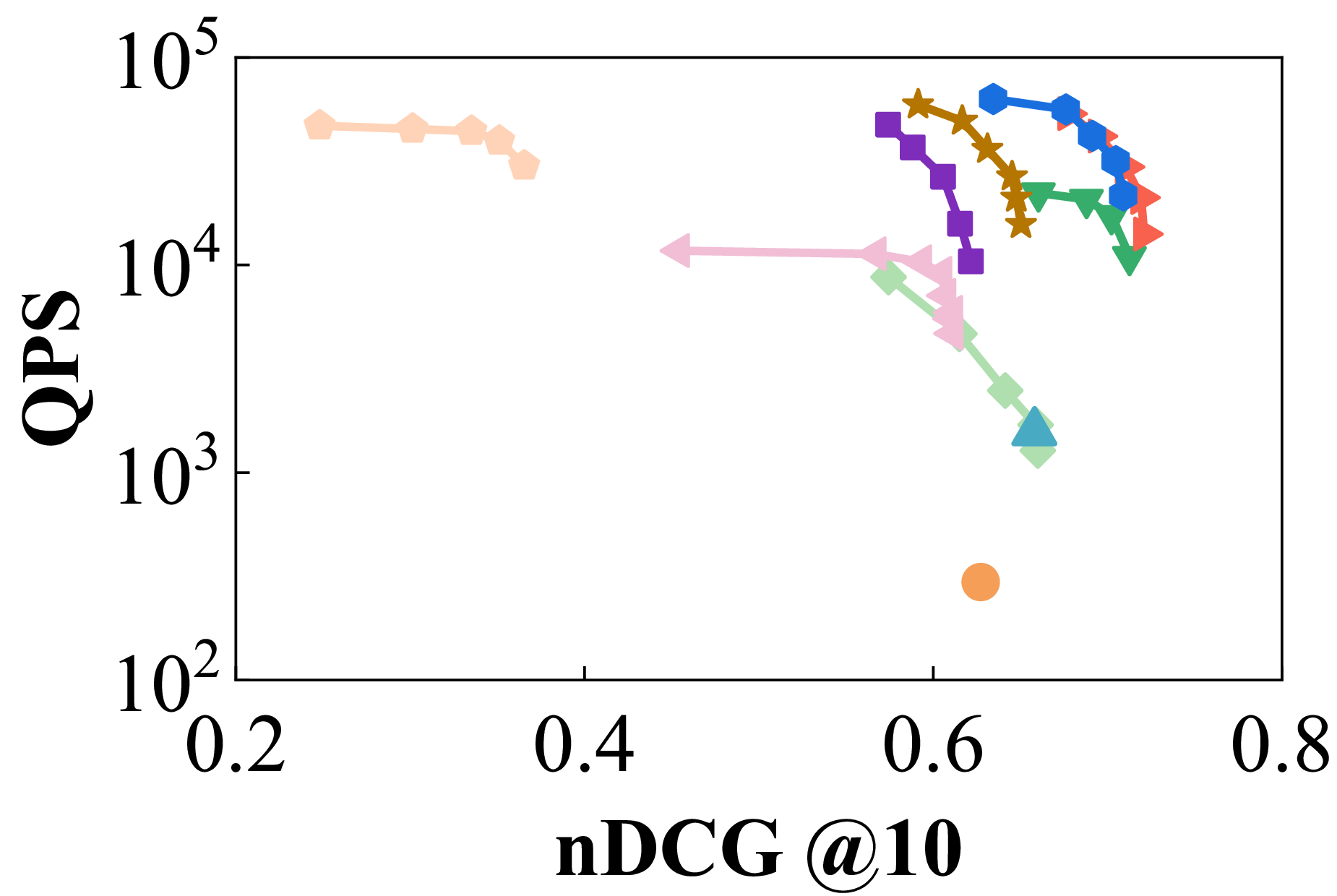}}
    \hspace{-1.4mm}
    \subfigure[\textit{MS}]{
    \includegraphics[width=0.245\textwidth]{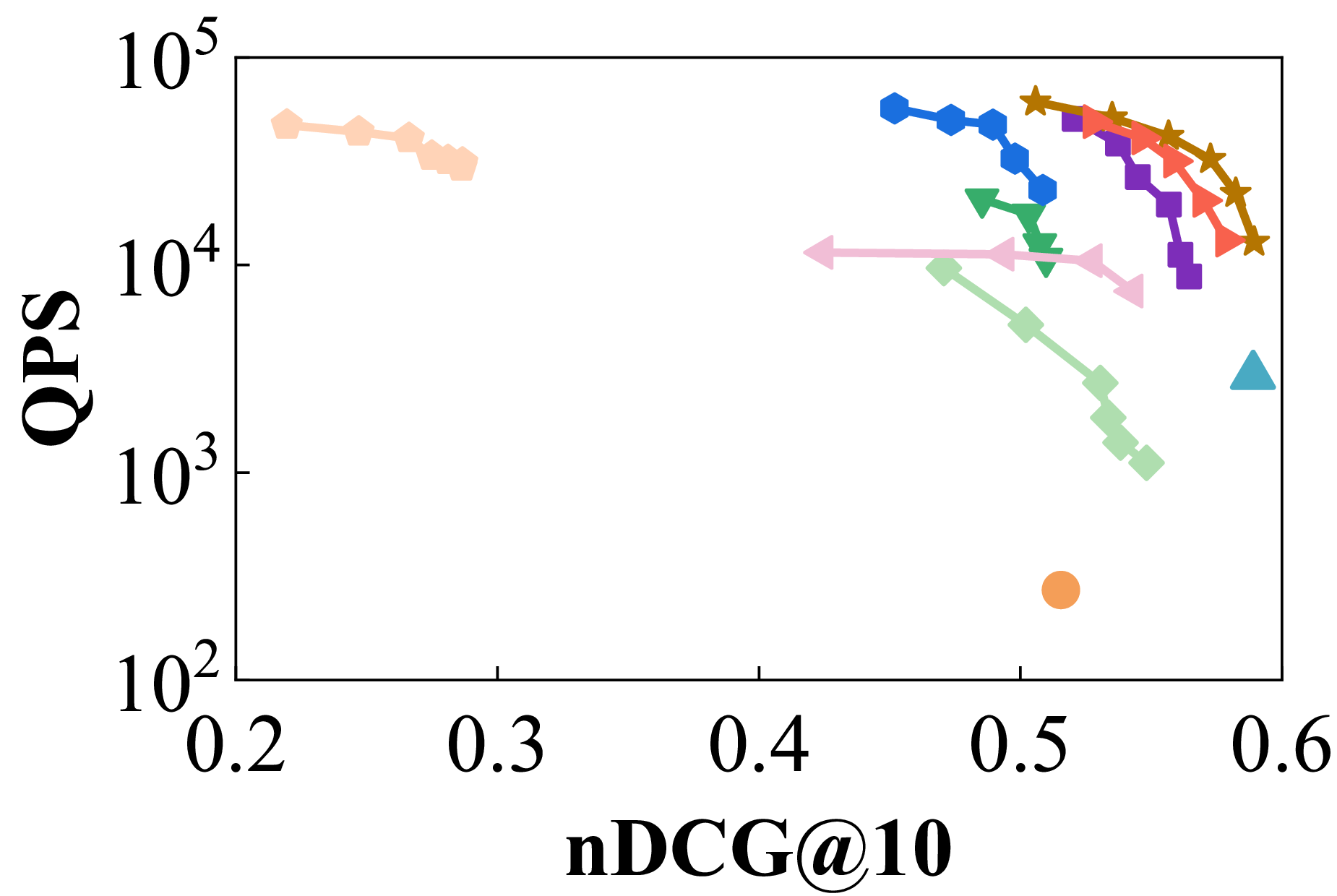}}
    \hspace{-1.4mm}
    \subfigure[\textit{WM}]{
    \includegraphics[width=0.245\textwidth]{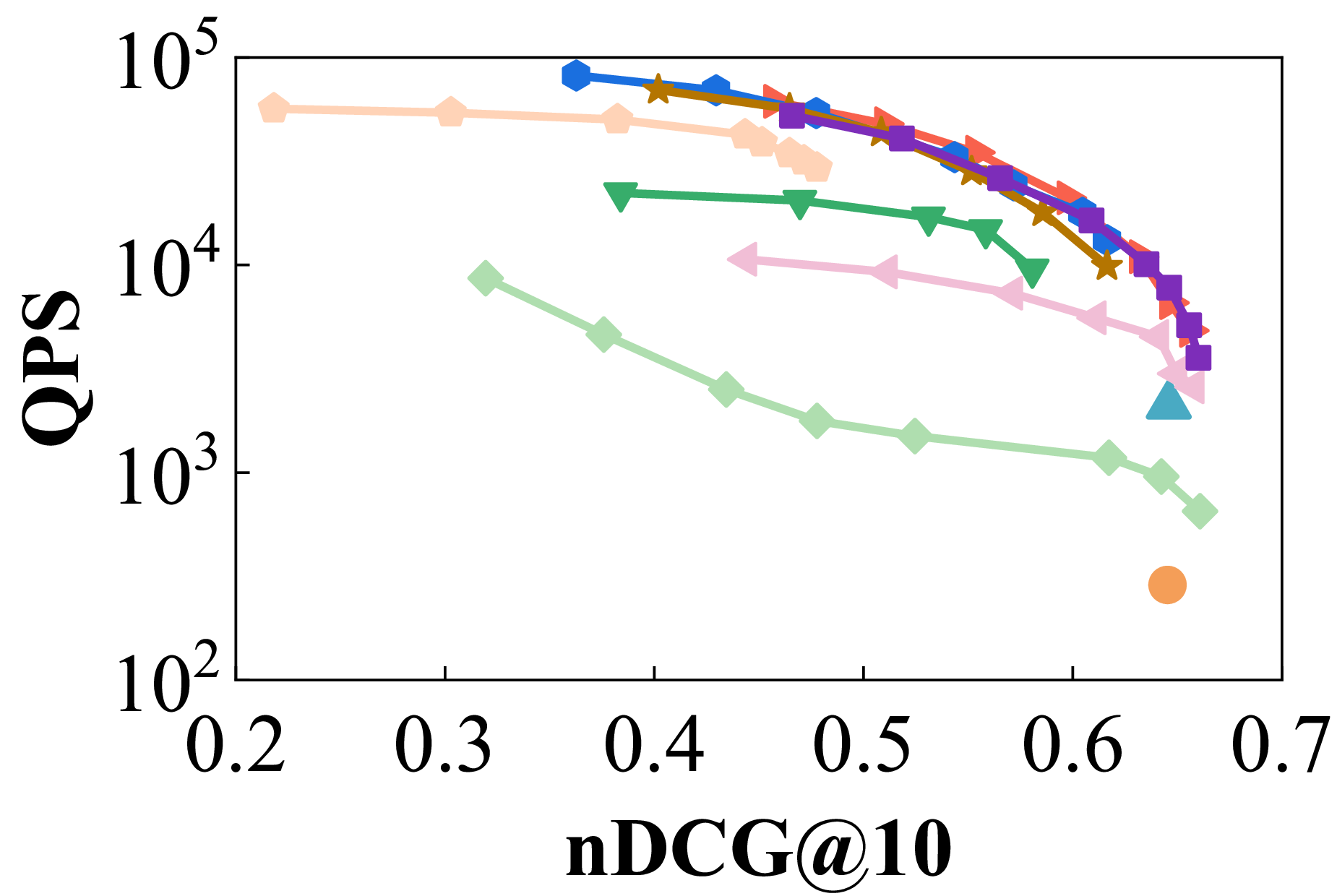}}
    \hspace{-1.4mm}
    \subfigure[\textit{HP}]{
    \includegraphics[width=0.245\textwidth]{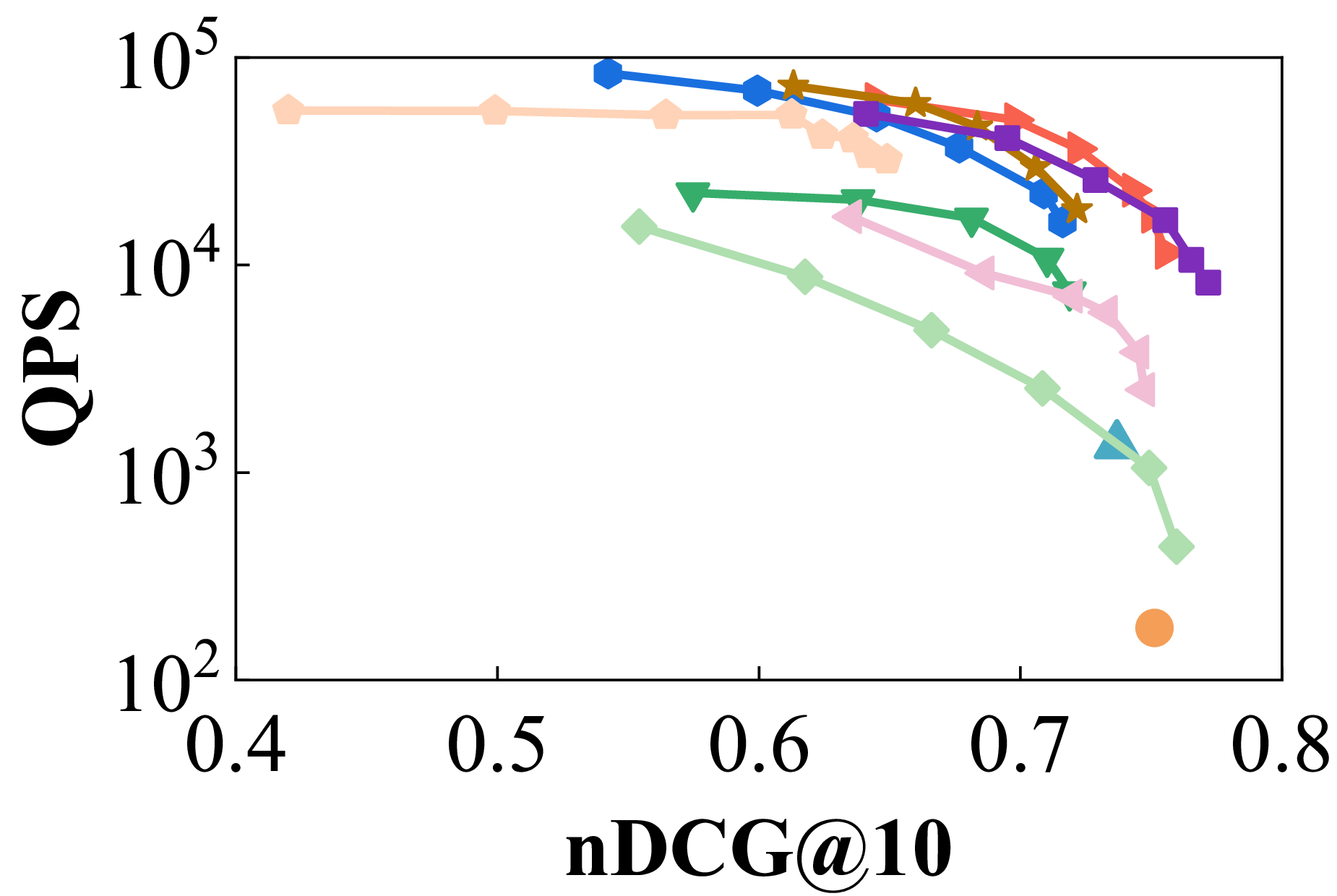}}
    \vspace{-6mm}
    \caption{Comparison of all methods on 4 real-world datasets. Paths in hybrid search methods share equal weights.}
    \label{fig:over_performance}
    \vspace{-4mm}
\end{figure*}

\vspace{-2.8mm}
\subsubsection{\textbf{Methods}.} We evaluate {\sf Allan-Poe} against 6 state-of-the-art competitors with both hybrid and single-path retrieval paradigms:
\begin{itemize}
\vspace{-1mm}
    \item \textbf{\textsf{SEISMIC}}~\cite{bruch2024efficient} is the state-of-the-art method supporting only the sparse vector search.
    \item \textbf{\textsf{CAGRA}}~\cite{ootomo2024cagra} is the state-of-the-art GPU-based method supporting only the dense vector search.
    \item \textbf{\textsf{IVF-Fusion}}~\cite{bruch2024bridging} is a hybrid search method adopting the fusion retrieval paradigm. It combines dense vectors and sparse vectors using the Johnson-Lindenstrauss (JL) transformation~\cite{johnson1984extensions} and utilizes an inverted index to accelerate the search. We implemented a GPU version of \textsf{IVF-Fusion} for fair comparison. 
    \item \textbf{\textsf{Infinity}}~\cite{infinity_hybrid,wang2025balancing} is a modern database featuring with efficient hybrid search. It adopts the multi-path retrieval paradigm, which supports combinations of dense vector, sparse vector, and full-text search. 
    \item \textbf{\textsf{ThreeRouteGPU}} is our implemented GPU-based hybrid search method adopting the multi-path retrieval paradigm. It constructs separate {\sf CAGRA} indexes for dense vectors, learned sparse vectors, and lexical sparse vectors (reducing dimension via JL transformation). Results retrieved from the three routes are then fused using fixed weights. 
    \item \textbf{\textsf{HippoRAG}}~\cite{gutierrez2025rag} is a state-of-the-art GraphRAG algorithm. 
    Following evaluations in recent studies~\cite{zhou2025depth,jimenez2024hipporag}, we evaluate knowledge graph augmentation on two benchmark datasets, \textit{NQ-9633} (50k entities, 68k relations) and \textit{WM-6119} (30k entities, 31k relations). 
    \item \textbf{\textsf{Allan-Poe}} is our proposed hybrid search method, with configurations denoted as: {\sf dense} (dense vectors only), {\sf sparse} (sparse vectors only), {\sf full} (full-text only), {\sf TwoPath} (dense and sparse vectors), and {\sf ThreePath} (all three retrieval paths combined). All configurations leverage the same hybrid index without reconstruction. 
\end{itemize}

\vspace{-2mm}
\subsubsection{\textbf{Metrics}.} We evaluate the indexing efficiency by measuring the construction time, the retrieval efficiency by \textit{Queries Per Second} (QPS), and the retrieval accuracy by nDCG@$k$ (higher is better). Without additional explanation, the value of $k$ is set to 10. 

\vspace{-1.4mm}
\subsubsection{\textbf{Platforms}.} All experiments are conducted on a server featuring an Intel Xeon Silver 4310 CPU@2.10GHz, 125GB RAM, and a Nvidia GeForce RTX 3090 GPU (24G). We implement Allan-Poe in C++/CUDA under CUDA 12.2. 

\subsection{Overall Performance for Query Processing}
In this section, we compare the QPS and nDCG@10 across all methods on real-world datasets. All hybrid search methods employ equal weighting for all retrieval paths unless specified otherwise. The CPU-based methods, i.e., {\sf SEISMIC} and {\sf Infinity}, utilize 48 threads during query processing. 
The results are depicted in Figure \ref{fig:over_performance}. For {\sf Infinity} and {\sf SEISMIC}, we report single data points since nDCG values show minimal variation with parameter tuning (e.g., $k'$).

Experimental results shown in Figure \ref{fig:over_performance} demonstrate that the corresponding retrieval configurations of {\sf Allan-Poe} outperform {\sf SEISMIC}, {\sf CAGRA}, {\sf IVF-Fusion}, {\sf Infinity}, and {\sf ThreeRouteGPU} by 4.4-13.6$\times$, 1.5-2.5$\times$, 17.6-41.8$\times$, 27.1-186.4$\times$, and 1.6-4.8$\times$, respectively. 
On datasets \textit{WM} and \textit{HP}, {\sf Allan-Poe-ThreePath} achieves the highest accuracy (nDCG@10). Despite requiring more distance calculations, {\sf Allan-Poe-ThreePath} maintains higher nDCG@10 than other methods at equivalent QPS levels on these datasets, demonstrating the benefit of complementary information from multiple retrieval paths.
From another perspective, {\sf Allan-Poe-dense} and {\sf CAGRA} solely employ dense vectors to retrieve documents, which consistently underperform the hybrid search paradigm across most datasets in nDCG@10 regardless of the QPS. 
For example, on \textit{MS}, the nDCG@10 of {\sf Allan-Poe-dense} and {\sf CAGRA} reach only 0.5, compared to {\sf Allan-Poe-ThreePath}'s nDCG@10 of 0.56.  
This confirms that the retrieval path using dense vectors alone is insufficient for end-to-end retrieval. 
Consequently, introducing more retrieval paths to complement the lost information of dense vectors is a promising way to enhance the end-to-end query efficiency. 

\begin{table}[tbp]
    \centering
    \small
    \caption{Comparison of index build time and index size.}
    \vspace{-4mm}
    \setlength{\tabcolsep}{.01\linewidth}{
    \begin{tabular}{lcccccccc}
        \toprule
        \multirow{2}{*}{\textbf{Methods}}  & \multicolumn{4}{c}{\textbf{Build Time (s)}} & \multicolumn{4}{c}{\textbf{Index Size (MB)}}\\
        \cmidrule(r){2-5} \cmidrule(r){6-9}
          & \textit{NQ} & \textit{MS} & \textit{WM} & \textit{HP} & \textit{NQ} & \textit{MS} & \textit{WM} & \textit{HP}\\
          \midrule
        {\sf SEISMIC} & 98.09 & 114.34 & 50.83 & 87.38 & 2921 & 1993 & 1526 & 1904 \\
        {\sf CAGRA} & 16.29 & 17.62 & 7.45 & 8.96 & 126 & 126 & 52 & 64 \\
        {\sf IVF-Fusion} & 2.63 & 2.42 & 1.77 & 1.41 & 136 & 136 & 131 & 134 \\
        {\sf Infinity} & 487.04 & 440.93 & 186.54 & 263.19 & 5738 & 4541 & 1962 & 2701 \\
        {\sf ThreeRouteGPU} & 49.20 & 48.89 & 22.56 & 27.18 & 378 & 378 & 156 & 192\\
        {\sf Allan-Poe} & 45.83 & 40.63 & 21.86 & 26.36 & 186 & 186 & 78 & 95\\
        \bottomrule
    \end{tabular}
    }
    \vspace{-6mm}
    \label{tab:indexing_overhead}
\end{table}

However, as noted in Section \ref{subsec:motivation}, additional retrieval paths do not guarantee improved accuracy in all scenarios. For instance, on datasets \textit{NQ} and \textit{MS}, {\sf Allan-Poe-TwoPath} and {\sf Allan-Poe-sparse} achieve optimal performance, respectively. 
This occurs because all participating paths influence the final accuracy—on \textit{NQ} and \textit{MS}. While dense and sparse retrieval perform well, full-text retrieval underperforms and reduces overall accuracy. 
Moreover, for dataset \textit{MS}, {\sf Allan-Poe-sparse} using a single retrieval path outperforms the hybrid retrieval methods because all the documents in \textit{MS} have a short length with simple semantics, which can be efficiently handled by sparse vectors. 
These results emphasize the importance of supporting flexible path combinations without index reconstruction, corresponding to the flexibility dimension in Figure~\ref{fig:trillemma}. 

For different retrieval paradigms adopted by existing methods (fusion retrieval and separate multi-path retrieval), the fusion retrieval paradigm (represented by {\sf Allan-Poe-ThreePath}) outperforms the separate multi-path retrieval paradigm (represented by {\sf ThreeRouteGPU} and {\sf Infinity}) on \textit{MS} and \textit{HP} in nDCG@10 regardless of the QPS, while achieving comparable nDCG@10 on \textit{NQ} and \textit{WM}. As discussed in Section \ref{subsec:motivation}, separate multi-path retrieval can miss relevant documents, reducing accuracy, which underscores the advantage of fusion retrieval. 

In terms of efficiency, the different configurations of \textsf{Allan-Poe} achieve the highest QPS among all methods, even with multiple retrieval paths. This performance stems from our GPU optimizations, particularly the hybrid distance calculation that addresses the primary retrieval overhead.

\begin{figure}
    \centering
    \includegraphics[width=0.47\textwidth]{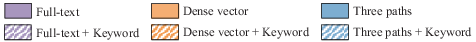}\\
    \subfigcapskip = -4pt
    \vspace{-2mm}
    \subfigure[Retrieval accuracy on \textit{NQ}.]{
    \includegraphics[width=0.23\textwidth]{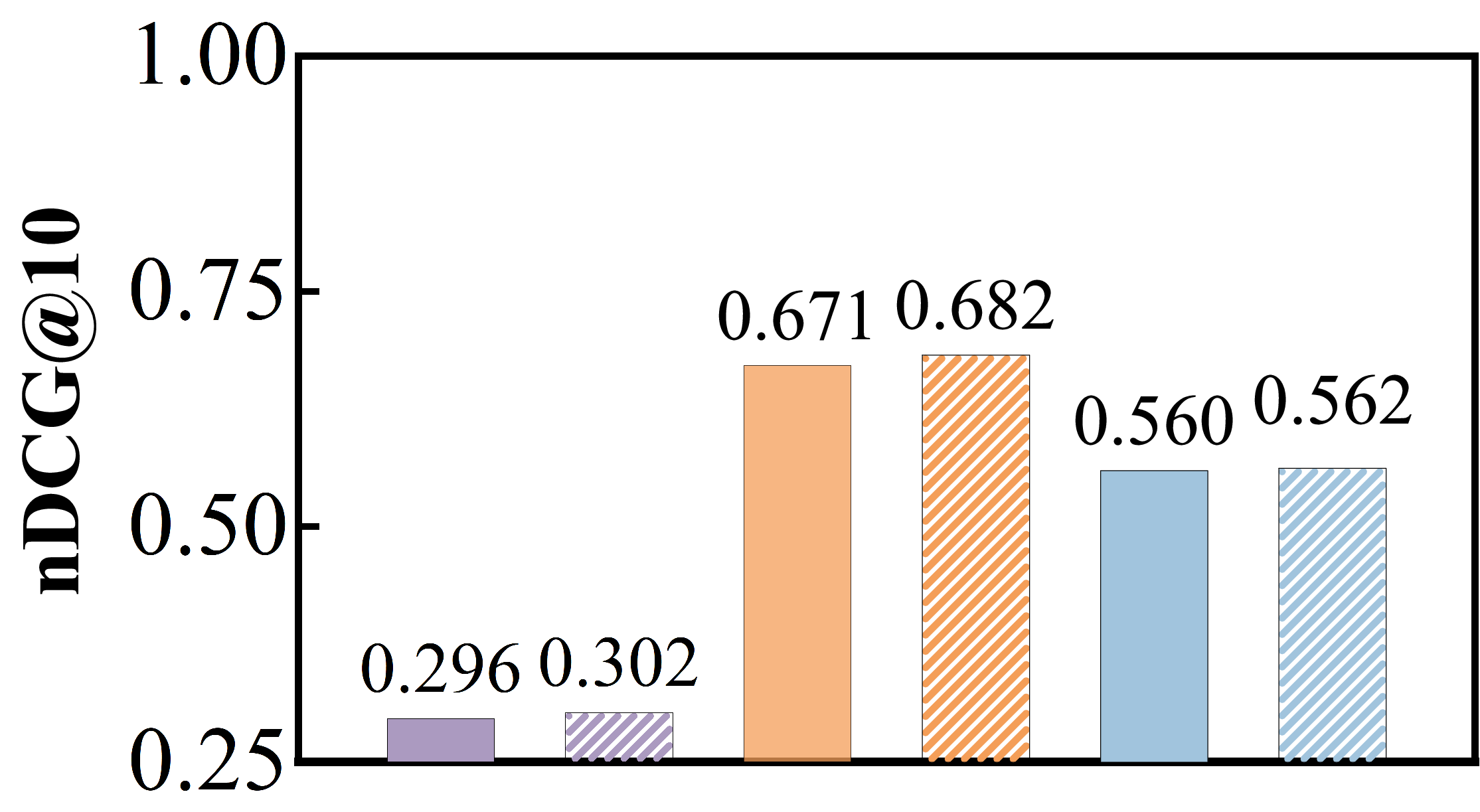}}
    \subfigure[Retrieval efficiency on \textit{NQ}.]{
    \includegraphics[width=0.23\textwidth]{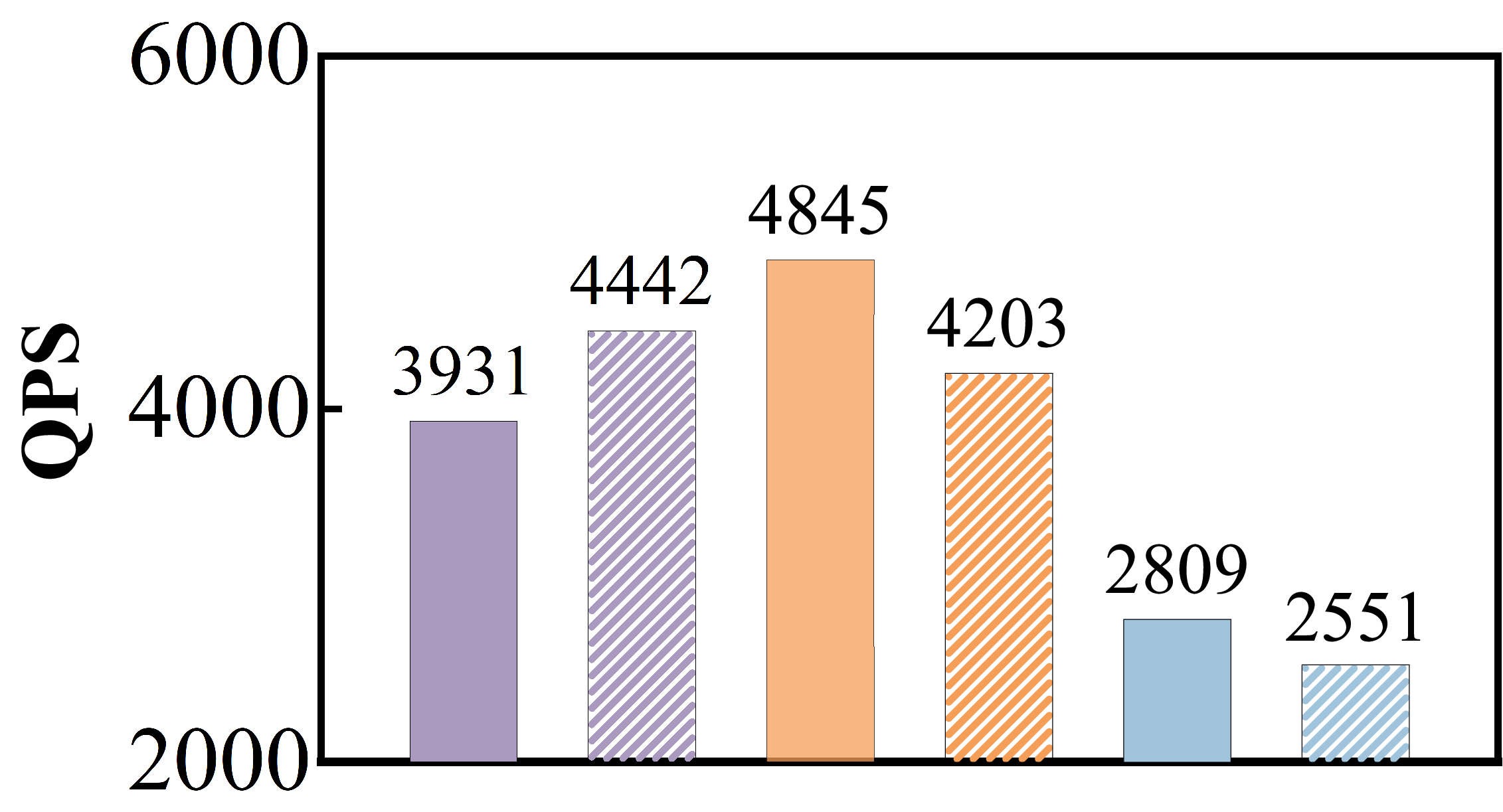}}\\
    \vspace{-3mm}
    \subfigure[Retrieval accuracy on \textit{WM}.]{
    \includegraphics[width=0.23\textwidth]{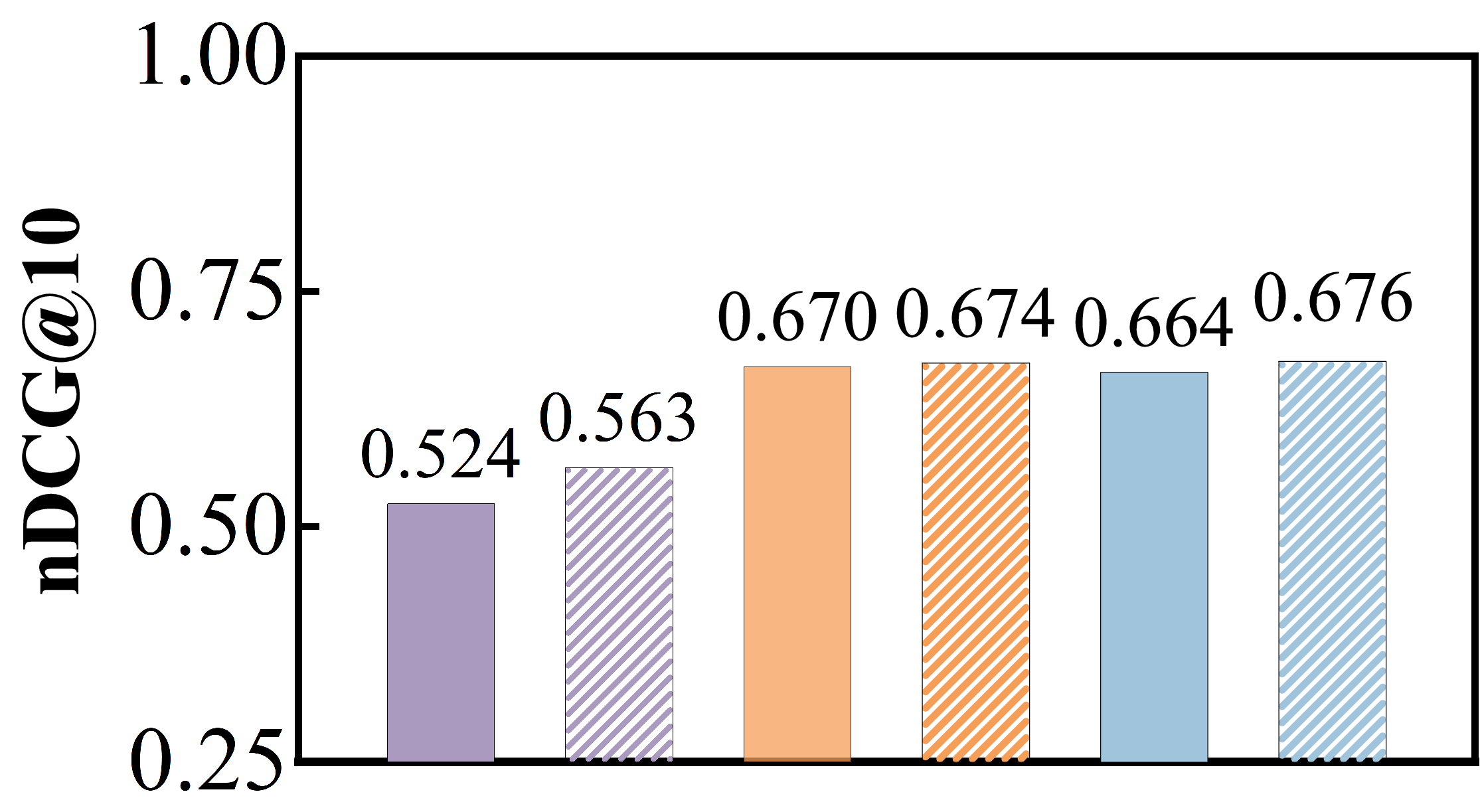}}
    \subfigure[Retrieval efficiency on \textit{WM}.]{
    \includegraphics[width=0.23\textwidth]{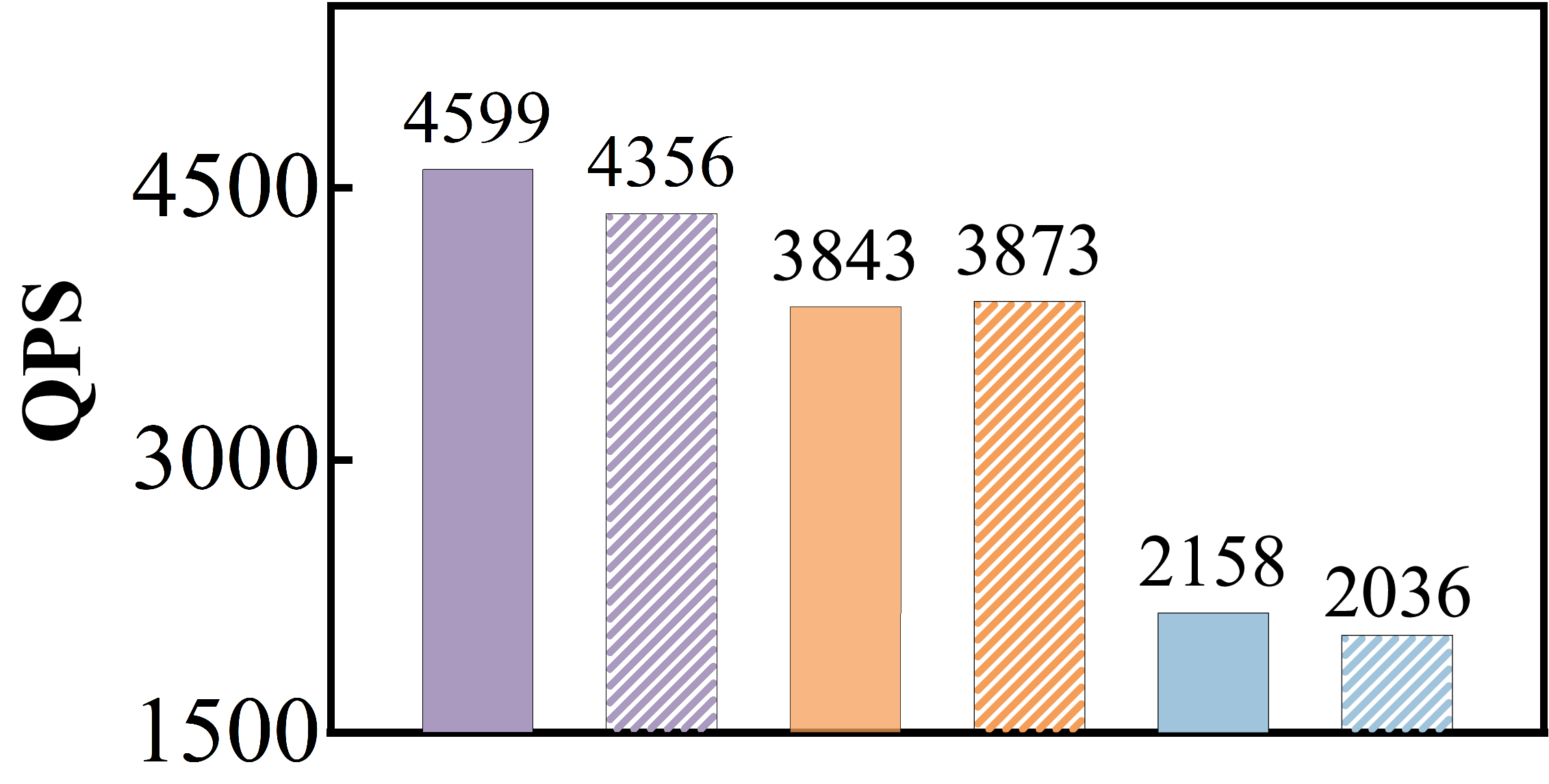}}
    \vspace{-5mm}
    \caption{Comparison of retrieval methods w/o keywords. }
    \vspace{-4mm}
    \label{fig:keyword_eva}
\end{figure}

\vspace{-2mm}
\subsection{Evaluations of Indexing Efficiency}

In Table \ref{tab:indexing_overhead}, we report the index construction time as well as the index size across all the compared methods. 
Among three-path retrieval methods, {\sf Allan-Poe} achieves the fastest build time while maintaining a compact index size. 
Specifically, compared to the GPU-based method {\sf ThreeRouteGPU}, {\sf Allan-Poe} demonstrates slightly faster construction and 2.0$\times$ smaller index size. 
These advantages are even more pronounced against {\sf Infinity}, with 10.0$\times$ faster build time and 21.0$\times$ reduction in index size. 
These results validate the effectiveness of our GPU-accelerated construction optimizations in Section \ref{subsec:construction}, which includes the hybrid distance acceleration and the parallel pruning implementation for heterogeneous edges. 
The compact index size further demonstrates the effectiveness of our unified design, which significantly reduces storage overhead and system complexity compared to separate index paradigms. 
Among all methods, \textsf{IVF-Fusion} achieves the fastest construction due to its simple inverted index structure, but suffers from consistently poor accuracy. \textsf{CAGRA} maintains a small index by supporting only single-path retrieval for dense vectors, but its retrieval performance lags behind hybrid approaches.

\vspace{-2mm}
\subsection{Evaluations of Keyword Retrieval}

To evaluate the effectiveness of keyword specification in queries, we employ the Qwen3 LLM~\cite{yang2025qwen3} to simulate users specifying required keywords for the first 100 queries from the representative datasets \textit{NQ} and \textit{WM}. 
Figure~\ref{fig:keyword_eva} compares results with and without keyword supplementation, reporting the highest nDCG@10 and corresponding QPS for each method.
Keyword supplementation improves nDCG@10 by 1.2\% on average across retrieval paths, with only a 3.2\% QPS reduction, demonstrating its effectiveness.
However, some methods show minimal accuracy gains (e.g., 0.2\% for three-path retrieval on \textit{NQ}), as many relevant documents already contain the required keywords, limiting the filtering impact. 
Interestingly, certain retrieval paths with keyword supplementation achieve higher QPS than their non-supplemented counterparts (e.g., full-text search on \textit{NQ} and dense vector search on \textit{WM}). This suggests that keyword constraints enable earlier convergence to optimal accuracy by focusing the search on more relevant nodes within a small candidate pool in the query process. 

\vspace{-2mm}
\subsection{Evaluations of Logical Augmentation}

\begin{table}[tbp]
\renewcommand{\arraystretch}{1.1}
\belowrulesep=0pt
\aboverulesep=0pt
    \centering
    \small
    \caption{Comparison of retrieval methods on small datasets. `KG' denotes the methods augmented by knowledge graphs.}
    \vspace{-4mm}
    \setlength{\tabcolsep}{.01\linewidth}{
    \begin{tabular}{lcccc}
    \toprule
        \multirow{2}{*}{\textbf{Methods}} & \multicolumn{2}{c}{\textbf{\textit{NQ-9633}}} & \multicolumn{2}{c}{\textbf{\textit{WM-6119}}} \\
        \cmidrule(r){2-3} \cmidrule(r){4-5}
          & nDCG@10 & QPS & nDCG@10 & QPS\\
          \midrule
          {\sf SEISIMC} & 0.732 & 838.10 & 0.765 & 659.40\\
          {\sf CAGRA} & 0.732 & 1254.82 & 0.761 & 1070.75\\
          {\sf IVF-Fusion} & 0.740 & 4030.35 & 0.780 & 3316.88\\
          {\sf Infinity} & 0.749 & 211.43 & 0.741 & 232.37\\
          {\sf ThreeRouteGPU} & 0.736 & 1167.36 & 0.765 & 1085.5\\
          {\sf HippoRAG} & 0.747 & 1.28 & 0.805 & 1.04\\
          {\sf Allan-Poe-full} & 0.605 & \textbf{14716.81} & 0.699 & 6306.72\\
          {\sf Allan-Poe-full} (+KG) & 0.666 & 9212.35 & 0.817 & 4572.75\\
          {\sf Allan-Poe-dense} & 0.728 & 13581.5 & 0.759 & \textbf{9487.67}\\
          {\sf Allan-Poe-dense} (+KG) & 0.738 & 9428.36 & 0.818 & 5941.49\\
          {\sf Allan-Poe-sparse} & 0.730 & 13303.75 & 0.759 & 6553.34\\
          {\sf Allan-Poe-sparse} (+KG) & 0.744 & 8318.01 & 0.832 & 5240.93\\
          {\sf Allan-Poe-ThreePath} & \textbf{0.751} & 12098.23 & 0.770 & 5508.97\\
          {\sf Allan-Poe-ThreePath} (+KG) & \textbf{0.751} & 8520.28 & \textbf{0.834} & 4126.26\\
        \bottomrule
    \end{tabular}
    }
    \vspace{-5mm}
    \label{tab:kg_eva}
\end{table}

To evaluate knowledge graph augmentation while managing the knowledge graph construction costs, we use Qwen3 to construct knowledge graphs for the two benchmark datasets \textit{NQ-9633} and \textit{WM-6119}, and simulate user queries by specifying required entities for the first 100 queries. 
As mentioned in Section \ref{subsec:keyword_edge}, logical edge augmentation is a trade-off between improvements in accuracy and costs of constructing the knowledge graph. 
Applying it to small-scale, high-value data is a cost-effective option. 
Table~\ref{tab:kg_eva} presents the evaluation results.
Overall, {\sf Allan-Poe-ThreePath} (+KG) exhibits the highest nDCG@10 while maintaining competitive QPS. On \textit{NQ-9633}, which contains simple queries without multi-hop reasoning, knowledge graph augmentation provides modest accuracy improvements—particularly for single-path {\sf Allan-Poe} variants, with slight gains for the three-path configuration. 
Conversely, the accuracy improvements on \textit{WM-6119} are significant due to the complex multi-hop queries in this dataset. 
While \textsf{HippoRAG} (a GraphRAG method) achieves higher nDCG@10 than approaches without knowledge graphs, it still underperforms compared to \textsf{Allan-Poe}. This gap occurs because {\sf HippoRAG} does not effectively integrate knowledge graph information with document-level semantic similarity, and its community search can introduce redundant documents that impair query efficiency despite being useful for global queries. 

To evaluate the effectiveness of logical edge augmentation at scale, we assess \textsf{Allan-Poe} with and without knowledge graph enhancement on the larger \textit{NQ} and \textit{WM} datasets. The results, summarized in Table~\ref{tab:kg_large_eva}, show consistent accuracy gains with knowledge graph integration. Notably, improvements are more pronounced on the \textit{WM} dataset, which contains complex, multi-hop queries. These findings underscore the significant potential of enhancing document-level semantic search with fine-grained, entity-level logical structures from knowledge graphs.

\begin{table}[tbp]
\renewcommand{\arraystretch}{1.1}
\belowrulesep=0pt
\aboverulesep=0pt
    \centering
    \small
    \caption{Effectiveness of KG augmentation on large datasets.}
    \vspace{-4mm}
    \setlength{\tabcolsep}{.01\linewidth}{
    \begin{tabular}{lcccc}
    \toprule
        \multirow{2}{*}{\textbf{Methods}} & \multicolumn{2}{c}{\textbf{\textit{NQ}}} & \multicolumn{2}{c}{\textbf{\textit{WM}}} \\
        \cmidrule(r){2-3} \cmidrule(r){4-5}
          & nDCG@10 & QPS & nDCG@10 & QPS\\
          \midrule
          {\sf Allan-Poe-ThreePath} & 0.580 & 28952.61 & 0.649 & 28335.84\\
          {\sf Allan-Poe-ThreePath} (+KG) & 0.631 & 26503.97 & 0.769 & 25010.07\\
        \bottomrule
    \end{tabular}
    }
    \vspace{-5mm}
    \label{tab:kg_large_eva}
\end{table}

\vspace{-2mm}
\subsection{Effectiveness of Heterogeneous Edges}
In this subsection, we evaluate the effectiveness of the three types of heterogeneous edges in our hybrid graph-based index, which are established in Section \ref{subsec:construction}. 

\vspace{-2mm}
\subsubsection{\textbf{Effectiveness of RNG-IP Joint Pruning}.}
\begin{figure}
    \centering
    \includegraphics[width=0.47\textwidth]{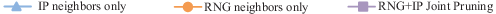}\\
    \subfigcapskip = -4pt
    \vspace{-2mm}
    \subfigure[Performance on \textit{NQ}.]{
    \includegraphics[width=0.235\textwidth]{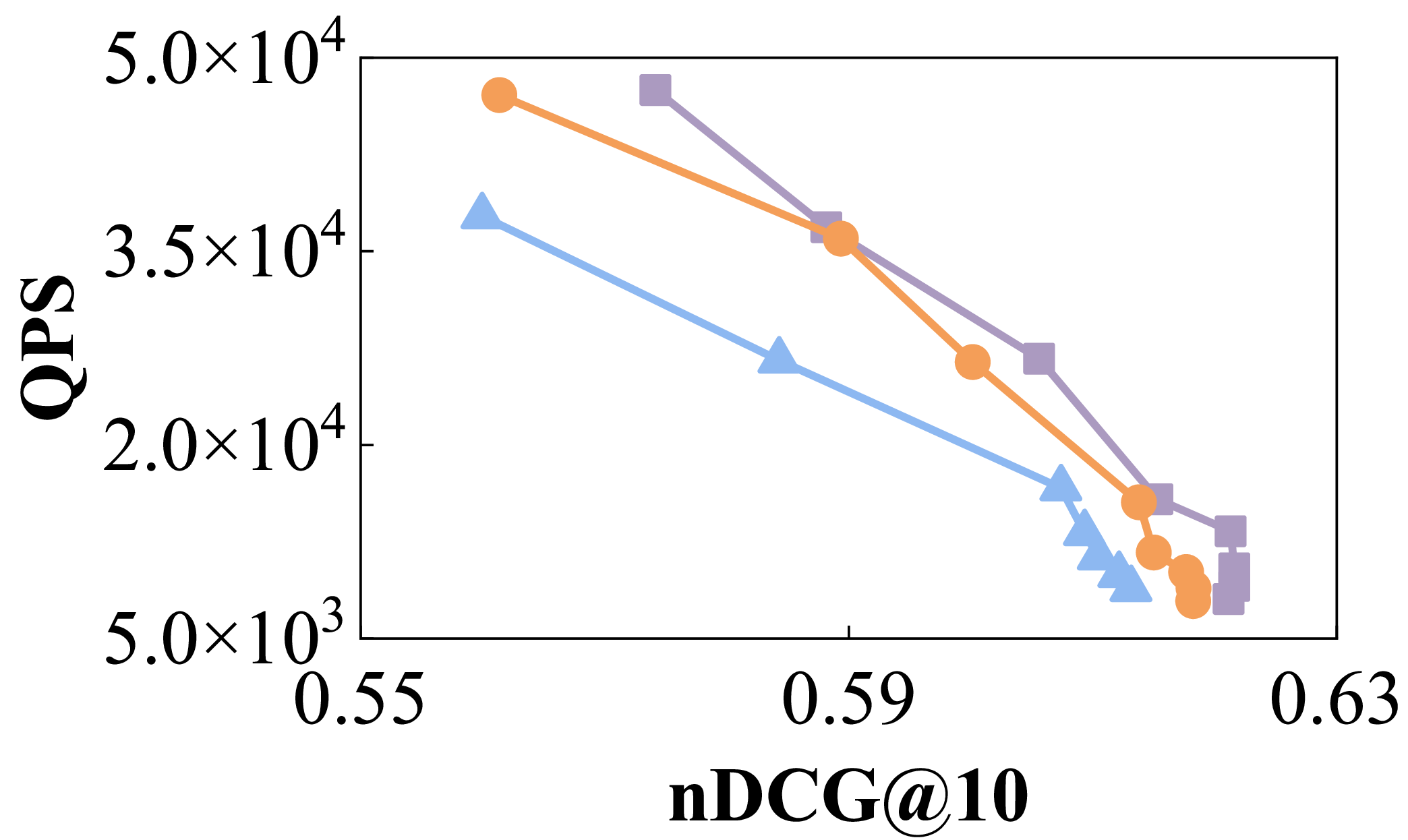}}
    \hspace{-2mm}
    \subfigure[Performance on \textit{WM}.]{
    \includegraphics[width=0.23\textwidth]{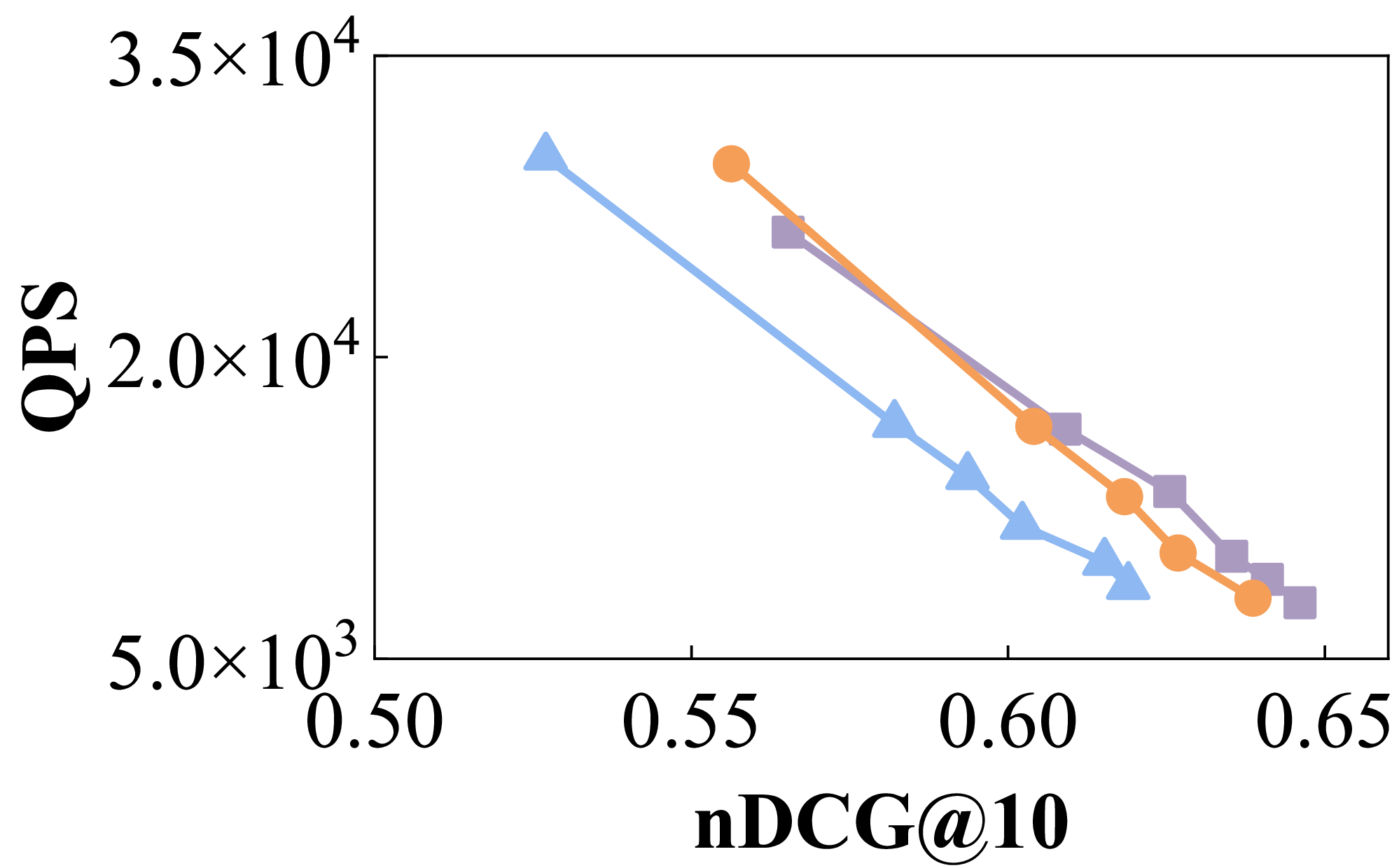}}
    \vspace{-5mm}
    \caption{Performance w/o RNG-IP joint pruning. }
    \vspace{-5mm}
    \label{fig:ip_eva}
\end{figure}

Figure \ref{fig:ip_eva} compares the performance of {\sf Allan-Poe-ThreePath} with and without the RNG-IP joint pruning strategy on two representative datasets. 
The joint RNG-IP pruning strategy improves both retrieval efficiency and accuracy compared to using RNG pruning alone, demonstrating its effectiveness in enhancing index quality. 
Notably, while the distance metric is Inner Product, using IP pruning alone (without RNG) yields lower performance than RNG pruning alone because IP pruning eliminates fewer candidate neighbors, providing limited reduction in search computation cost.  

\vspace{-2mm}
\subsubsection{\textbf{Effectiveness of Keyword Edges}.} 
\begin{figure}
    \centering
    \includegraphics[width=0.4\textwidth]{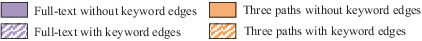}\\
    \subfigcapskip = -4pt
    \vspace{-2mm}
    \subfigure[\textit{NQ}]{
    \includegraphics[width=0.12\textwidth]{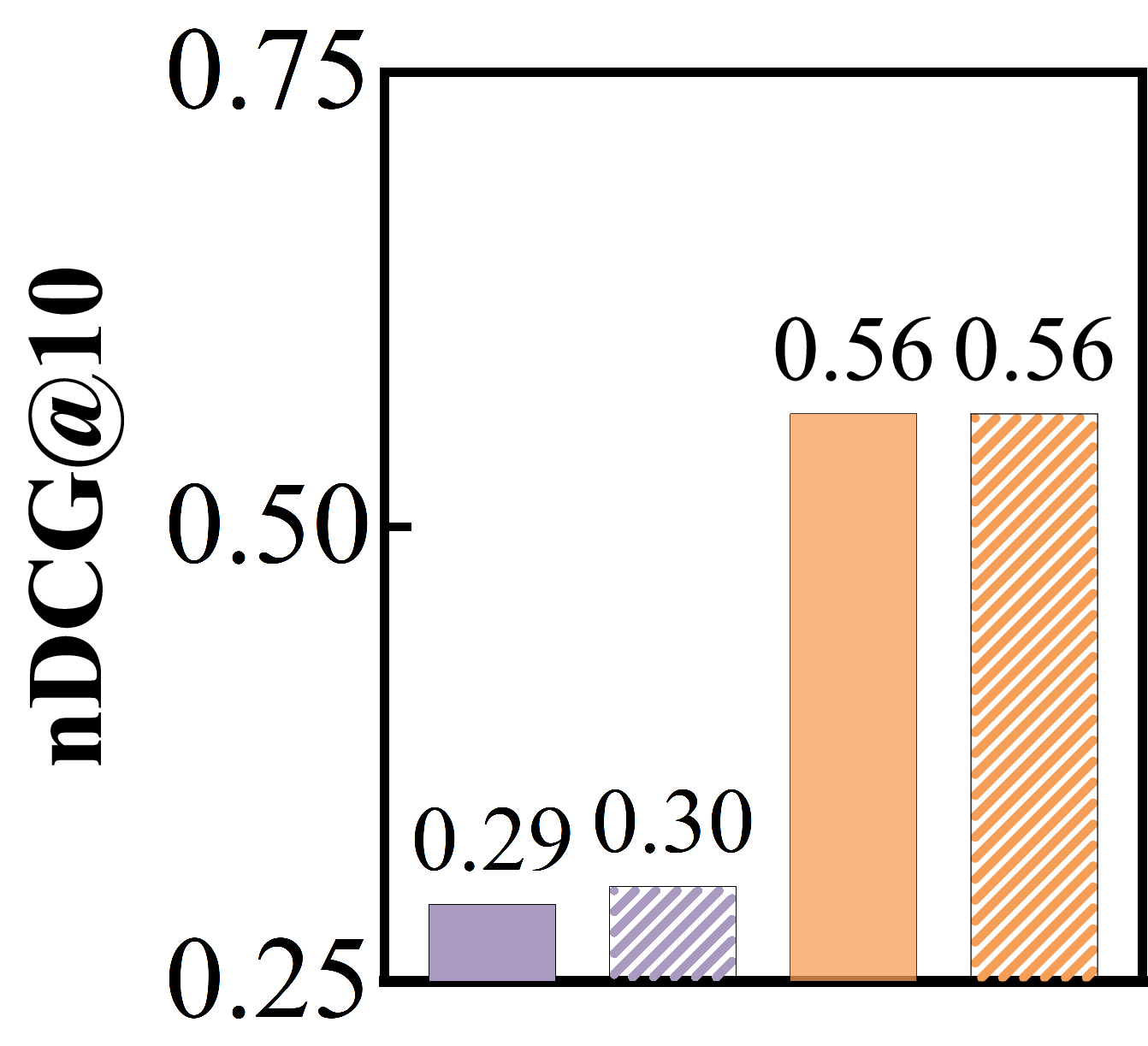}
    \includegraphics[width=0.105\textwidth]{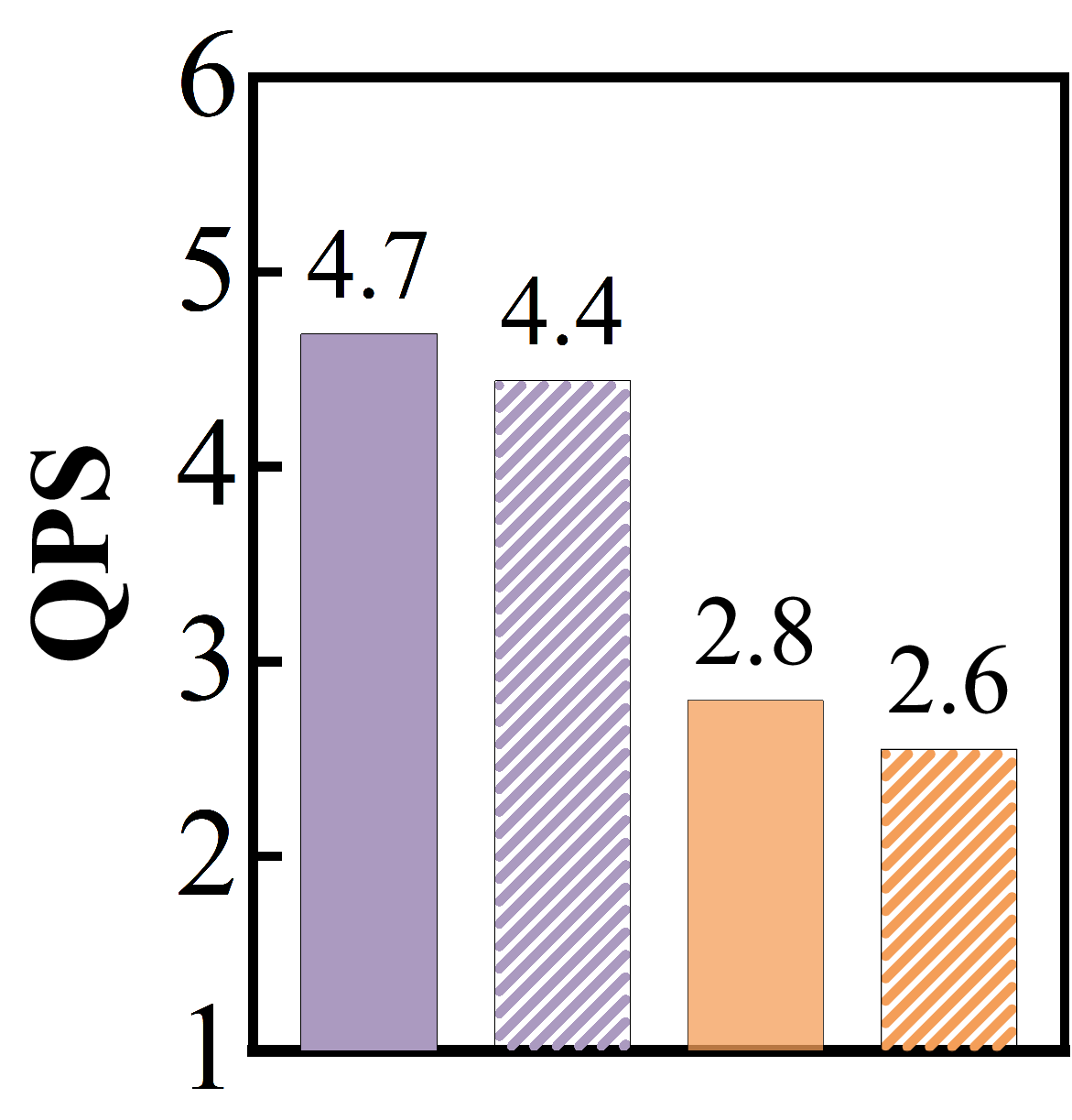}}
    \subfigure[\textit{WM}]{
    \includegraphics[width=0.12\textwidth]{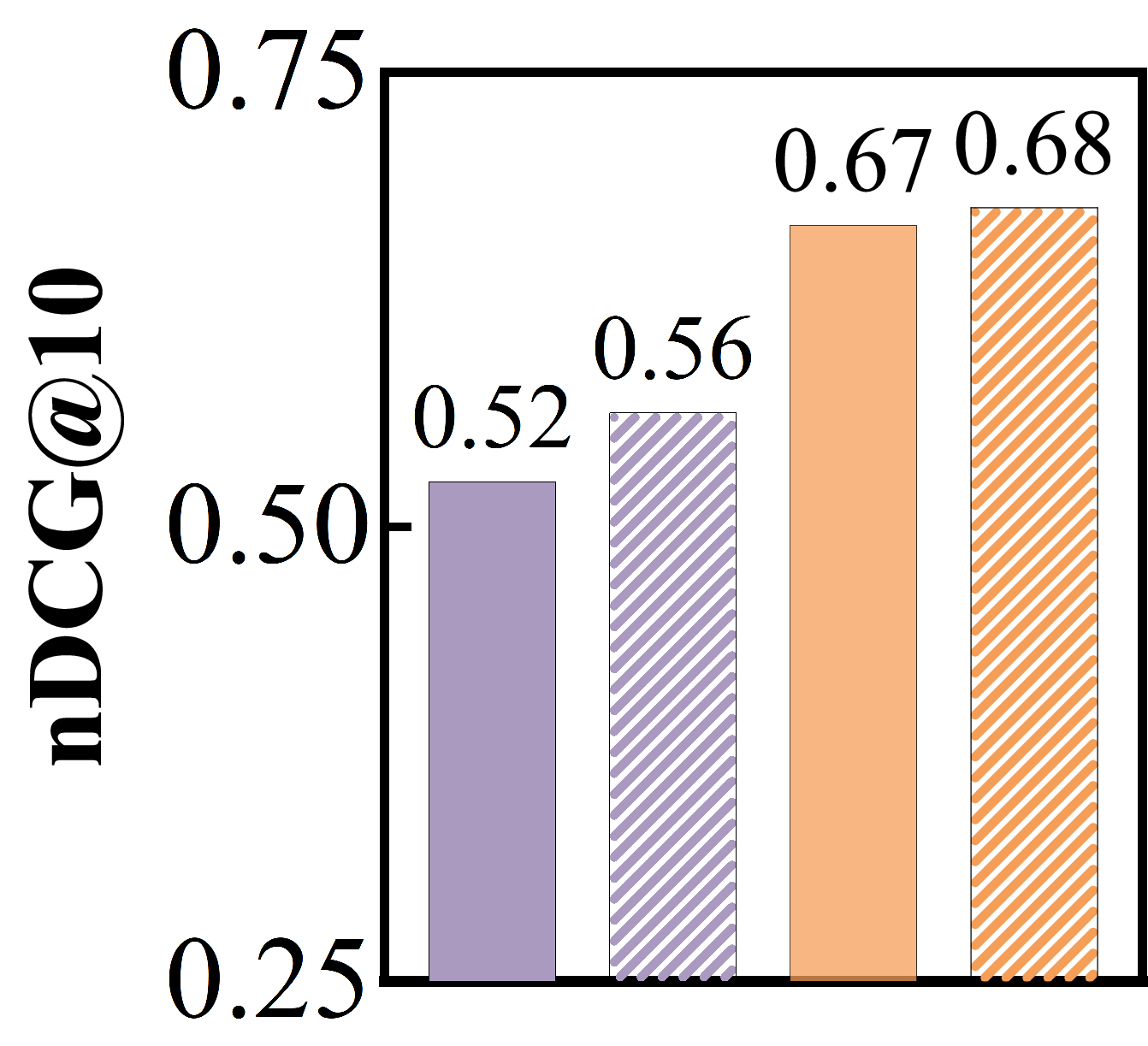}
    \includegraphics[width=0.105\textwidth]{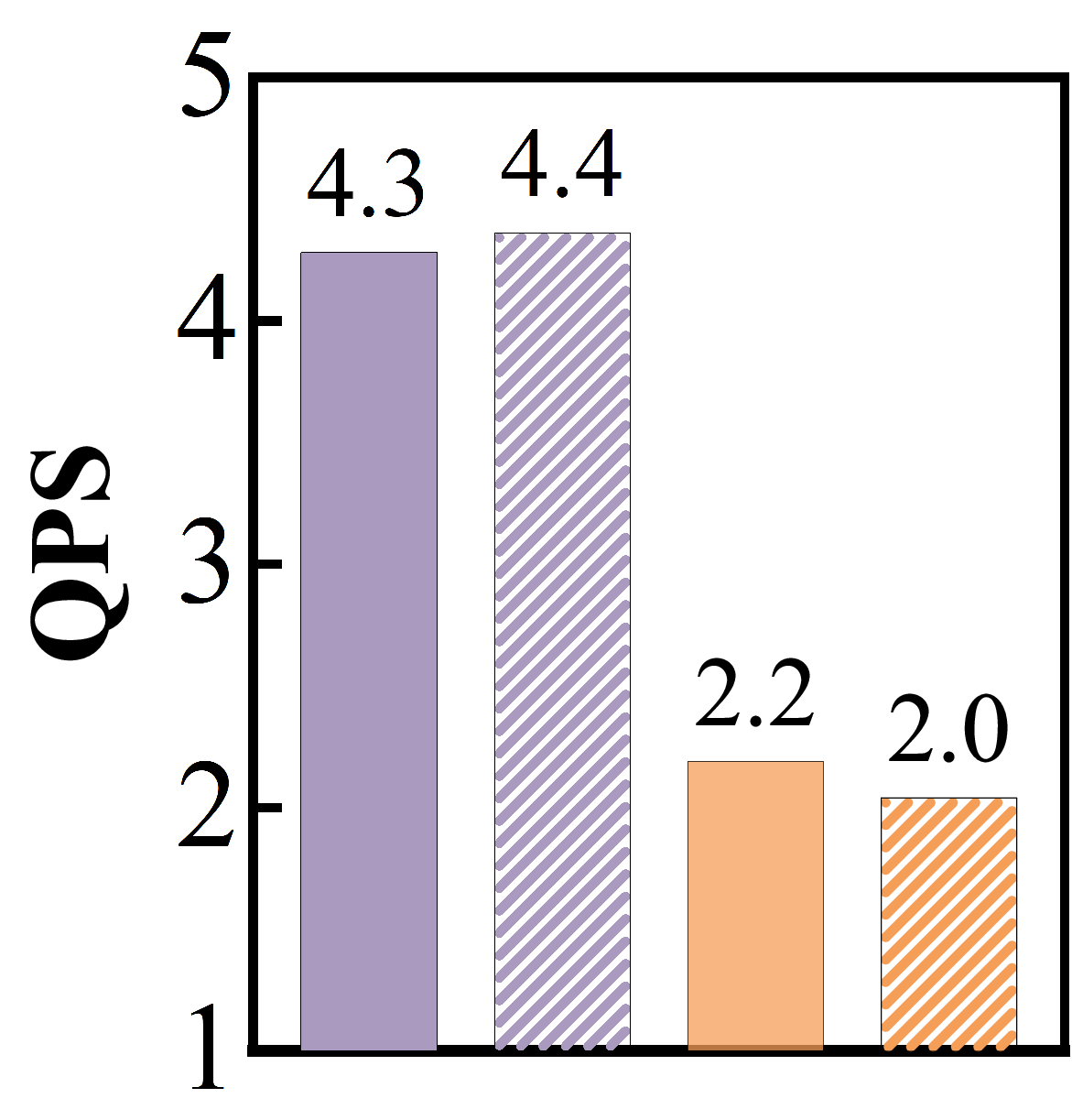}}
    \vspace{-6mm}
    \caption{Performance w/o keyword edges. }
    \vspace{-6mm}
    \label{fig:ip_keyword_edge}
\end{figure}

Figure \ref{fig:ip_keyword_edge} compares {\sf Allan-Poe} with three retrieval paths and its full-text search configuration with and without keyword edges. As discussed in Section \ref{subsec:keyword_edge}, the introduction of keyword edges is to restore keyword-based retrieval capability lost in full-text search during vector fusion. As shown in Figure \ref{fig:ip_keyword_edge}, keyword edges improve nDCG@10 for full-text search by 1\% and 4\% on \textit{NQ} and \textit{WM}, respectively. 
These improvements extend to three-path search on \textit{WM}, but are less pronounced on \textit{NQ}, where baseline full-text search accuracy is substantially lower than other retrieval paths. 

\vspace{-2mm}
\subsubsection{\textbf{Effectiveness of Logical Edges}.} Without logical edges, \textsf{Allan-Poe} achieves nDCG@10 of 0.655 (full-text), 0.737 (dense), 0.734 (sparse), and 0.751 (three-path) on \textit{NQ-9633}, and 0.727, 0.746, 0.739, and 0.760 on \textit{WM-6119} according to our experiments. Compared to the results in Table \ref{tab:kg_eva}, these values show significant degradation, indicating that logical edges effectively compensate for semantic edge limitations by enhancing query-document relevance. 

\begin{figure}
    \centering
    \subfigcapskip = -4pt
    \subfigure[\textit{NQ}]{
    \includegraphics[width=0.22\textwidth]{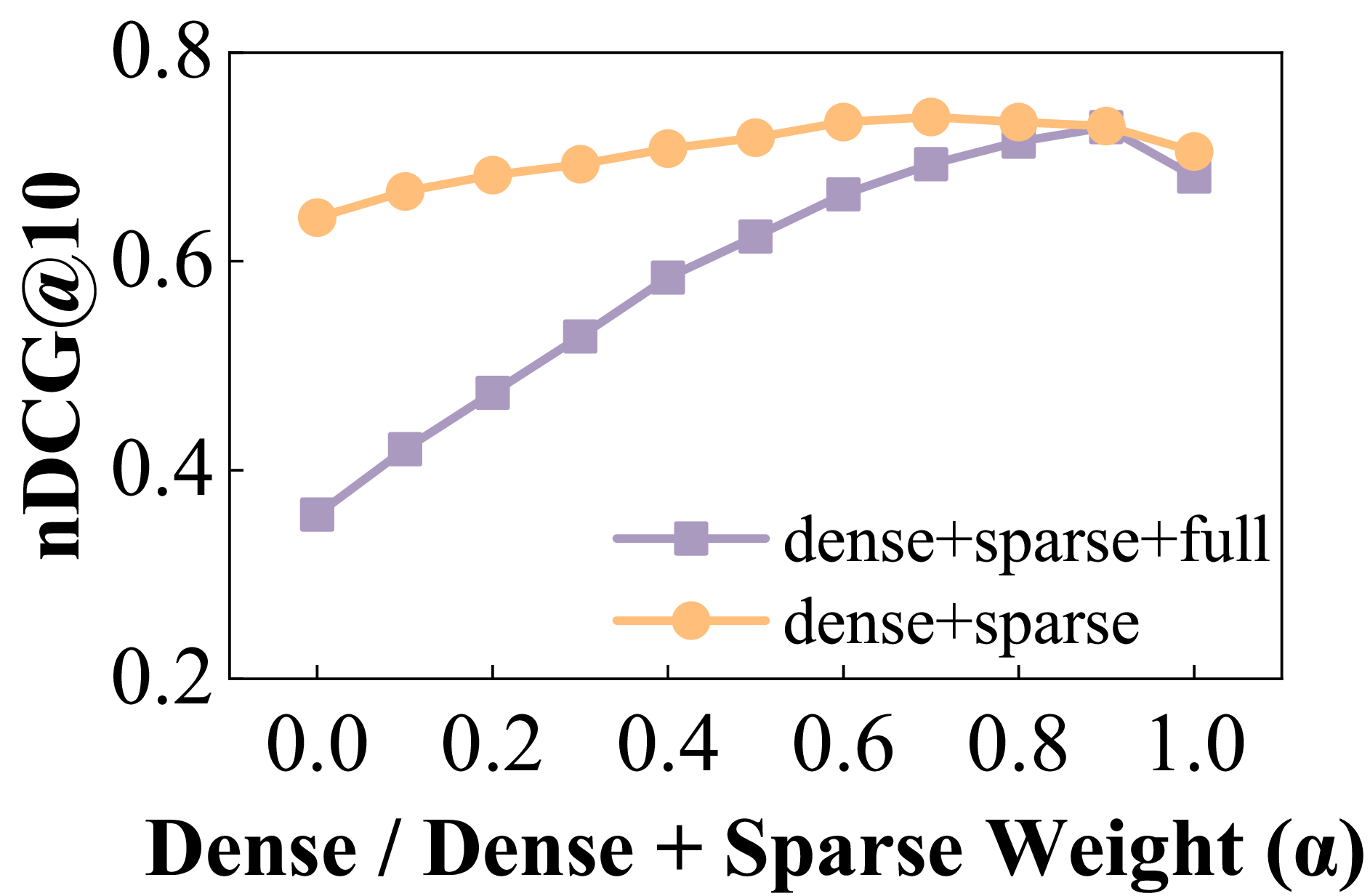}}
    \hspace{2mm}
    \subfigure[\textit{WM}]{
    \includegraphics[width=0.22\textwidth]{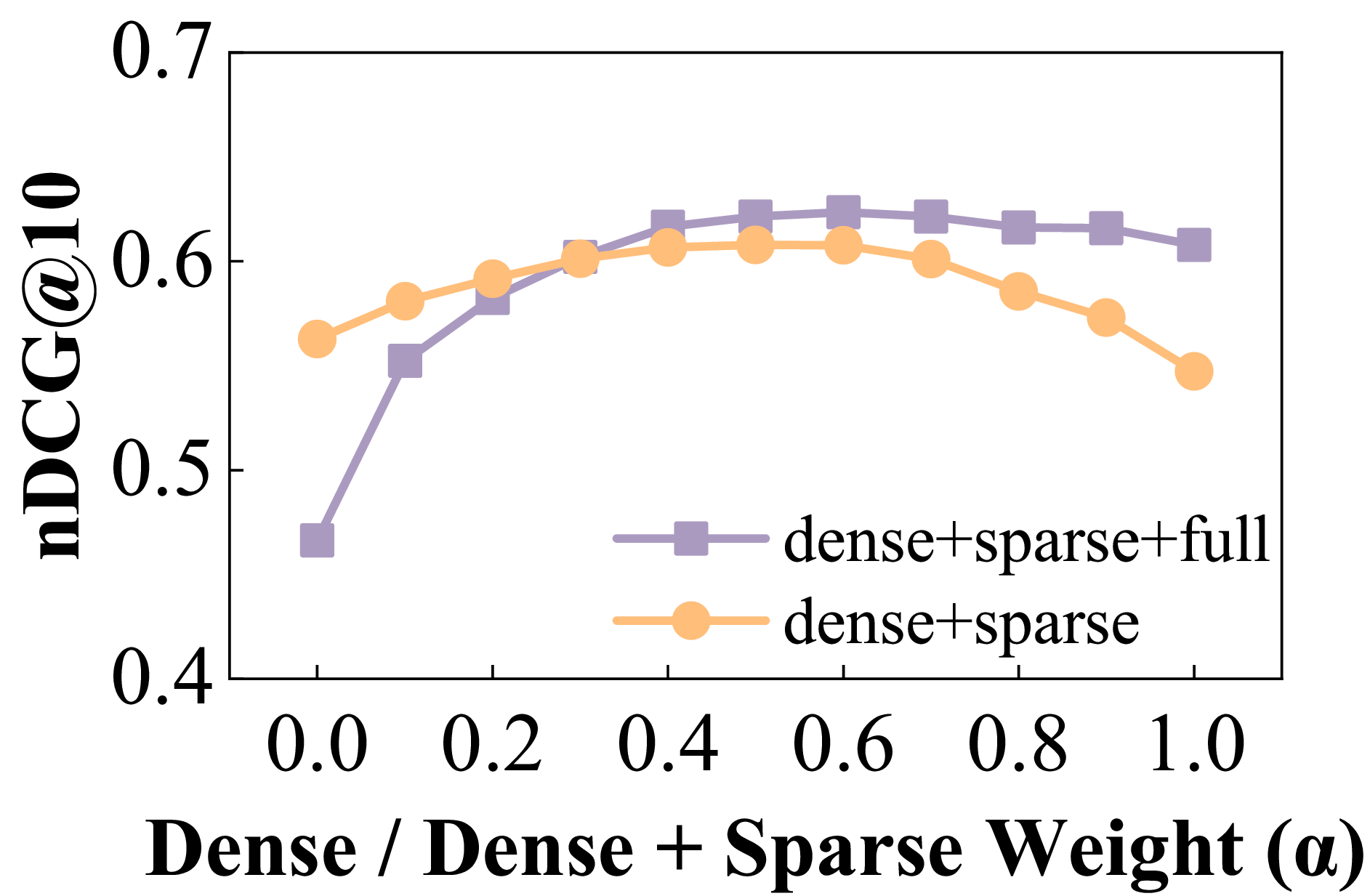}}
    \vspace{-6mm}
    \caption{Performance of various weights for retrieval paths.}
    \vspace{-4mm}
    \label{fig:weight}
\end{figure}

\vspace{-2mm}
\subsection{Weights of Retrieval Paths}

To investigate the impact of retrieval path weighting and the robustness of the hybrid index for arbitrary weights, we evaluate \textsf{Allan-Poe-TwoPath} and \textsf{Allan-Poe-ThreePath} under various weight configurations. For the two-path configuration (dense + sparse vectors), the fused distance is computed as $\alpha \cdot \text{sim}_d(q,d) + (1-\alpha) \cdot \text{sim}_s(q,d)$, where $\alpha \in [0,1]$, and $q$, $d$ denote query and document respectively. For the three-path configuration, the distance function is $\alpha \cdot [\text{sim}_d(q,d) + w_{\text{opt}} \cdot \text{sim}_s(q,d)] + (1-\alpha) \cdot \text{sim}_f(q,d)$, where $w_{\text{opt}}$ represents the optimal dense-sparse weight derived from \textsf{Allan-Poe-TwoPath} evaluations. Results are presented in Figure~\ref{fig:weight}. 

Optimal weights correlate strongly with individual path performance: higher-accuracy paths warrant greater weighting to achieve overall high accuracy. For instance, on the \textit{NQ} dataset, sole dense vector retrieval ($\alpha=1$ for the line of dense+sparse) achieves higher nDCG@10 than sparse retrieval ($\alpha=0$ for the line of dense+sparse), resulting in an optimal $\alpha=0.7$ that favors dense vectors. Similarly, since dense+sparse retrieval substantially outperforms full-text search on \textit{NQ}, the optimal three-path configuration allocates 0.9 weight to dense+sparse and 0.1 to full-text.  
The evaluation results in Figure \ref{fig:weight} also demonstrate that three-path retrieval can surpass or at least have comparable accuracy with two-path retrieval under appropriate weight selection. 
Based on these findings, we derive an empirical weighting criterion based on the nDCG gap of two paths:
\begin{itemize}
    \item \textbf{nDCG gap < 5\%}: Equal weighting ($\alpha \in [0.4,0.6]$);
    \item \textbf{nDCG gap 5-10\%}: Favor higher-accuracy path ($\alpha \in [0.7, 0.8]$);
    \item \textbf{nDCG gap > 10\%}: Strongly favor higher-accuracy path ($\alpha\in [0.9,1)$).
\end{itemize}

\vspace{-3mm}
\subsection{Evaluations of Data Insertion}
\begin{figure}
    \centering
    \subfigcapskip = -4pt
    \subfigure[\textit{NQ}]{
    \includegraphics[width=0.23\textwidth]{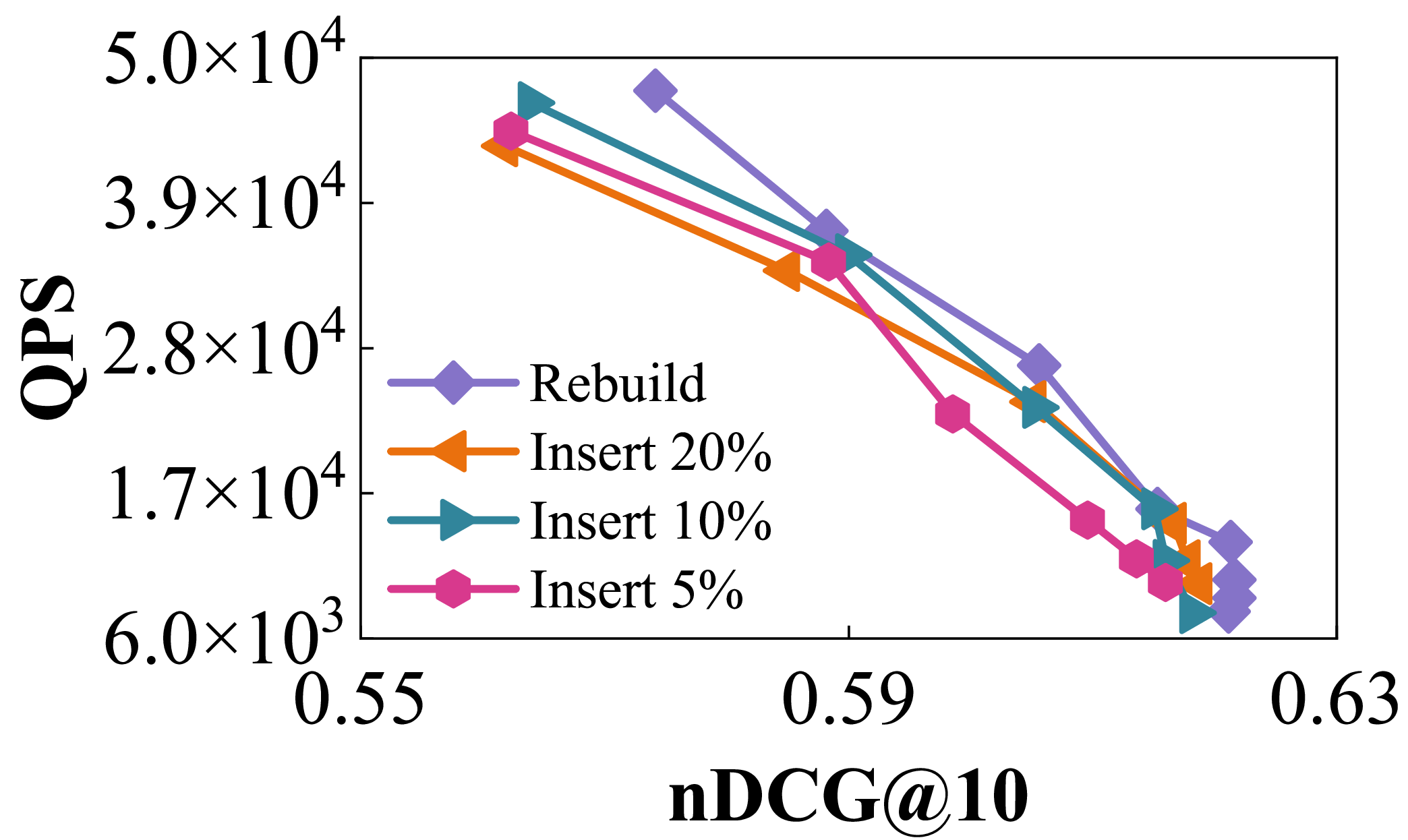}}
    \hspace{-2.2mm}
    \subfigure[\textit{WM}]{
    \includegraphics[width=0.23\textwidth]{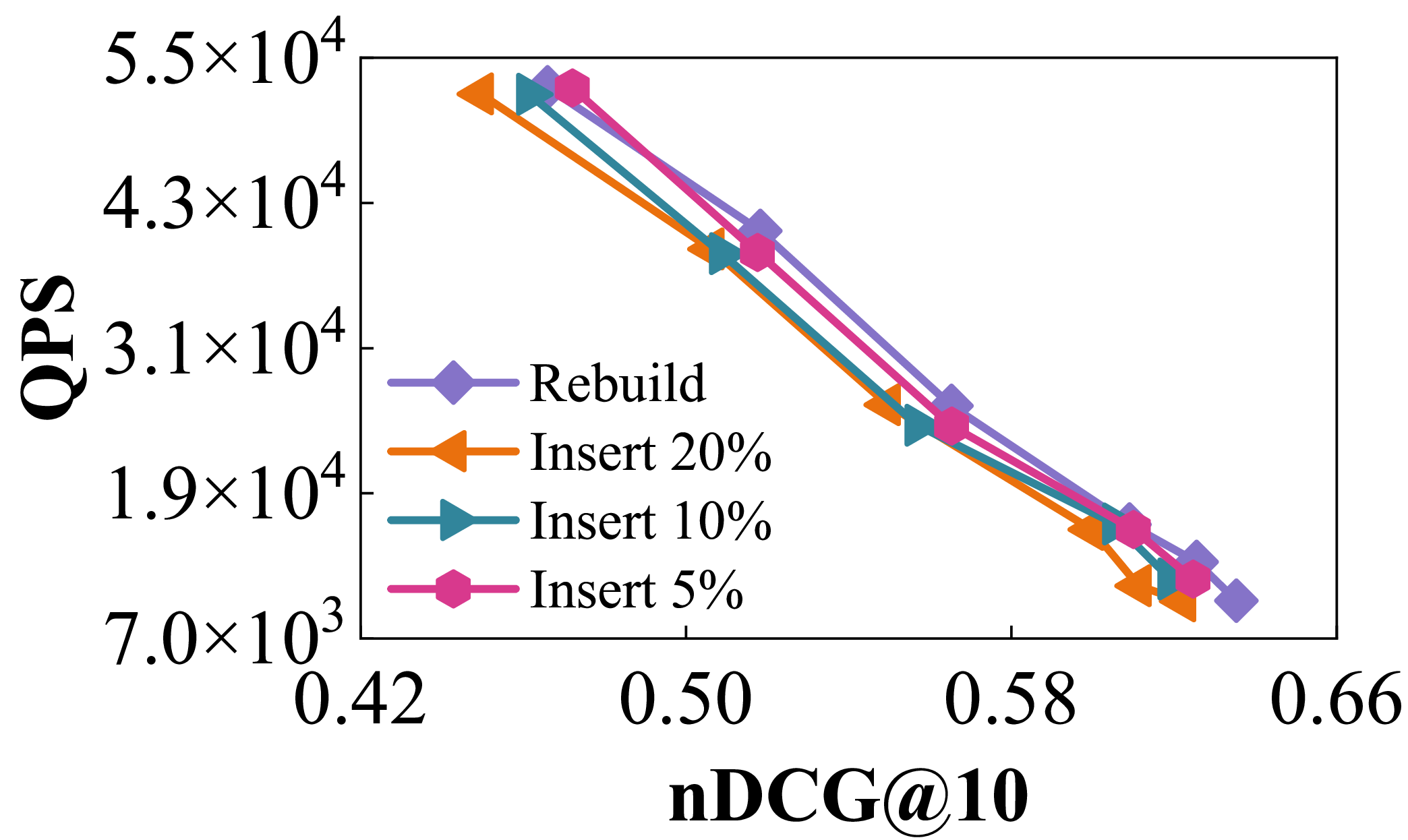}}
    \vspace{-5mm}
    \caption{Comparison of inserting various data volumes.}
    \vspace{-6mm}
    \label{fig:insert}
\end{figure}

As established in Section \ref{subsec:construction}, {\sf Allan-Poe} supports efficient data insertion to accommodate data updates. We evaluate both insertion efficiency and its impact on retrieval quality by inserting varying data volumes into pre-built hybrid indexes and measuring subsequent search performance. 
As shown in Table~\ref{tab:insert_overhead}, our insertion strategy incorporates 20\% new data with only 13.7\% of the computational overhead required for a full index rebuild.
Furthermore, Figure~\ref{fig:insert} shows that the updated index experiences only marginal performance degradation compared to a complete rebuild, with at most 1\% reduction in nDCG@10 while maintaining equivalent QPS. These results demonstrate \textsf{Allan-Poe}'s capability to handle data updates in dynamic environments efficiently.

\begin{table}[tbp]
    \centering
    \small
    \caption{Comparison of indexing overhead.}
    \vspace{-4mm}
    \setlength{\tabcolsep}{.03\linewidth}{
    \begin{tabular}{ccccc}
        \toprule
        \textbf{Datasets} & \textbf{Rebuild} & \textbf{Insert 20\%} & \textbf{Insert 10\%} & \textbf{Insert 5\%} \\
        \midrule
        \textit{NQ} & 45.83s & 6.97s & 3.43s & 1.69s \\
        \textit{WM} & 21.86s & 2.67s & 1.30s & 0.64s \\
        \bottomrule
    \end{tabular}
    }
    \vspace{-5mm}
    \label{tab:insert_overhead}
\end{table}

\vspace{-2mm}
\subsection{Scalability}
\begin{figure}
    \centering
    \subfigcapskip = -4pt
    \subfigure[Index size.]{
    \includegraphics[width=0.148\textwidth]{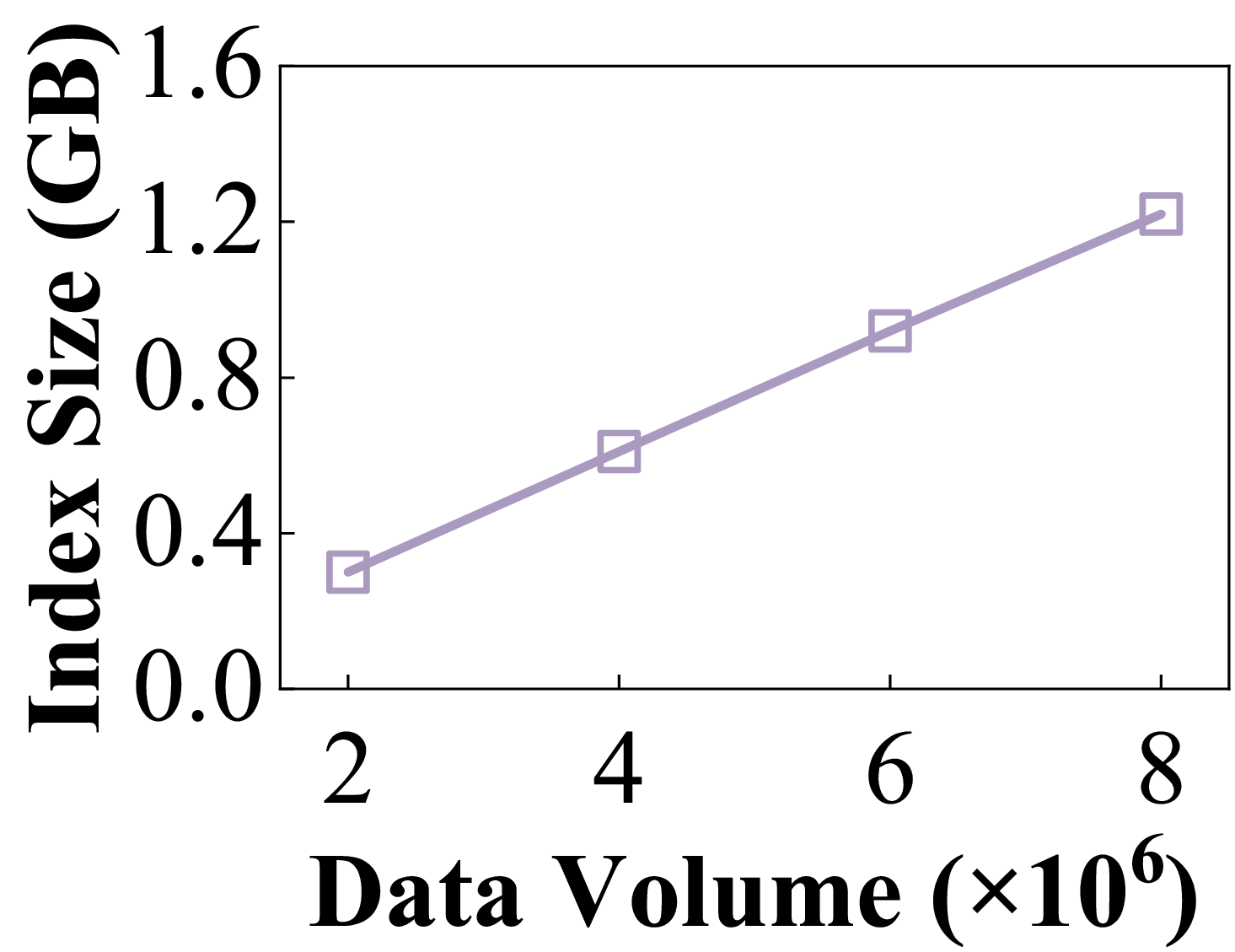}}
    \hspace{-2.2mm}
    \subfigure[Index build time.]{
    \includegraphics[width=0.15\textwidth]{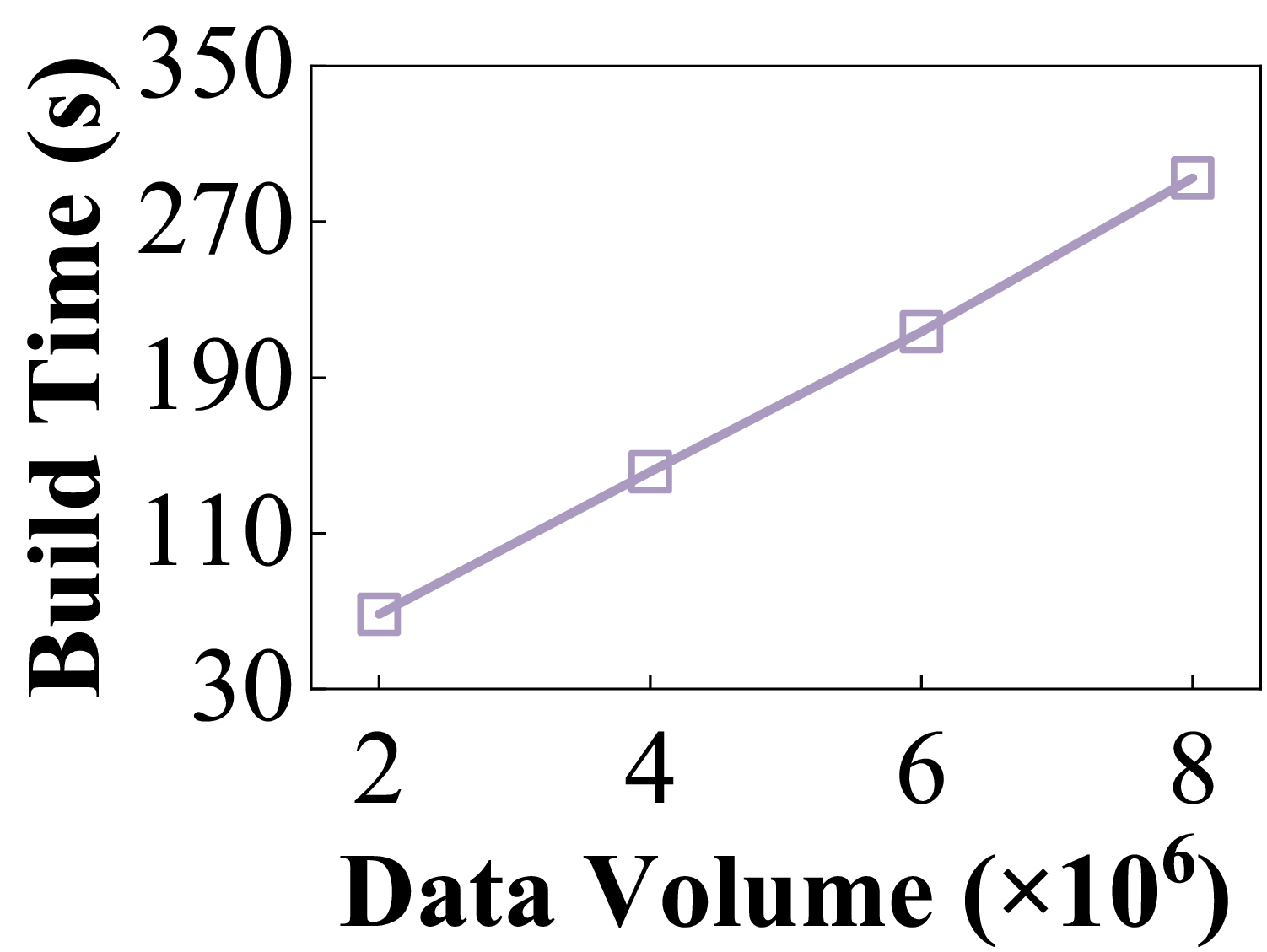}}
    \hspace{-2.2mm}
    \subfigure[Latency at nDCG of 0.55.]{
    \includegraphics[width=0.15\textwidth]{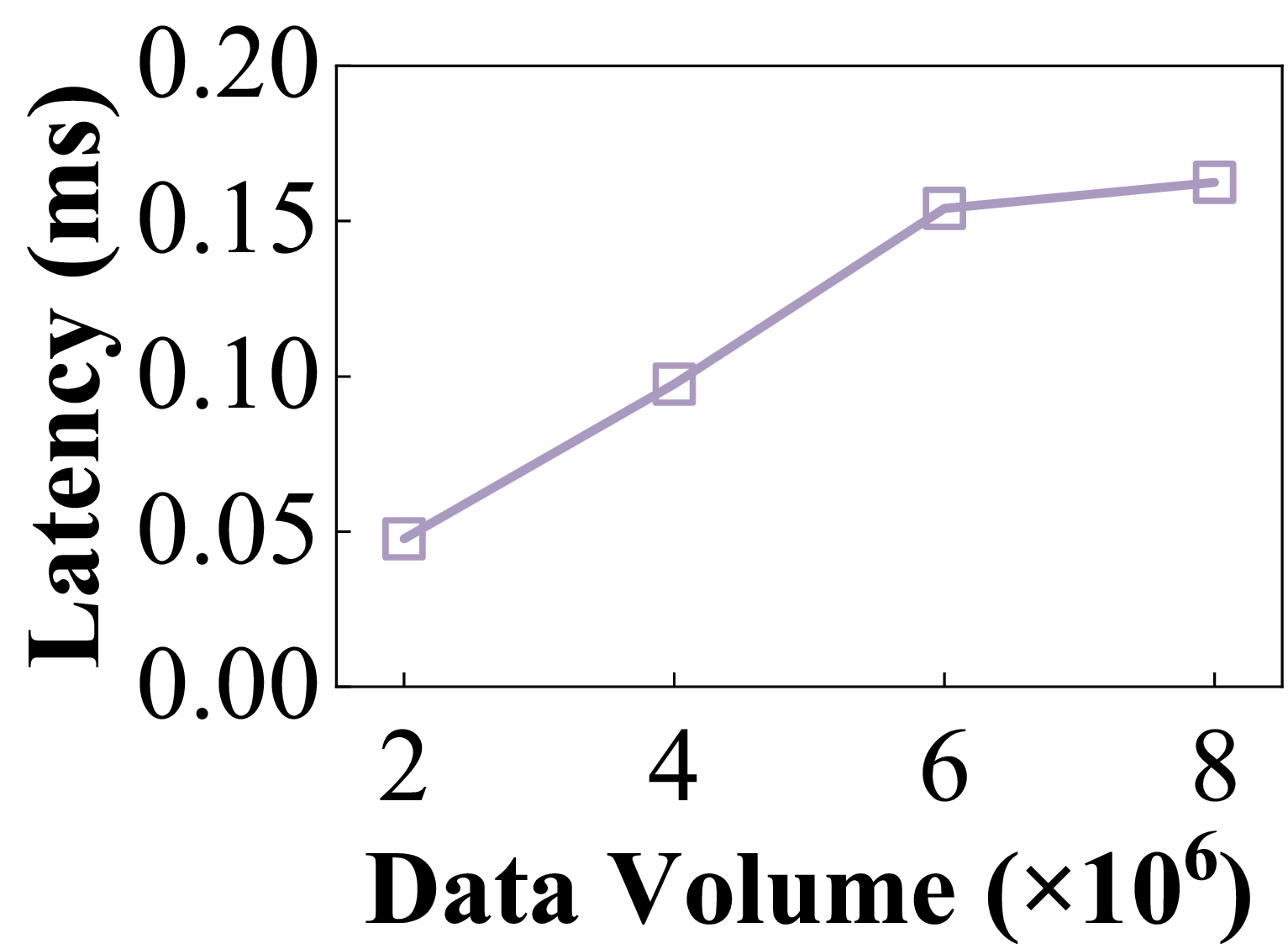}}
    \vspace{-5mm}
    \caption{Impact of varying data size on dataset \textit{MS}.}
    \vspace{-4mm}
    \label{fig:scalability}
\end{figure}

To assess the scalability of \textsf{Allan-Poe}, we conduct experiments on the \textit{MS} dataset (\textasciitilde 8.8M documents in total), scaling from 2 to 8 million documents using an NVIDIA A100 (40GB) GPU. As shown in Figure~\ref{fig:scalability}, both index size and construction time scale nearly linearly with dataset size. Notably, query latency exhibits a more favorable sub-linear (logarithmic) growth. These results demonstrate {\sf Allan-Poe}'s robust scalability for both indexing and search.
For datasets where the hybrid vectors exceed single-GPU memory, the graph-based architecture of \textsf{Allan-Poe} remains compatible with established horizontal and vertical scaling strategies~\cite{li2025scalable, wang2024starling, wang2021milvus}, ensuring practical applicability to very large corpora. 

\vspace{-2mm}
\section{Conclusion}
\label{sec:conclusion}
This paper presents \textsf{Allan-Poe}, a unified, GPU-accelerated hybrid index integrating dense vector, sparse vector, and full-text retrieval. 
We derive design principles for hybrid indexing and build an all-in-one graph-based index with isolated heterogeneous edge storage, enabling flexible path combinations with low overhead. 
The construction pipeline is optimized via hybrid distance computation, RNG-IP joint pruning, and keyword-aware recycling, fully leveraging GPU parallelism.
At query time, we propose a dynamic fusion framework that combines all retrieval paths and incorporates knowledge graph structures to handle complex queries. 
Extensive experiments show that {\sf Allan-Poe} outperforms state-of-the-art methods in both accuracy and efficiency.


\bibliographystyle{ACM-Reference-Format}
\balance
\bibliography{sample}


\begin{thebibliography}{100}


\ifx \showCODEN    \undefined \def \showCODEN     #1{\unskip}     \fi
\ifx \showDOI      \undefined \def \showDOI       #1{#1}\fi
\ifx \showISBNx    \undefined \def \showISBNx     #1{\unskip}     \fi
\ifx \showISBNxiii \undefined \def \showISBNxiii  #1{\unskip}     \fi
\ifx \showISSN     \undefined \def \showISSN      #1{\unskip}     \fi
\ifx \showLCCN     \undefined \def \showLCCN      #1{\unskip}     \fi
\ifx \shownote     \undefined \def \shownote      #1{#1}          \fi
\ifx \showarticletitle \undefined \def \showarticletitle #1{#1}   \fi
\ifx \showURL      \undefined \def \showURL       {\relax}        \fi
\providecommand\bibfield[2]{#2}
\providecommand\bibinfo[2]{#2}
\providecommand\natexlab[1]{#1}
\providecommand\showeprint[2][]{arXiv:#2}

\bibitem[\protect\citeauthoryear{??}{ela}{2025}]%
        {elastic_hybrid}
 \bibinfo{year}{2025}\natexlab{}.
\newblock \showarticletitle{Elastic Hybrid Search}.
\newblock
\newblock
\shownote{https://www.elastic.co/what-is/hybrid-search.}


\bibitem[\protect\citeauthoryear{??}{inf}{2025}]%
        {infinity_hybrid}
 \bibinfo{year}{2025}\natexlab{}.
\newblock \showarticletitle{Infinity Hybrid Search}.
\newblock
\newblock
\shownote{https://infiniflow.org/blog/best-hybrid-search-solution.}


\bibitem[\protect\citeauthoryear{??}{mil}{2025a}]%
        {milvus_hybrid}
 \bibinfo{year}{2025}\natexlab{a}.
\newblock \showarticletitle{Milvus Hybrid Search}.
\newblock
\newblock
\shownote{https://milvus.io/docs/hybrid\_search\_with\_milvus.md.}


\bibitem[\protect\citeauthoryear{??}{mil}{2025b}]%
        {milvus_rerank}
 \bibinfo{year}{2025}\natexlab{b}.
\newblock \showarticletitle{Milvus Rerank Methods}.
\newblock
\newblock
\shownote{https://milvus.io/docs/reranking.md.}


\bibitem[\protect\citeauthoryear{??}{wea}{2025}]%
        {weaviate_hybrid}
 \bibinfo{year}{2025}\natexlab{}.
\newblock \showarticletitle{Weaviate Hybrid Search}.
\newblock
\newblock
\shownote{https://docs.weaviate.io/weaviate/search/hybrid.}


\bibitem[\protect\citeauthoryear{Ait~Aomar, Echihabi, Arnaboldi, Alagiannis, Hilloulin, and Cherkaoui}{Ait~Aomar et~al\mbox{.}}{2025}]%
        {ait2025rwalks}
\bibfield{author}{\bibinfo{person}{Anas Ait~Aomar}, \bibinfo{person}{Karima Echihabi}, \bibinfo{person}{Marco Arnaboldi}, \bibinfo{person}{Ioannis Alagiannis}, \bibinfo{person}{Damien Hilloulin}, {and} \bibinfo{person}{Manal Cherkaoui}.} \bibinfo{year}{2025}\natexlab{}.
\newblock \showarticletitle{RWalks: Random walks as attribute diffusers for filtered vector search}.
\newblock \bibinfo{journal}{\emph{PACMMOD (SIGMOD)}} \bibinfo{volume}{3}, \bibinfo{number}{3} (\bibinfo{year}{2025}), \bibinfo{pages}{1--26}.
\newblock


\bibitem[\protect\citeauthoryear{Ang, Bao, Huang, Tung, and Huang}{Ang et~al\mbox{.}}{2024}]%
        {ang2024tsgassist}
\bibfield{author}{\bibinfo{person}{Yihao Ang}, \bibinfo{person}{Yifan Bao}, \bibinfo{person}{Qiang Huang}, \bibinfo{person}{Anthony~KH Tung}, {and} \bibinfo{person}{Zhiyong Huang}.} \bibinfo{year}{2024}\natexlab{}.
\newblock \showarticletitle{Tsgassist: An interactive assistant harnessing llms and rag for time series generation recommendations and benchmarking}.
\newblock \bibinfo{journal}{\emph{PVLDB}} \bibinfo{volume}{17}, \bibinfo{number}{12} (\bibinfo{year}{2024}), \bibinfo{pages}{4309--4312}.
\newblock


\bibitem[\protect\citeauthoryear{Blanco and Boldi}{Blanco and Boldi}{2012}]%
        {blanco2012extending}
\bibfield{author}{\bibinfo{person}{Roi Blanco} {and} \bibinfo{person}{Paolo Boldi}.} \bibinfo{year}{2012}\natexlab{}.
\newblock \showarticletitle{Extending BM25 with multiple query operators}. In \bibinfo{booktitle}{\emph{SIGIR}}. \bibinfo{pages}{921--930}.
\newblock


\bibitem[\protect\citeauthoryear{Bruch, Gai, and Ingber}{Bruch et~al\mbox{.}}{2023}]%
        {bruch2023analysis}
\bibfield{author}{\bibinfo{person}{Sebastian Bruch}, \bibinfo{person}{Siyu Gai}, {and} \bibinfo{person}{Amir Ingber}.} \bibinfo{year}{2023}\natexlab{}.
\newblock \showarticletitle{An analysis of fusion functions for hybrid retrieval}.
\newblock \bibinfo{journal}{\emph{TOIS}} \bibinfo{volume}{42}, \bibinfo{number}{1} (\bibinfo{year}{2023}), \bibinfo{pages}{1--35}.
\newblock


\bibitem[\protect\citeauthoryear{Bruch, Nardini, Ingber, and Liberty}{Bruch et~al\mbox{.}}{2024a}]%
        {bruch2024bridging}
\bibfield{author}{\bibinfo{person}{Sebastian Bruch}, \bibinfo{person}{Franco~Maria Nardini}, \bibinfo{person}{Amir Ingber}, {and} \bibinfo{person}{Edo Liberty}.} \bibinfo{year}{2024}\natexlab{a}.
\newblock \showarticletitle{Bridging dense and sparse maximum inner product search}.
\newblock \bibinfo{journal}{\emph{TOIS}} \bibinfo{volume}{42}, \bibinfo{number}{6} (\bibinfo{year}{2024}), \bibinfo{pages}{1--38}.
\newblock


\bibitem[\protect\citeauthoryear{Bruch, Nardini, Rulli, and Venturini}{Bruch et~al\mbox{.}}{2024b}]%
        {bruch2024efficient}
\bibfield{author}{\bibinfo{person}{Sebastian Bruch}, \bibinfo{person}{Franco~Maria Nardini}, \bibinfo{person}{Cosimo Rulli}, {and} \bibinfo{person}{Rossano Venturini}.} \bibinfo{year}{2024}\natexlab{b}.
\newblock \showarticletitle{Efficient inverted indexes for approximate retrieval over learned sparse representations}. In \bibinfo{booktitle}{\emph{SIGIR}}. \bibinfo{pages}{152--162}.
\newblock


\bibitem[\protect\citeauthoryear{Cai, Shi, Chen, and Zheng}{Cai et~al\mbox{.}}{2024}]%
        {cai2024navigating}
\bibfield{author}{\bibinfo{person}{Yuzheng Cai}, \bibinfo{person}{Jiayang Shi}, \bibinfo{person}{Yizhuo Chen}, {and} \bibinfo{person}{Weiguo Zheng}.} \bibinfo{year}{2024}\natexlab{}.
\newblock \showarticletitle{Navigating labels and vectors: A unified approach to filtered approximate nearest neighbor search}.
\newblock \bibinfo{journal}{\emph{PACMMOD (SIGMOD)}} \bibinfo{volume}{2}, \bibinfo{number}{6} (\bibinfo{year}{2024}), \bibinfo{pages}{1--27}.
\newblock


\bibitem[\protect\citeauthoryear{Cao, Gao, Li, Xie, Zhou, and Xu}{Cao et~al\mbox{.}}{2025}]%
        {cao2024lego}
\bibfield{author}{\bibinfo{person}{Yukun Cao}, \bibinfo{person}{Zengyi Gao}, \bibinfo{person}{Zhiyang Li}, \bibinfo{person}{Xike Xie}, \bibinfo{person}{S~Kevin Zhou}, {and} \bibinfo{person}{Jianliang Xu}.} \bibinfo{year}{2025}\natexlab{}.
\newblock \showarticletitle{Lego-graphrag: Modularizing graph-based retrieval-augmented generation for design space exploration}.
\newblock \bibinfo{journal}{\emph{PVLDB}} \bibinfo{volume}{18}, \bibinfo{number}{10} (\bibinfo{year}{2025}), \bibinfo{pages}{3269--3283}.
\newblock


\bibitem[\protect\citeauthoryear{Chen, Xiao, Zhang, Luo, Lian, and Liu}{Chen et~al\mbox{.}}{2024}]%
        {chen2024bge}
\bibfield{author}{\bibinfo{person}{Jianlv Chen}, \bibinfo{person}{Shitao Xiao}, \bibinfo{person}{Peitian Zhang}, \bibinfo{person}{Kun Luo}, \bibinfo{person}{Defu Lian}, {and} \bibinfo{person}{Zheng Liu}.} \bibinfo{year}{2024}\natexlab{}.
\newblock \showarticletitle{Bge m3-embedding: Multi-lingual, multi-functionality, multi-granularity text embeddings through self-knowledge distillation}.
\newblock \bibinfo{journal}{\emph{arXiv}} (\bibinfo{year}{2024}).
\newblock


\bibitem[\protect\citeauthoryear{Chen, Fan, Wu, Tang, Deng, Wang, Li, Tan, et~al\mbox{.}}{Chen et~al\mbox{.}}{2025a}]%
        {chen2025automatic}
\bibfield{author}{\bibinfo{person}{Sibei Chen}, \bibinfo{person}{Ju Fan}, \bibinfo{person}{Bin Wu}, \bibinfo{person}{Nan Tang}, \bibinfo{person}{Chao Deng}, \bibinfo{person}{Pengyi Wang}, \bibinfo{person}{Ye Li}, \bibinfo{person}{Jian Tan}, {et~al\mbox{.}}} \bibinfo{year}{2025}\natexlab{a}.
\newblock \showarticletitle{Automatic database configuration debugging using retrieval-augmented language models}.
\newblock \bibinfo{journal}{\emph{PACMMOD (SIGMOD)}} \bibinfo{volume}{3}, \bibinfo{number}{1} (\bibinfo{year}{2025}), \bibinfo{pages}{1--27}.
\newblock


\bibitem[\protect\citeauthoryear{Chen, Fu, Ke, Gao, Ni, and Zeng}{Chen et~al\mbox{.}}{2025b}]%
        {chen2025stitching}
\bibfield{author}{\bibinfo{person}{Tingyang Chen}, \bibinfo{person}{Cong Fu}, \bibinfo{person}{Xiangyu Ke}, \bibinfo{person}{Yunjun Gao}, \bibinfo{person}{Yabo Ni}, {and} \bibinfo{person}{Anxiang Zeng}.} \bibinfo{year}{2025}\natexlab{b}.
\newblock \showarticletitle{Stitching inner product and euclidean metrics for topology-aware maximum inner product search}. In \bibinfo{booktitle}{\emph{SIGIR}}. \bibinfo{pages}{2341--2350}.
\newblock


\bibitem[\protect\citeauthoryear{Chen, Fu, Wang, Ke, Gao, Zhou, Ni, and Zeng}{Chen et~al\mbox{.}}{2025c}]%
        {chen2025maximum}
\bibfield{author}{\bibinfo{person}{Tingyang Chen}, \bibinfo{person}{Cong Fu}, \bibinfo{person}{Kun Wang}, \bibinfo{person}{Xiangyu Ke}, \bibinfo{person}{Yunjun Gao}, \bibinfo{person}{Wenchao Zhou}, \bibinfo{person}{Yabo Ni}, {and} \bibinfo{person}{Anxiang Zeng}.} \bibinfo{year}{2025}\natexlab{c}.
\newblock \showarticletitle{Maximum inner product is query-scaled nearest neighbor}.
\newblock \bibinfo{journal}{\emph{PVLDB}} \bibinfo{volume}{18}, \bibinfo{number}{6} (\bibinfo{year}{2025}), \bibinfo{pages}{1770--1783}.
\newblock


\bibitem[\protect\citeauthoryear{Cormack, Clarke, and Buettcher}{Cormack et~al\mbox{.}}{2009}]%
        {cormack2009reciprocal}
\bibfield{author}{\bibinfo{person}{Gordon~V Cormack}, \bibinfo{person}{Charles~LA Clarke}, {and} \bibinfo{person}{Stefan Buettcher}.} \bibinfo{year}{2009}\natexlab{}.
\newblock \showarticletitle{Reciprocal rank fusion outperforms condorcet and individual rank learning methods}. In \bibinfo{booktitle}{\emph{SIGIR}}. \bibinfo{pages}{758--759}.
\newblock


\bibitem[\protect\citeauthoryear{Devlin, Chang, Lee, and Toutanova}{Devlin et~al\mbox{.}}{2019}]%
        {devlin2019bert}
\bibfield{author}{\bibinfo{person}{Jacob Devlin}, \bibinfo{person}{Ming-Wei Chang}, \bibinfo{person}{Kenton Lee}, {and} \bibinfo{person}{Kristina Toutanova}.} \bibinfo{year}{2019}\natexlab{}.
\newblock \showarticletitle{Bert: Pre-training of deep bidirectional transformers for language understanding}. In \bibinfo{booktitle}{\emph{NAACL}}. \bibinfo{pages}{4171--4186}.
\newblock


\bibitem[\protect\citeauthoryear{Dong, Moses, and Li}{Dong et~al\mbox{.}}{2011}]%
        {dong2011efficient}
\bibfield{author}{\bibinfo{person}{Wei Dong}, \bibinfo{person}{Charikar Moses}, {and} \bibinfo{person}{Kai Li}.} \bibinfo{year}{2011}\natexlab{}.
\newblock \showarticletitle{Efficient k-nearest neighbor graph construction for generic similarity measures}. In \bibinfo{booktitle}{\emph{WWW}}. \bibinfo{pages}{577--586}.
\newblock


\bibitem[\protect\citeauthoryear{Echihabi, Fatourou, Zoumpatianos, Palpanas, and Benbrahim}{Echihabi et~al\mbox{.}}{2022}]%
        {echihabi2022hercules}
\bibfield{author}{\bibinfo{person}{Karima Echihabi}, \bibinfo{person}{Panagiota Fatourou}, \bibinfo{person}{Kostas Zoumpatianos}, \bibinfo{person}{Themis Palpanas}, {and} \bibinfo{person}{Houda Benbrahim}.} \bibinfo{year}{2022}\natexlab{}.
\newblock \showarticletitle{Hercules against data series similarity search}.
\newblock \bibinfo{journal}{\emph{PVLDB}} \bibinfo{volume}{15}, \bibinfo{number}{10} (\bibinfo{year}{2022}), \bibinfo{pages}{2005--2018}.
\newblock


\bibitem[\protect\citeauthoryear{Edge, Trinh, Cheng, Bradley, Chao, Mody, Truitt, Metropolitansky, Ness, and Larson}{Edge et~al\mbox{.}}{2024}]%
        {edge2024local}
\bibfield{author}{\bibinfo{person}{Darren Edge}, \bibinfo{person}{Ha Trinh}, \bibinfo{person}{Newman Cheng}, \bibinfo{person}{Joshua Bradley}, \bibinfo{person}{Alex Chao}, \bibinfo{person}{Apurva Mody}, \bibinfo{person}{Steven Truitt}, \bibinfo{person}{Dasha Metropolitansky}, \bibinfo{person}{Robert~Osazuwa Ness}, {and} \bibinfo{person}{Jonathan Larson}.} \bibinfo{year}{2024}\natexlab{}.
\newblock \showarticletitle{From local to global: A graph rag approach to query-focused summarization}.
\newblock \bibinfo{journal}{\emph{arXiv}} (\bibinfo{year}{2024}).
\newblock


\bibitem[\protect\citeauthoryear{Formal, Clinchant, D{\'e}jean, and Lassance}{Formal et~al\mbox{.}}{2024a}]%
        {formal2024splate}
\bibfield{author}{\bibinfo{person}{Thibault Formal}, \bibinfo{person}{St{\'e}phane Clinchant}, \bibinfo{person}{Herv{\'e} D{\'e}jean}, {and} \bibinfo{person}{Carlos Lassance}.} \bibinfo{year}{2024}\natexlab{a}.
\newblock \showarticletitle{Splate: Sparse late interaction retrieval}. In \bibinfo{booktitle}{\emph{SIGIR}}. \bibinfo{pages}{2635--2640}.
\newblock


\bibitem[\protect\citeauthoryear{Formal, Lassance, Piwowarski, and Clinchant}{Formal et~al\mbox{.}}{2024b}]%
        {formal2024towards}
\bibfield{author}{\bibinfo{person}{Thibault Formal}, \bibinfo{person}{Carlos Lassance}, \bibinfo{person}{Benjamin Piwowarski}, {and} \bibinfo{person}{St{\'e}phane Clinchant}.} \bibinfo{year}{2024}\natexlab{b}.
\newblock \showarticletitle{Towards effective and efficient sparse neural information retrieval}.
\newblock \bibinfo{journal}{\emph{TOIS}} \bibinfo{volume}{42}, \bibinfo{number}{5} (\bibinfo{year}{2024}), \bibinfo{pages}{1--46}.
\newblock


\bibitem[\protect\citeauthoryear{Formal, Piwowarski, and Clinchant}{Formal et~al\mbox{.}}{2021}]%
        {formal2021splade}
\bibfield{author}{\bibinfo{person}{Thibault Formal}, \bibinfo{person}{Benjamin Piwowarski}, {and} \bibinfo{person}{St{\'e}phane Clinchant}.} \bibinfo{year}{2021}\natexlab{}.
\newblock \showarticletitle{SPLADE: Sparse lexical and expansion model for first stage ranking}. In \bibinfo{booktitle}{\emph{SIGIR}}. \bibinfo{pages}{2288--2292}.
\newblock


\bibitem[\protect\citeauthoryear{Fu, Xiang, Wang, and Cai}{Fu et~al\mbox{.}}{2019}]%
        {fu2017nsg}
\bibfield{author}{\bibinfo{person}{Cong Fu}, \bibinfo{person}{Chao Xiang}, \bibinfo{person}{Changxu Wang}, {and} \bibinfo{person}{Deng Cai}.} \bibinfo{year}{2019}\natexlab{}.
\newblock \showarticletitle{Fast approximate nearest neighbor search with the navigating spreading-out graph}.
\newblock \bibinfo{journal}{\emph{PVLDB}} \bibinfo{volume}{12}, \bibinfo{number}{5} (\bibinfo{year}{2019}), \bibinfo{pages}{461--474}.
\newblock


\bibitem[\protect\citeauthoryear{Fu, Tang, Khan, Mehrotra, Ke, and Gao}{Fu et~al\mbox{.}}{2025}]%
        {fu2026llm}
\bibfield{author}{\bibinfo{person}{Jiajie Fu}, \bibinfo{person}{Haitong Tang}, \bibinfo{person}{Arijit Khan}, \bibinfo{person}{Sharad Mehrotra}, \bibinfo{person}{Xiangyu Ke}, {and} \bibinfo{person}{Yunjun Gao}.} \bibinfo{year}{2025}\natexlab{}.
\newblock \showarticletitle{In-context clustering-based entity resolution with large language models: A design space exploration}.
\newblock \bibinfo{journal}{\emph{PACMMOD (SIGMOD)}} \bibinfo{volume}{3}, \bibinfo{number}{4} (\bibinfo{year}{2025}), \bibinfo{pages}{1--28}.
\newblock


\bibitem[\protect\citeauthoryear{Gale, Zaharia, Young, and Elsen}{Gale et~al\mbox{.}}{2020}]%
        {gale2020sparse}
\bibfield{author}{\bibinfo{person}{Trevor Gale}, \bibinfo{person}{Matei Zaharia}, \bibinfo{person}{Cliff Young}, {and} \bibinfo{person}{Erich Elsen}.} \bibinfo{year}{2020}\natexlab{}.
\newblock \showarticletitle{Sparse gpu kernels for deep learning}. In \bibinfo{booktitle}{\emph{SC}}. \bibinfo{pages}{1--14}.
\newblock


\bibitem[\protect\citeauthoryear{Gao, Gou, Xu, Yang, Long, and Wong}{Gao et~al\mbox{.}}{2025}]%
        {gao2025practical}
\bibfield{author}{\bibinfo{person}{Jianyang Gao}, \bibinfo{person}{Yutong Gou}, \bibinfo{person}{Yuexuan Xu}, \bibinfo{person}{Yongyi Yang}, \bibinfo{person}{Cheng Long}, {and} \bibinfo{person}{Raymond Chi-Wing Wong}.} \bibinfo{year}{2025}\natexlab{}.
\newblock \showarticletitle{Practical and asymptotically optimal quantization of high-dimensional vectors in euclidean space for approximate nearest neighbor search}.
\newblock \bibinfo{journal}{\emph{PACMMOD (SIGMOD)}} \bibinfo{volume}{3}, \bibinfo{number}{3} (\bibinfo{year}{2025}), \bibinfo{pages}{1--26}.
\newblock


\bibitem[\protect\citeauthoryear{Gao and Long}{Gao and Long}{2024}]%
        {gao2024rabitq}
\bibfield{author}{\bibinfo{person}{Jianyang Gao} {and} \bibinfo{person}{Cheng Long}.} \bibinfo{year}{2024}\natexlab{}.
\newblock \showarticletitle{Rabitq: Quantizing high-dimensional vectors with a theoretical error bound for approximate nearest neighbor search}.
\newblock \bibinfo{journal}{\emph{PACMMOD (SIGMOD)}} \bibinfo{volume}{2}, \bibinfo{number}{3} (\bibinfo{year}{2024}), \bibinfo{pages}{1--27}.
\newblock


\bibitem[\protect\citeauthoryear{Gollapudi, Karia, Sivashankar, Krishnaswamy, Begwani, Raz, Lin, Zhang, Mahapatro, Srinivasan, et~al\mbox{.}}{Gollapudi et~al\mbox{.}}{2023}]%
        {gollapudi2023filtered}
\bibfield{author}{\bibinfo{person}{Siddharth Gollapudi}, \bibinfo{person}{Neel Karia}, \bibinfo{person}{Varun Sivashankar}, \bibinfo{person}{Ravishankar Krishnaswamy}, \bibinfo{person}{Nikit Begwani}, \bibinfo{person}{Swapnil Raz}, \bibinfo{person}{Yiyong Lin}, \bibinfo{person}{Yin Zhang}, \bibinfo{person}{Neelam Mahapatro}, \bibinfo{person}{Premkumar Srinivasan}, {et~al\mbox{.}}} \bibinfo{year}{2023}\natexlab{}.
\newblock \showarticletitle{Filtered-diskann: Graph algorithms for approximate nearest neighbor search with filters}. In \bibinfo{booktitle}{\emph{WWW}}. \bibinfo{pages}{3406--3416}.
\newblock


\bibitem[\protect\citeauthoryear{Gou, Dong, Wu, and Ke}{Gou et~al\mbox{.}}{2024}]%
        {gou2024semantic}
\bibfield{author}{\bibinfo{person}{Qianwen Gou}, \bibinfo{person}{Yunwei Dong}, \bibinfo{person}{YuJiao Wu}, {and} \bibinfo{person}{Qiao Ke}.} \bibinfo{year}{2024}\natexlab{}.
\newblock \showarticletitle{Semantic similarity-based program retrieval: a multi-relational graph perspective}.
\newblock \bibinfo{journal}{\emph{FCS}} \bibinfo{volume}{18}, \bibinfo{number}{3} (\bibinfo{year}{2024}), \bibinfo{pages}{183209}.
\newblock


\bibitem[\protect\citeauthoryear{Gou, Gao, Xu, and Long}{Gou et~al\mbox{.}}{2025}]%
        {gou2025symphonyqg}
\bibfield{author}{\bibinfo{person}{Yutong Gou}, \bibinfo{person}{Jianyang Gao}, \bibinfo{person}{Yuexuan Xu}, {and} \bibinfo{person}{Cheng Long}.} \bibinfo{year}{2025}\natexlab{}.
\newblock \showarticletitle{SymphonyQG: Towards symphonious integration of quantization and graph for approximate nearest neighbor search}.
\newblock \bibinfo{journal}{\emph{PACMMOD (SIGMOD)}} \bibinfo{volume}{3}, \bibinfo{number}{1} (\bibinfo{year}{2025}), \bibinfo{pages}{1--26}.
\newblock


\bibitem[\protect\citeauthoryear{Greathouse and Daga}{Greathouse and Daga}{2014}]%
        {greathouse2014efficient}
\bibfield{author}{\bibinfo{person}{Joseph~L Greathouse} {and} \bibinfo{person}{Mayank Daga}.} \bibinfo{year}{2014}\natexlab{}.
\newblock \showarticletitle{Efficient sparse matrix-vector multiplication on GPUs using the CSR storage format}. In \bibinfo{booktitle}{\emph{SC}}. \bibinfo{pages}{769--780}.
\newblock


\bibitem[\protect\citeauthoryear{Guti{\'e}rrez, Shu, Qi, Zhou, and Su}{Guti{\'e}rrez et~al\mbox{.}}{2025}]%
        {gutierrez2025rag}
\bibfield{author}{\bibinfo{person}{Bernal~Jim{\'e}nez Guti{\'e}rrez}, \bibinfo{person}{Yiheng Shu}, \bibinfo{person}{Weijian Qi}, \bibinfo{person}{Sizhe Zhou}, {and} \bibinfo{person}{Yu Su}.} \bibinfo{year}{2025}\natexlab{}.
\newblock \showarticletitle{From rag to memory: Non-parametric continual learning for large language models}.
\newblock \bibinfo{journal}{\emph{arXiv}} (\bibinfo{year}{2025}).
\newblock


\bibitem[\protect\citeauthoryear{Han, Shomer, Wang, Lei, Guo, Hua, Long, Liu, and Tang}{Han et~al\mbox{.}}{2025}]%
        {han2025rag}
\bibfield{author}{\bibinfo{person}{Haoyu Han}, \bibinfo{person}{Harry Shomer}, \bibinfo{person}{Yu Wang}, \bibinfo{person}{Yongjia Lei}, \bibinfo{person}{Kai Guo}, \bibinfo{person}{Zhigang Hua}, \bibinfo{person}{Bo Long}, \bibinfo{person}{Hui Liu}, {and} \bibinfo{person}{Jiliang Tang}.} \bibinfo{year}{2025}\natexlab{}.
\newblock \showarticletitle{Rag vs. graphrag: A systematic evaluation and key insights}.
\newblock \bibinfo{journal}{\emph{arXiv}} (\bibinfo{year}{2025}).
\newblock


\bibitem[\protect\citeauthoryear{Ho, Duong~Nguyen, Sugawara, and Aizawa}{Ho et~al\mbox{.}}{2020}]%
        {ho-etal-2020-constructing}
\bibfield{author}{\bibinfo{person}{Xanh Ho}, \bibinfo{person}{Anh-Khoa Duong~Nguyen}, \bibinfo{person}{Saku Sugawara}, {and} \bibinfo{person}{Akiko Aizawa}.} \bibinfo{year}{2020}\natexlab{}.
\newblock \showarticletitle{Constructing a multi-hop {QA} dataset for comprehensive evaluation of reasoning steps}. In \bibinfo{booktitle}{\emph{COLING}}. \bibinfo{pages}{6609--6625}.
\newblock


\bibitem[\protect\citeauthoryear{Hu, Zou, Yu, Wang, and Zhao}{Hu et~al\mbox{.}}{2017}]%
        {hu2017answering}
\bibfield{author}{\bibinfo{person}{Sen Hu}, \bibinfo{person}{Lei Zou}, \bibinfo{person}{Jeffrey~Xu Yu}, \bibinfo{person}{Haixun Wang}, {and} \bibinfo{person}{Dongyan Zhao}.} \bibinfo{year}{2017}\natexlab{}.
\newblock \showarticletitle{Answering natural language questions by subgraph matching over knowledge graphs}.
\newblock \bibinfo{journal}{\emph{TKDE}} \bibinfo{volume}{30}, \bibinfo{number}{5} (\bibinfo{year}{2017}), \bibinfo{pages}{824--837}.
\newblock


\bibitem[\protect\citeauthoryear{Jiang, Yang, Zhang, Hou, Shi, Zhou, Li, and Wang}{Jiang et~al\mbox{.}}{2025}]%
        {jiang2025digra}
\bibfield{author}{\bibinfo{person}{Mengxu Jiang}, \bibinfo{person}{Zhi Yang}, \bibinfo{person}{Fangyuan Zhang}, \bibinfo{person}{Guanhao Hou}, \bibinfo{person}{Jieming Shi}, \bibinfo{person}{Wenchao Zhou}, \bibinfo{person}{Feifei Li}, {and} \bibinfo{person}{Sibo Wang}.} \bibinfo{year}{2025}\natexlab{}.
\newblock \showarticletitle{DIGRA: A dynamic graph indexing for approximate nearest neighbor search with range filter}.
\newblock \bibinfo{journal}{\emph{PACMMOD (SIGMOD)}} \bibinfo{volume}{3}, \bibinfo{number}{3} (\bibinfo{year}{2025}), \bibinfo{pages}{1--26}.
\newblock


\bibitem[\protect\citeauthoryear{Jiang, Zeller, Waleffe, Hoefler, and Alonso}{Jiang et~al\mbox{.}}{2024}]%
        {jiang2023chameleon}
\bibfield{author}{\bibinfo{person}{Wenqi Jiang}, \bibinfo{person}{Marco Zeller}, \bibinfo{person}{Roger Waleffe}, \bibinfo{person}{Torsten Hoefler}, {and} \bibinfo{person}{Gustavo Alonso}.} \bibinfo{year}{2024}\natexlab{}.
\newblock \showarticletitle{Chameleon: A heterogeneous and disaggregated accelerator system for retrieval-augmented language models}.
\newblock \bibinfo{journal}{\emph{PVLDB}} \bibinfo{volume}{18}, \bibinfo{number}{1} (\bibinfo{year}{2024}), \bibinfo{pages}{42--52}.
\newblock


\bibitem[\protect\citeauthoryear{Jimenez~Gutierrez, Shu, Gu, Yasunaga, and Su}{Jimenez~Gutierrez et~al\mbox{.}}{2024}]%
        {jimenez2024hipporag}
\bibfield{author}{\bibinfo{person}{Bernal Jimenez~Gutierrez}, \bibinfo{person}{Yiheng Shu}, \bibinfo{person}{Yu Gu}, \bibinfo{person}{Michihiro Yasunaga}, {and} \bibinfo{person}{Yu Su}.} \bibinfo{year}{2024}\natexlab{}.
\newblock \showarticletitle{Hipporag: Neurobiologically inspired long-term memory for large language models}.
\newblock \bibinfo{journal}{\emph{NeurIPS}}  \bibinfo{volume}{37} (\bibinfo{year}{2024}), \bibinfo{pages}{59532--59569}.
\newblock


\bibitem[\protect\citeauthoryear{Johnson, Lindenstrauss, et~al\mbox{.}}{Johnson et~al\mbox{.}}{1984}]%
        {johnson1984extensions}
\bibfield{author}{\bibinfo{person}{William~B Johnson}, \bibinfo{person}{Joram Lindenstrauss}, {et~al\mbox{.}}} \bibinfo{year}{1984}\natexlab{}.
\newblock \showarticletitle{Extensions of Lipschitz mappings into a Hilbert space}.
\newblock \bibinfo{journal}{\emph{Contemporary mathematics}} \bibinfo{volume}{26}, \bibinfo{number}{189-206} (\bibinfo{year}{1984}), \bibinfo{pages}{1}.
\newblock


\bibitem[\protect\citeauthoryear{Khattab, Hammoud, and Elsayed}{Khattab et~al\mbox{.}}{2020}]%
        {khattab2020finding}
\bibfield{author}{\bibinfo{person}{Omar Khattab}, \bibinfo{person}{Mohammad Hammoud}, {and} \bibinfo{person}{Tamer Elsayed}.} \bibinfo{year}{2020}\natexlab{}.
\newblock \showarticletitle{Finding the best of both worlds: Faster and more robust top-k document retrieval}. In \bibinfo{booktitle}{\emph{SIGIR}}. \bibinfo{pages}{1031--1040}.
\newblock


\bibitem[\protect\citeauthoryear{Khattab and Zaharia}{Khattab and Zaharia}{2020}]%
        {khattab2020colbert}
\bibfield{author}{\bibinfo{person}{Omar Khattab} {and} \bibinfo{person}{Matei Zaharia}.} \bibinfo{year}{2020}\natexlab{}.
\newblock \showarticletitle{Colbert: Efficient and effective passage search via contextualized late interaction over bert}. In \bibinfo{booktitle}{\emph{SIGIR}}. \bibinfo{pages}{39--48}.
\newblock


\bibitem[\protect\citeauthoryear{Kong, Dudek, Li, Zhang, and Bendersky}{Kong et~al\mbox{.}}{2023}]%
        {kong2023sparseembed}
\bibfield{author}{\bibinfo{person}{Weize Kong}, \bibinfo{person}{Jeffrey~M Dudek}, \bibinfo{person}{Cheng Li}, \bibinfo{person}{Mingyang Zhang}, {and} \bibinfo{person}{Michael Bendersky}.} \bibinfo{year}{2023}\natexlab{}.
\newblock \showarticletitle{Sparseembed: Learning sparse lexical representations with contextual embeddings for retrieval}. In \bibinfo{booktitle}{\emph{SIGIR}}. \bibinfo{pages}{2399--2403}.
\newblock


\bibitem[\protect\citeauthoryear{Li, Feng, Wang, and Zhou}{Li et~al\mbox{.}}{2008}]%
        {li2008effective}
\bibfield{author}{\bibinfo{person}{Guoliang Li}, \bibinfo{person}{Jianhua Feng}, \bibinfo{person}{Jianyong Wang}, {and} \bibinfo{person}{Lizhu Zhou}.} \bibinfo{year}{2008}\natexlab{}.
\newblock \showarticletitle{An effective and versatile keyword search engine on heterogenous data sources}.
\newblock \bibinfo{journal}{\emph{PVLDB}} \bibinfo{volume}{1}, \bibinfo{number}{2} (\bibinfo{year}{2008}), \bibinfo{pages}{1452--1455}.
\newblock


\bibitem[\protect\citeauthoryear{Li, Liu, Gui, Chen, Ni, Wang, and Chen}{Li et~al\mbox{.}}{2018}]%
        {li2018design}
\bibfield{author}{\bibinfo{person}{Jie Li}, \bibinfo{person}{Haifeng Liu}, \bibinfo{person}{Chuanghua Gui}, \bibinfo{person}{Jianyu Chen}, \bibinfo{person}{Zhenyuan Ni}, \bibinfo{person}{Ning Wang}, {and} \bibinfo{person}{Yuan Chen}.} \bibinfo{year}{2018}\natexlab{}.
\newblock \showarticletitle{The design and implementation of a real time visual search system on JD E-commerce platform}. In \bibinfo{booktitle}{\emph{Middleware}}. \bibinfo{pages}{9--16}.
\newblock


\bibitem[\protect\citeauthoryear{Li, Chen, Yuan, Fu, Shen, Yang, Wang, Ai, Zhang, Wen, et~al\mbox{.}}{Li et~al\mbox{.}}{2025a}]%
        {li2025neutronrag}
\bibfield{author}{\bibinfo{person}{Peizheng Li}, \bibinfo{person}{Chaoyi Chen}, \bibinfo{person}{Hao Yuan}, \bibinfo{person}{Zhenbo Fu}, \bibinfo{person}{Hang Shen}, \bibinfo{person}{Xinbo Yang}, \bibinfo{person}{Qiange Wang}, \bibinfo{person}{Xin Ai}, \bibinfo{person}{Yanfeng Zhang}, \bibinfo{person}{Yingyou Wen}, {et~al\mbox{.}}} \bibinfo{year}{2025}\natexlab{a}.
\newblock \showarticletitle{NeutronRAG: Towards understanding the effectiveness of RAG from a data retrieval perspective}. In \bibinfo{booktitle}{\emph{SIGMOD}}. \bibinfo{pages}{163--166}.
\newblock


\bibitem[\protect\citeauthoryear{Li, Liu, and Zhu}{Li et~al\mbox{.}}{2014}]%
        {li2014enterprise}
\bibfield{author}{\bibinfo{person}{Yunyao Li}, \bibinfo{person}{Ziyang Liu}, {and} \bibinfo{person}{Huaiyu Zhu}.} \bibinfo{year}{2014}\natexlab{}.
\newblock \showarticletitle{Enterprise search in the big data era: Recent developments and open challenges}.
\newblock \bibinfo{journal}{\emph{PVLDB}} \bibinfo{volume}{7}, \bibinfo{number}{13} (\bibinfo{year}{2014}), \bibinfo{pages}{1717--1718}.
\newblock


\bibitem[\protect\citeauthoryear{Li, Ke, Zhu, Yu, Zheng, and Gao}{Li et~al\mbox{.}}{2025b}]%
        {li2025scalable}
\bibfield{author}{\bibinfo{person}{Zhonggen Li}, \bibinfo{person}{Xiangyu Ke}, \bibinfo{person}{Yifan Zhu}, \bibinfo{person}{Bocheng Yu}, \bibinfo{person}{Baihua Zheng}, {and} \bibinfo{person}{Yunjun Gao}.} \bibinfo{year}{2025}\natexlab{b}.
\newblock \showarticletitle{Scalable graph indexing using GPUs for approximate nearest neighbor search}.
\newblock \bibinfo{journal}{\emph{PACMMOD (SIGMOD)}} \bibinfo{volume}{3}, \bibinfo{number}{6} (\bibinfo{year}{2025}), \bibinfo{pages}{1--27}.
\newblock


\bibitem[\protect\citeauthoryear{Liang, Zhang, Yao, Chen, Song, and Cheng}{Liang et~al\mbox{.}}{2025}]%
        {liang2024unify}
\bibfield{author}{\bibinfo{person}{Anqi Liang}, \bibinfo{person}{Pengcheng Zhang}, \bibinfo{person}{Bin Yao}, \bibinfo{person}{Zhongpu Chen}, \bibinfo{person}{Yitong Song}, {and} \bibinfo{person}{Guangxu Cheng}.} \bibinfo{year}{2025}\natexlab{}.
\newblock \showarticletitle{UNIFY: Unified index for range filtered approximate nearest neighbors search}.
\newblock \bibinfo{journal}{\emph{PVLDB}} \bibinfo{volume}{18}, \bibinfo{number}{4} (\bibinfo{year}{2025}), \bibinfo{pages}{1118--1130}.
\newblock


\bibitem[\protect\citeauthoryear{Liu, Yu, Meng, and Chowdhury}{Liu et~al\mbox{.}}{2006}]%
        {liu2006effective}
\bibfield{author}{\bibinfo{person}{Fang Liu}, \bibinfo{person}{Clement Yu}, \bibinfo{person}{Weiyi Meng}, {and} \bibinfo{person}{Abdur Chowdhury}.} \bibinfo{year}{2006}\natexlab{}.
\newblock \showarticletitle{Effective keyword search in relational databases}. In \bibinfo{booktitle}{\emph{SIGMOD}}. \bibinfo{pages}{563--574}.
\newblock


\bibitem[\protect\citeauthoryear{Liu, Wang, Chen, Li, Xiong, Yu, and Zhang}{Liu et~al\mbox{.}}{2025}]%
        {liu-etal-2025-hoprag}
\bibfield{author}{\bibinfo{person}{Hao Liu}, \bibinfo{person}{Zhengren Wang}, \bibinfo{person}{Xi Chen}, \bibinfo{person}{Zhiyu Li}, \bibinfo{person}{Feiyu Xiong}, \bibinfo{person}{Qinhan Yu}, {and} \bibinfo{person}{Wentao Zhang}.} \bibinfo{year}{2025}\natexlab{}.
\newblock \showarticletitle{HopRAG: Multi-hop reasoning for logic-aware retrieval-augmented generation}. In \bibinfo{booktitle}{\emph{ACL}}. \bibinfo{pages}{1897--1913}.
\newblock


\bibitem[\protect\citeauthoryear{Lu, Wang, Wang, and Kudo}{Lu et~al\mbox{.}}{2020}]%
        {lu2020vhp}
\bibfield{author}{\bibinfo{person}{Kejing Lu}, \bibinfo{person}{Hongya Wang}, \bibinfo{person}{Wei Wang}, {and} \bibinfo{person}{Mineichi Kudo}.} \bibinfo{year}{2020}\natexlab{}.
\newblock \showarticletitle{VHP: Approximate nearest neighbor search via virtual hypersphere partitioning}.
\newblock \bibinfo{journal}{\emph{PVLDB}} \bibinfo{volume}{13}, \bibinfo{number}{9} (\bibinfo{year}{2020}), \bibinfo{pages}{1443--1455}.
\newblock


\bibitem[\protect\citeauthoryear{Lv and Zhai}{Lv and Zhai}{2011}]%
        {lv2011documents}
\bibfield{author}{\bibinfo{person}{Yuanhua Lv} {and} \bibinfo{person}{ChengXiang Zhai}.} \bibinfo{year}{2011}\natexlab{}.
\newblock \showarticletitle{When documents are very long, bm25 fails!}. In \bibinfo{booktitle}{\emph{SIGIR}}. \bibinfo{pages}{1103--1104}.
\newblock


\bibitem[\protect\citeauthoryear{Mala, Gezici, and Giannotti}{Mala et~al\mbox{.}}{2025}]%
        {mala2025hybrid}
\bibfield{author}{\bibinfo{person}{Chandana~Sree Mala}, \bibinfo{person}{Gizem Gezici}, {and} \bibinfo{person}{Fosca Giannotti}.} \bibinfo{year}{2025}\natexlab{}.
\newblock \showarticletitle{Hybrid retrieval for hallucination mitigation in large language models: A comparative analysis}.
\newblock \bibinfo{journal}{\emph{arXiv}} (\bibinfo{year}{2025}).
\newblock


\bibitem[\protect\citeauthoryear{Malkov and Yashunin}{Malkov and Yashunin}{2018}]%
        {malkov2018efficient}
\bibfield{author}{\bibinfo{person}{Yu~A Malkov} {and} \bibinfo{person}{Dmitry~A Yashunin}.} \bibinfo{year}{2018}\natexlab{}.
\newblock \showarticletitle{Efficient and robust approximate nearest neighbor search using hierarchical navigable small world graphs}.
\newblock \bibinfo{journal}{\emph{TPAMI}} \bibinfo{volume}{42}, \bibinfo{number}{4} (\bibinfo{year}{2018}), \bibinfo{pages}{824--836}.
\newblock


\bibitem[\protect\citeauthoryear{Mallia, Ottaviano, Porciani, Tonellotto, and Venturini}{Mallia et~al\mbox{.}}{2017}]%
        {mallia2017faster}
\bibfield{author}{\bibinfo{person}{Antonio Mallia}, \bibinfo{person}{Giuseppe Ottaviano}, \bibinfo{person}{Elia Porciani}, \bibinfo{person}{Nicola Tonellotto}, {and} \bibinfo{person}{Rossano Venturini}.} \bibinfo{year}{2017}\natexlab{}.
\newblock \showarticletitle{Faster BlockMax WAND with variable-sized blocks}. In \bibinfo{booktitle}{\emph{SIGIR}}. \bibinfo{pages}{625--634}.
\newblock


\bibitem[\protect\citeauthoryear{Mallia, Suel, and Tonellotto}{Mallia et~al\mbox{.}}{2024}]%
        {mallia2024faster}
\bibfield{author}{\bibinfo{person}{Antonio Mallia}, \bibinfo{person}{Torsten Suel}, {and} \bibinfo{person}{Nicola Tonellotto}.} \bibinfo{year}{2024}\natexlab{}.
\newblock \showarticletitle{Faster learned sparse retrieval with block-max pruning}. In \bibinfo{booktitle}{\emph{SIGIR}}. \bibinfo{pages}{2411--2415}.
\newblock


\bibitem[\protect\citeauthoryear{Miao, Shi, Zhang, Zhang, Nie, Yang, and Cui}{Miao et~al\mbox{.}}{2022}]%
        {miao2022het}
\bibfield{author}{\bibinfo{person}{Xupeng Miao}, \bibinfo{person}{Yining Shi}, \bibinfo{person}{Hailin Zhang}, \bibinfo{person}{Xin Zhang}, \bibinfo{person}{Xiaonan Nie}, \bibinfo{person}{Zhi Yang}, {and} \bibinfo{person}{Bin Cui}.} \bibinfo{year}{2022}\natexlab{}.
\newblock \showarticletitle{HET-GMP: A graph-based system approach to scaling large embedding model training}. In \bibinfo{booktitle}{\emph{SIGMOD}}. \bibinfo{pages}{470--480}.
\newblock


\bibitem[\protect\citeauthoryear{Nguyen, Rosenberg, Song, Gao, Tiwary, Majumder, and Deng}{Nguyen et~al\mbox{.}}{2016}]%
        {nguyen2016ms}
\bibfield{author}{\bibinfo{person}{Tri Nguyen}, \bibinfo{person}{Mir Rosenberg}, \bibinfo{person}{Xia Song}, \bibinfo{person}{Jianfeng Gao}, \bibinfo{person}{Saurabh Tiwary}, \bibinfo{person}{Rangan Majumder}, {and} \bibinfo{person}{Li Deng}.} \bibinfo{year}{2016}\natexlab{}.
\newblock \showarticletitle{Ms marco: A human-generated machine reading comprehension dataset}.
\newblock \bibinfo{journal}{\emph{arXiv}} (\bibinfo{year}{2016}).
\newblock


\bibitem[\protect\citeauthoryear{Ootomo, Naruse, Nolet, Wang, Feher, and Wang}{Ootomo et~al\mbox{.}}{2024}]%
        {ootomo2024cagra}
\bibfield{author}{\bibinfo{person}{Hiroyuki Ootomo}, \bibinfo{person}{Akira Naruse}, \bibinfo{person}{Corey Nolet}, \bibinfo{person}{Ray Wang}, \bibinfo{person}{Tamas Feher}, {and} \bibinfo{person}{Yong Wang}.} \bibinfo{year}{2024}\natexlab{}.
\newblock \showarticletitle{Cagra: Highly parallel graph construction and approximate nearest neighbor search for gpus}. In \bibinfo{booktitle}{\emph{ICDE}}. \bibinfo{pages}{4236--4247}.
\newblock


\bibitem[\protect\citeauthoryear{Ouyang, Hong, Zhao, Zhou, Wu, Lv, and Chen}{Ouyang et~al\mbox{.}}{2025}]%
        {ouyang2025adarag}
\bibfield{author}{\bibinfo{person}{Tao Ouyang}, \bibinfo{person}{Guihang Hong}, \bibinfo{person}{Kongyange Zhao}, \bibinfo{person}{Zhi Zhou}, \bibinfo{person}{Weigang Wu}, \bibinfo{person}{Zhaobiao Lv}, {and} \bibinfo{person}{Xu Chen}.} \bibinfo{year}{2025}\natexlab{}.
\newblock \showarticletitle{AdaRAG: Adaptive optimization for retrieval augmented generation with multilevel retrievers at the edge}. In \bibinfo{booktitle}{\emph{INFOCOM}}. \bibinfo{pages}{1--10}.
\newblock


\bibitem[\protect\citeauthoryear{Paik}{Paik}{2013}]%
        {paik2013novel}
\bibfield{author}{\bibinfo{person}{Jiaul~H Paik}.} \bibinfo{year}{2013}\natexlab{}.
\newblock \showarticletitle{A novel TF-IDF weighting scheme for effective ranking}. In \bibinfo{booktitle}{\emph{SIGIR}}. \bibinfo{pages}{343--352}.
\newblock


\bibitem[\protect\citeauthoryear{Pan, Wang, and Li}{Pan et~al\mbox{.}}{2024}]%
        {pan2024survey}
\bibfield{author}{\bibinfo{person}{James~Jie Pan}, \bibinfo{person}{Jianguo Wang}, {and} \bibinfo{person}{Guoliang Li}.} \bibinfo{year}{2024}\natexlab{}.
\newblock \showarticletitle{Survey of vector database management systems}.
\newblock \bibinfo{journal}{\emph{VLDBJ}} \bibinfo{volume}{33}, \bibinfo{number}{5} (\bibinfo{year}{2024}), \bibinfo{pages}{1591--1615}.
\newblock


\bibitem[\protect\citeauthoryear{Parchas, Naamad, Van~Bouwel, Faloutsos, and Petropoulos}{Parchas et~al\mbox{.}}{2020}]%
        {parchas2020fast}
\bibfield{author}{\bibinfo{person}{Panos Parchas}, \bibinfo{person}{Yonatan Naamad}, \bibinfo{person}{Peter Van~Bouwel}, \bibinfo{person}{Christos Faloutsos}, {and} \bibinfo{person}{Michalis Petropoulos}.} \bibinfo{year}{2020}\natexlab{}.
\newblock \showarticletitle{Fast and effective distribution-key recommendation for amazon redshift}.
\newblock \bibinfo{journal}{\emph{PVLDB}} \bibinfo{volume}{13}, \bibinfo{number}{12} (\bibinfo{year}{2020}), \bibinfo{pages}{2411--2423}.
\newblock


\bibitem[\protect\citeauthoryear{Patel, Kraft, Guestrin, and Zaharia}{Patel et~al\mbox{.}}{2024}]%
        {patel2024acorn}
\bibfield{author}{\bibinfo{person}{Liana Patel}, \bibinfo{person}{Peter Kraft}, \bibinfo{person}{Carlos Guestrin}, {and} \bibinfo{person}{Matei Zaharia}.} \bibinfo{year}{2024}\natexlab{}.
\newblock \showarticletitle{Acorn: Performant and predicate-agnostic search over vector embeddings and structured data}.
\newblock \bibinfo{journal}{\emph{PACMMOD (SIGMOD)}} \bibinfo{volume}{2}, \bibinfo{number}{3} (\bibinfo{year}{2024}), \bibinfo{pages}{1--27}.
\newblock


\bibitem[\protect\citeauthoryear{Peng, Choi, Chan, Yang, and Xu}{Peng et~al\mbox{.}}{2023}]%
        {peng2023efficient}
\bibfield{author}{\bibinfo{person}{Yun Peng}, \bibinfo{person}{Byron Choi}, \bibinfo{person}{Tsz~Nam Chan}, \bibinfo{person}{Jianye Yang}, {and} \bibinfo{person}{Jianliang Xu}.} \bibinfo{year}{2023}\natexlab{}.
\newblock \showarticletitle{Efficient approximate nearest neighbor search in multi-dimensional databases}.
\newblock \bibinfo{journal}{\emph{PACMMOD (SIGMOD)}} \bibinfo{volume}{1}, \bibinfo{number}{1} (\bibinfo{year}{2023}), \bibinfo{pages}{1--27}.
\newblock


\bibitem[\protect\citeauthoryear{Robertson}{Robertson}{2025}]%
        {robertson2025bm25}
\bibfield{author}{\bibinfo{person}{Stephen Robertson}.} \bibinfo{year}{2025}\natexlab{}.
\newblock \showarticletitle{BM25 and all that--a look back}. In \bibinfo{booktitle}{\emph{SIGIR}}. \bibinfo{pages}{5--8}.
\newblock


\bibitem[\protect\citeauthoryear{Sawarkar, Mangal, and Solanki}{Sawarkar et~al\mbox{.}}{2024}]%
        {sawarkar2024blended}
\bibfield{author}{\bibinfo{person}{Kunal Sawarkar}, \bibinfo{person}{Abhilasha Mangal}, {and} \bibinfo{person}{Shivam~Raj Solanki}.} \bibinfo{year}{2024}\natexlab{}.
\newblock \showarticletitle{Blended rag: Improving rag (retriever-augmented generation) accuracy with semantic search and hybrid query-based retrievers}. In \bibinfo{booktitle}{\emph{MIPR}}. \bibinfo{pages}{155--161}.
\newblock


\bibitem[\protect\citeauthoryear{Sun, Sun, Luo, and He}{Sun et~al\mbox{.}}{2022}]%
        {sun2022depth}
\bibfield{author}{\bibinfo{person}{Xibo Sun}, \bibinfo{person}{Shixuan Sun}, \bibinfo{person}{Qiong Luo}, {and} \bibinfo{person}{Bingsheng He}.} \bibinfo{year}{2022}\natexlab{}.
\newblock \showarticletitle{An in-depth study of continuous subgraph matching}.
\newblock \bibinfo{journal}{\emph{PVLDB}} \bibinfo{volume}{15}, \bibinfo{number}{7} (\bibinfo{year}{2022}), \bibinfo{pages}{1403--1416}.
\newblock


\bibitem[\protect\citeauthoryear{Tan, Zhou, Xu, and Li}{Tan et~al\mbox{.}}{2019}]%
        {tan2019efficient}
\bibfield{author}{\bibinfo{person}{Shulong Tan}, \bibinfo{person}{Zhixin Zhou}, \bibinfo{person}{Zhaozhuo Xu}, {and} \bibinfo{person}{Ping Li}.} \bibinfo{year}{2019}\natexlab{}.
\newblock \showarticletitle{On efficient retrieval of top similarity vectors}. In \bibinfo{booktitle}{\emph{EMNLP}}. \bibinfo{pages}{5236--5246}.
\newblock


\bibitem[\protect\citeauthoryear{Voruganti and {\"O}zsu}{Voruganti and {\"O}zsu}{2025}]%
        {voruganti2025mirage}
\bibfield{author}{\bibinfo{person}{Sairaj Voruganti} {and} \bibinfo{person}{M~Tamer {\"O}zsu}.} \bibinfo{year}{2025}\natexlab{}.
\newblock \showarticletitle{MIRAGE-ANNS: Mixed approach graph-based indexing for approximate nearest neighbor search}.
\newblock \bibinfo{journal}{\emph{PACMMOD (SIGMOD)}} \bibinfo{volume}{3}, \bibinfo{number}{3} (\bibinfo{year}{2025}), \bibinfo{pages}{1--27}.
\newblock


\bibitem[\protect\citeauthoryear{Wang, Zhao, Zeng, and Yang}{Wang et~al\mbox{.}}{2021c}]%
        {wang2021fast}
\bibfield{author}{\bibinfo{person}{Hui Wang}, \bibinfo{person}{Wan-Lei Zhao}, \bibinfo{person}{Xiangxiang Zeng}, {and} \bibinfo{person}{Jianye Yang}.} \bibinfo{year}{2021}\natexlab{c}.
\newblock \showarticletitle{Fast k-nn graph construction by gpu based nn-descent}. In \bibinfo{booktitle}{\emph{CIKM}}. \bibinfo{pages}{1929--1938}.
\newblock


\bibitem[\protect\citeauthoryear{Wang, Lin, Papakonstantinou, and Swanson}{Wang et~al\mbox{.}}{2017a}]%
        {wang2017experimental}
\bibfield{author}{\bibinfo{person}{Jianguo Wang}, \bibinfo{person}{Chunbin Lin}, \bibinfo{person}{Yannis Papakonstantinou}, {and} \bibinfo{person}{Steven Swanson}.} \bibinfo{year}{2017}\natexlab{a}.
\newblock \showarticletitle{An experimental study of bitmap compression vs. inverted list compression}. In \bibinfo{booktitle}{\emph{SIGMOD}}. \bibinfo{pages}{993--1008}.
\newblock


\bibitem[\protect\citeauthoryear{Wang, Yi, Guo, Jin, Xu, Li, Wang, Guo, Li, Xu, et~al\mbox{.}}{Wang et~al\mbox{.}}{2021b}]%
        {wang2021milvus}
\bibfield{author}{\bibinfo{person}{Jianguo Wang}, \bibinfo{person}{Xiaomeng Yi}, \bibinfo{person}{Rentong Guo}, \bibinfo{person}{Hai Jin}, \bibinfo{person}{Peng Xu}, \bibinfo{person}{Shengjun Li}, \bibinfo{person}{Xiangyu Wang}, \bibinfo{person}{Xiangzhou Guo}, \bibinfo{person}{Chengming Li}, \bibinfo{person}{Xiaohai Xu}, {et~al\mbox{.}}} \bibinfo{year}{2021}\natexlab{b}.
\newblock \showarticletitle{Milvus: A purpose-built vector data management system}. In \bibinfo{booktitle}{\emph{SIGMOD}}. \bibinfo{pages}{2614--2627}.
\newblock


\bibitem[\protect\citeauthoryear{Wang, Ke, Xu, Chen, Gao, Huang, and Zhu}{Wang et~al\mbox{.}}{2024a}]%
        {wang2024must}
\bibfield{author}{\bibinfo{person}{Mengzhao Wang}, \bibinfo{person}{Xiangyu Ke}, \bibinfo{person}{Xiaoliang Xu}, \bibinfo{person}{Lu Chen}, \bibinfo{person}{Yunjun Gao}, \bibinfo{person}{Pinpin Huang}, {and} \bibinfo{person}{Runkai Zhu}.} \bibinfo{year}{2024}\natexlab{a}.
\newblock \showarticletitle{Must: An effective and scalable framework for multimodal search of target modality}. In \bibinfo{booktitle}{\emph{ICDE}}. \bibinfo{pages}{4747--4759}.
\newblock


\bibitem[\protect\citeauthoryear{Wang, Tan, Gao, Jin, Zhang, Ke, Xu, and Zhu}{Wang et~al\mbox{.}}{2025}]%
        {wang2025balancing}
\bibfield{author}{\bibinfo{person}{Mengzhao Wang}, \bibinfo{person}{Boyu Tan}, \bibinfo{person}{Yunjun Gao}, \bibinfo{person}{Hai Jin}, \bibinfo{person}{Yingfeng Zhang}, \bibinfo{person}{Xiangyu Ke}, \bibinfo{person}{Xiaoliang Xu}, {and} \bibinfo{person}{Yifan Zhu}.} \bibinfo{year}{2025}\natexlab{}.
\newblock \showarticletitle{Balancing the blend: An experimental analysis of trade-offs in hybrid search}.
\newblock \bibinfo{journal}{\emph{arXiv}} (\bibinfo{year}{2025}).
\newblock


\bibitem[\protect\citeauthoryear{Wang, Xu, Yi, Wu, Peng, Ke, Gao, Xu, Guo, and Xie}{Wang et~al\mbox{.}}{2024c}]%
        {wang2024starling}
\bibfield{author}{\bibinfo{person}{Mengzhao Wang}, \bibinfo{person}{Weizhi Xu}, \bibinfo{person}{Xiaomeng Yi}, \bibinfo{person}{Songlin Wu}, \bibinfo{person}{Zhangyang Peng}, \bibinfo{person}{Xiangyu Ke}, \bibinfo{person}{Yunjun Gao}, \bibinfo{person}{Xiaoliang Xu}, \bibinfo{person}{Rentong Guo}, {and} \bibinfo{person}{Charles Xie}.} \bibinfo{year}{2024}\natexlab{c}.
\newblock \showarticletitle{Starling: An i/o-efficient disk-resident graph index framework for high-dimensional vector similarity search on data segment}.
\newblock \bibinfo{journal}{\emph{PACMMOD (SIGMOD)}} \bibinfo{volume}{2}, \bibinfo{number}{1} (\bibinfo{year}{2024}), \bibinfo{pages}{1--27}.
\newblock


\bibitem[\protect\citeauthoryear{Wang, Xu, Yue, and Wang}{Wang et~al\mbox{.}}{2021a}]%
        {wang2021comprehensive}
\bibfield{author}{\bibinfo{person}{Mengzhao Wang}, \bibinfo{person}{Xiaoliang Xu}, \bibinfo{person}{Qiang Yue}, {and} \bibinfo{person}{Yuxiang Wang}.} \bibinfo{year}{2021}\natexlab{a}.
\newblock \showarticletitle{A comprehensive survey and experimental comparison of graph-based approximate nearest neighbor search}.
\newblock \bibinfo{journal}{\emph{PVLDB}} \bibinfo{volume}{14}, \bibinfo{number}{11} (\bibinfo{year}{2021}), \bibinfo{pages}{1964--1978}.
\newblock


\bibitem[\protect\citeauthoryear{Wang, Mao, Wang, and Guo}{Wang et~al\mbox{.}}{2017b}]%
        {wang2017knowledge}
\bibfield{author}{\bibinfo{person}{Quan Wang}, \bibinfo{person}{Zhendong Mao}, \bibinfo{person}{Bin Wang}, {and} \bibinfo{person}{Li Guo}.} \bibinfo{year}{2017}\natexlab{b}.
\newblock \showarticletitle{Knowledge graph embedding: A survey of approaches and applications}.
\newblock \bibinfo{journal}{\emph{TKDE}} \bibinfo{volume}{29}, \bibinfo{number}{12} (\bibinfo{year}{2017}), \bibinfo{pages}{2724--2743}.
\newblock


\bibitem[\protect\citeauthoryear{Wang, Ren, Li, Zhao, Liu, and Wen}{Wang et~al\mbox{.}}{2024b}]%
        {wang-etal-2024-rear}
\bibfield{author}{\bibinfo{person}{Yuhao Wang}, \bibinfo{person}{Ruiyang Ren}, \bibinfo{person}{Junyi Li}, \bibinfo{person}{Xin Zhao}, \bibinfo{person}{Jing Liu}, {and} \bibinfo{person}{Ji-Rong Wen}.} \bibinfo{year}{2024}\natexlab{b}.
\newblock \showarticletitle{{REAR}: A relevance-aware retrieval-augmented framework for open-domain question answering}. In \bibinfo{booktitle}{\emph{EMNLP}}. \bibinfo{pages}{5613--5626}.
\newblock


\bibitem[\protect\citeauthoryear{Wang, Wang, Li, He, and Liu}{Wang et~al\mbox{.}}{2013}]%
        {wang2013theoretical}
\bibfield{author}{\bibinfo{person}{Yining Wang}, \bibinfo{person}{Liwei Wang}, \bibinfo{person}{Yuanzhi Li}, \bibinfo{person}{Di He}, {and} \bibinfo{person}{Tie-Yan Liu}.} \bibinfo{year}{2013}\natexlab{}.
\newblock \showarticletitle{A theoretical analysis of NDCG type ranking measures}. In \bibinfo{booktitle}{\emph{PMLR}}. \bibinfo{pages}{25--54}.
\newblock


\bibitem[\protect\citeauthoryear{Weller, Boratko, Naim, and Lee}{Weller et~al\mbox{.}}{2025}]%
        {weller2025theoretical}
\bibfield{author}{\bibinfo{person}{Orion Weller}, \bibinfo{person}{Michael Boratko}, \bibinfo{person}{Iftekhar Naim}, {and} \bibinfo{person}{Jinhyuk Lee}.} \bibinfo{year}{2025}\natexlab{}.
\newblock \showarticletitle{On the theoretical limitations of embedding-based retrieval}.
\newblock \bibinfo{journal}{\emph{arXiv}} (\bibinfo{year}{2025}).
\newblock


\bibitem[\protect\citeauthoryear{Wu, Luk, Wong, and Kwok}{Wu et~al\mbox{.}}{2008}]%
        {wu2008interpreting}
\bibfield{author}{\bibinfo{person}{Ho~Chung Wu}, \bibinfo{person}{Robert Wing~Pong Luk}, \bibinfo{person}{Kam~Fai Wong}, {and} \bibinfo{person}{Kui~Lam Kwok}.} \bibinfo{year}{2008}\natexlab{}.
\newblock \showarticletitle{Interpreting TF-IDF term weights as making relevance decisions}.
\newblock \bibinfo{journal}{\emph{TOIS}} \bibinfo{volume}{26}, \bibinfo{number}{3} (\bibinfo{year}{2008}), \bibinfo{pages}{1--37}.
\newblock


\bibitem[\protect\citeauthoryear{Xu, Gao, Gou, Long, and Jensen}{Xu et~al\mbox{.}}{2024}]%
        {xu2024irangegraph}
\bibfield{author}{\bibinfo{person}{Yuexuan Xu}, \bibinfo{person}{Jianyang Gao}, \bibinfo{person}{Yutong Gou}, \bibinfo{person}{Cheng Long}, {and} \bibinfo{person}{Christian~S Jensen}.} \bibinfo{year}{2024}\natexlab{}.
\newblock \showarticletitle{irangegraph: Improvising range-dedicated graphs for range-filtering nearest neighbor search}.
\newblock \bibinfo{journal}{\emph{PACMMOD (SIGMOD)}} \bibinfo{volume}{2}, \bibinfo{number}{6} (\bibinfo{year}{2024}), \bibinfo{pages}{1--26}.
\newblock


\bibitem[\protect\citeauthoryear{Yahav, Shehory, and Schwartz}{Yahav et~al\mbox{.}}{2018}]%
        {yahav2018comments}
\bibfield{author}{\bibinfo{person}{Inbal Yahav}, \bibinfo{person}{Onn Shehory}, {and} \bibinfo{person}{David Schwartz}.} \bibinfo{year}{2018}\natexlab{}.
\newblock \showarticletitle{Comments mining with TF-IDF: the inherent bias and its removal}.
\newblock \bibinfo{journal}{\emph{TKDE}} \bibinfo{volume}{31}, \bibinfo{number}{3} (\bibinfo{year}{2018}), \bibinfo{pages}{437--450}.
\newblock


\bibitem[\protect\citeauthoryear{Yang, Li, Yang, Zhang, Hui, Zheng, Yu, Gao, Huang, Lv, et~al\mbox{.}}{Yang et~al\mbox{.}}{2025a}]%
        {yang2025qwen3}
\bibfield{author}{\bibinfo{person}{An Yang}, \bibinfo{person}{Anfeng Li}, \bibinfo{person}{Baosong Yang}, \bibinfo{person}{Beichen Zhang}, \bibinfo{person}{Binyuan Hui}, \bibinfo{person}{Bo Zheng}, \bibinfo{person}{Bowen Yu}, \bibinfo{person}{Chang Gao}, \bibinfo{person}{Chengen Huang}, \bibinfo{person}{Chenxu Lv}, {et~al\mbox{.}}} \bibinfo{year}{2025}\natexlab{a}.
\newblock \showarticletitle{Qwen3 technical report}.
\newblock \bibinfo{journal}{\emph{arXiv}} (\bibinfo{year}{2025}).
\newblock


\bibitem[\protect\citeauthoryear{Yang, Dai, Hou, Zhao, Xu, Song, and Zhu}{Yang et~al\mbox{.}}{2024}]%
        {yang2024revisitingkdd}
\bibfield{author}{\bibinfo{person}{Chen Yang}, \bibinfo{person}{Sunhao Dai}, \bibinfo{person}{Yupeng Hou}, \bibinfo{person}{Wayne~Xin Zhao}, \bibinfo{person}{Jun Xu}, \bibinfo{person}{Yang Song}, {and} \bibinfo{person}{Hengshu Zhu}.} \bibinfo{year}{2024}\natexlab{}.
\newblock \showarticletitle{Revisiting reciprocal recommender systems: Metrics, formulation, and method}. In \bibinfo{booktitle}{\emph{SIGKDD}}. \bibinfo{pages}{3714--3723}.
\newblock


\bibitem[\protect\citeauthoryear{Yang, Yang, Zhao, Gao, Feng, Ouyang, Blum, She, Jiang, Lecue, et~al\mbox{.}}{Yang et~al\mbox{.}}{2025c}]%
        {yang2025graphusion}
\bibfield{author}{\bibinfo{person}{Rui Yang}, \bibinfo{person}{Boming Yang}, \bibinfo{person}{Xinjie Zhao}, \bibinfo{person}{Fan Gao}, \bibinfo{person}{Aosong Feng}, \bibinfo{person}{Sixun Ouyang}, \bibinfo{person}{Moritz Blum}, \bibinfo{person}{Tianwei She}, \bibinfo{person}{Yuang Jiang}, \bibinfo{person}{Freddy Lecue}, {et~al\mbox{.}}} \bibinfo{year}{2025}\natexlab{c}.
\newblock \showarticletitle{Graphusion: A RAG framework for scientific knowledge graph construction with a global perspective}. In \bibinfo{booktitle}{\emph{WWW}}. \bibinfo{pages}{2579--2588}.
\newblock


\bibitem[\protect\citeauthoryear{Yang, Xie, Liu, Yu, Gao, Wang, Peng, and Cui}{Yang et~al\mbox{.}}{2025b}]%
        {yang2024revisiting}
\bibfield{author}{\bibinfo{person}{Shuo Yang}, \bibinfo{person}{Jiadong Xie}, \bibinfo{person}{Yingfan Liu}, \bibinfo{person}{Jeffrey~Xu Yu}, \bibinfo{person}{Xiyue Gao}, \bibinfo{person}{Qianru Wang}, \bibinfo{person}{Yanguo Peng}, {and} \bibinfo{person}{Jiangtao Cui}.} \bibinfo{year}{2025}\natexlab{b}.
\newblock \showarticletitle{Revisiting the index construction of proximity graph-based approximate nearest neighbor search}.
\newblock \bibinfo{journal}{\emph{PVLDB}} \bibinfo{volume}{18}, \bibinfo{number}{6} (\bibinfo{year}{2025}), \bibinfo{pages}{1825--1838}.
\newblock


\bibitem[\protect\citeauthoryear{Yang, Qi, Zhang, Bengio, Cohen, Salakhutdinov, and Manning}{Yang et~al\mbox{.}}{2018}]%
        {yang2018hotpotqa}
\bibfield{author}{\bibinfo{person}{Zhilin Yang}, \bibinfo{person}{Peng Qi}, \bibinfo{person}{Saizheng Zhang}, \bibinfo{person}{Yoshua Bengio}, \bibinfo{person}{William~W Cohen}, \bibinfo{person}{Ruslan Salakhutdinov}, {and} \bibinfo{person}{Christopher~D Manning}.} \bibinfo{year}{2018}\natexlab{}.
\newblock \showarticletitle{HotpotQA: A dataset for diverse, explainable multi-hop question answering}. In \bibinfo{booktitle}{\emph{EMNLP}}. \bibinfo{pages}{2369--2380}.
\newblock


\bibitem[\protect\citeauthoryear{Yu, Huang, Bai, Zhou, and Wu}{Yu et~al\mbox{.}}{2025}]%
        {yu2025aquapipe}
\bibfield{author}{\bibinfo{person}{Runjie Yu}, \bibinfo{person}{Weizhou Huang}, \bibinfo{person}{Shuhan Bai}, \bibinfo{person}{Jian Zhou}, {and} \bibinfo{person}{Fei Wu}.} \bibinfo{year}{2025}\natexlab{}.
\newblock \showarticletitle{AquaPipe: A quality-aware pipeline for knowledge retrieval and large language models}.
\newblock \bibinfo{journal}{\emph{PACMMOD (SIGMOD)}} \bibinfo{volume}{3}, \bibinfo{number}{1} (\bibinfo{year}{2025}), \bibinfo{pages}{1--26}.
\newblock


\bibitem[\protect\citeauthoryear{Yuan, Ma, Wen, Zhang, and Wang}{Yuan et~al\mbox{.}}{2021}]%
        {yuan2021subgraph}
\bibfield{author}{\bibinfo{person}{Ye Yuan}, \bibinfo{person}{Delong Ma}, \bibinfo{person}{Zhenyu Wen}, \bibinfo{person}{Zhiwei Zhang}, {and} \bibinfo{person}{Guoren Wang}.} \bibinfo{year}{2021}\natexlab{}.
\newblock \showarticletitle{Subgraph matching over graph federation}.
\newblock \bibinfo{journal}{\emph{PVLDB}} \bibinfo{volume}{15}, \bibinfo{number}{3} (\bibinfo{year}{2021}), \bibinfo{pages}{437--450}.
\newblock


\bibitem[\protect\citeauthoryear{Zeng, Tong, and Chen}{Zeng et~al\mbox{.}}{2023}]%
        {zeng2023litehst}
\bibfield{author}{\bibinfo{person}{Yuxiang Zeng}, \bibinfo{person}{Yongxin Tong}, {and} \bibinfo{person}{Lei Chen}.} \bibinfo{year}{2023}\natexlab{}.
\newblock \showarticletitle{Litehst: A tree embedding based method for similarity search}.
\newblock \bibinfo{journal}{\emph{PACMMOD (SIGMOD)}} \bibinfo{volume}{1}, \bibinfo{number}{1} (\bibinfo{year}{2023}), \bibinfo{pages}{1--26}.
\newblock


\bibitem[\protect\citeauthoryear{Zhang, Cao, Yan, Madden, and Rundensteiner}{Zhang et~al\mbox{.}}{2020}]%
        {zhang2020continuously}
\bibfield{author}{\bibinfo{person}{Huayi Zhang}, \bibinfo{person}{Lei Cao}, \bibinfo{person}{Yizhou Yan}, \bibinfo{person}{Samuel Madden}, {and} \bibinfo{person}{Elke~A Rundensteiner}.} \bibinfo{year}{2020}\natexlab{}.
\newblock \showarticletitle{Continuously adaptive similarity search}. In \bibinfo{booktitle}{\emph{SIGMOD}}. \bibinfo{pages}{2601--2616}.
\newblock


\bibitem[\protect\citeauthoryear{Zhang, Liu, Zhu, Zeng, Sheng, Yang, Dai, and Wang}{Zhang et~al\mbox{.}}{2024}]%
        {zhang2024efficient}
\bibfield{author}{\bibinfo{person}{Haoyu Zhang}, \bibinfo{person}{Jun Liu}, \bibinfo{person}{Zhenhua Zhu}, \bibinfo{person}{Shulin Zeng}, \bibinfo{person}{Maojia Sheng}, \bibinfo{person}{Tao Yang}, \bibinfo{person}{Guohao Dai}, {and} \bibinfo{person}{Yu Wang}.} \bibinfo{year}{2024}\natexlab{}.
\newblock \showarticletitle{Efficient and effective retrieval of dense-sparse hybrid vectors using graph-based approximate nearest neighbor search}.
\newblock \bibinfo{journal}{\emph{arXiv}} (\bibinfo{year}{2024}).
\newblock


\bibitem[\protect\citeauthoryear{Zheng, Song, Ma, Tan, Ye, Dong, Xiong, Zhang, and Karypis}{Zheng et~al\mbox{.}}{2020}]%
        {zheng2020dgl}
\bibfield{author}{\bibinfo{person}{Da Zheng}, \bibinfo{person}{Xiang Song}, \bibinfo{person}{Chao Ma}, \bibinfo{person}{Zeyuan Tan}, \bibinfo{person}{Zihao Ye}, \bibinfo{person}{Jin Dong}, \bibinfo{person}{Hao Xiong}, \bibinfo{person}{Zheng Zhang}, {and} \bibinfo{person}{George Karypis}.} \bibinfo{year}{2020}\natexlab{}.
\newblock \showarticletitle{Dgl-ke: Training knowledge graph embeddings at scale}. In \bibinfo{booktitle}{\emph{SIGIR}}. \bibinfo{pages}{739--748}.
\newblock


\bibitem[\protect\citeauthoryear{Zhou, Su, Sun, Wang, Wang, He, Zhang, Liang, Liu, Ma, et~al\mbox{.}}{Zhou et~al\mbox{.}}{2026}]%
        {zhou2025depth}
\bibfield{author}{\bibinfo{person}{Yingli Zhou}, \bibinfo{person}{Yaodong Su}, \bibinfo{person}{Youran Sun}, \bibinfo{person}{Shu Wang}, \bibinfo{person}{Taotao Wang}, \bibinfo{person}{Runyuan He}, \bibinfo{person}{Yongwei Zhang}, \bibinfo{person}{Sicong Liang}, \bibinfo{person}{Xilin Liu}, \bibinfo{person}{Yuchi Ma}, {et~al\mbox{.}}} \bibinfo{year}{2026}\natexlab{}.
\newblock \showarticletitle{In-depth analysis of graph-based RAG in a unified framework}.
\newblock \bibinfo{journal}{\emph{PVLDB}}  \bibinfo{volume}{19} (\bibinfo{year}{2026}).
\newblock


\bibitem[\protect\citeauthoryear{Zhu, Chen, Gao, Ma, Zheng, and Zhao}{Zhu et~al\mbox{.}}{2024}]%
        {zhu2024hjg}
\bibfield{author}{\bibinfo{person}{Yifan Zhu}, \bibinfo{person}{Lu Chen}, \bibinfo{person}{Yunjun Gao}, \bibinfo{person}{Ruiyao Ma}, \bibinfo{person}{Baihua Zheng}, {and} \bibinfo{person}{Jingwen Zhao}.} \bibinfo{year}{2024}\natexlab{}.
\newblock \showarticletitle{HJG: An effective hierarchical joint graph for ANNS in multi-metric spaces}. In \bibinfo{booktitle}{\emph{ICDE}}. \bibinfo{pages}{4275--4287}.
\newblock


\end{thebibliography}

\end{document}